\documentclass[twocolumn,trackchanges,tighten]{aastex62}

\usepackage{bm}
\usepackage{multirow}
\usepackage{booktabs}
\usepackage[normalem]{ulem}	
\usepackage{xspace}
\usepackage{amsmath}
\usepackage{tabularx}
\usepackage{lineno}
\usepackage[normalem]{ulem}
\usepackage{blkarray}

\newcommand{\Elat}{\text{\textsc{Elat}}}

\newcommand{\psrplus}[2]{PSR #1+#2\xspace}
\newcommand{\psrminus}[2]{PSR #1$-$#2\xspace}

\defcitealias{NG11a}{NG11a}
\defcitealias{NG12p5data}{NG12p5data}
\defcitealias{NG12p5gwb}{NG12p5gwb}

\graphicspath{{./}{./figures/}}

\submitjournal{ApJ}
\shorttitle{NANOGrav Pulsar Chromatic Noise Characterization}
\shortauthors{The NANOGrav Collaboration}

\begin{document}
\title{The NANOGrav $15$ yr Data Set: Customized Chromatic Noise Models}

\correspondingauthor{Jeremy G. Baier}
\email{baierj@oregonstate.edu}
\correspondingauthor{Bjorn Larsen}
\email{bjorn.larsen@yale.edu}
\author[0000-0001-6436-8216]{Bjorn Larsen}
\affiliation{Department of Physics, Yale University, New Haven, CT 06511, USA}
\author[0000-0002-4972-1525]{Jeremy G. Baier}
\affiliation{Department of Physics, Oregon State University, Corvallis, OR 97331, USA}
\author[0000-0002-7374-6925]{Daniel J. Oliver}
\altaffiliation{NANOGrav Physics Frontiers Center Postdoctoral Fellow}
\affiliation{Department of Physics, Oregon State University, Corvallis, OR 97331, USA}
\author[0000-0001-6630-5198]{Kalista Wayt}
\affiliation{Department of Physics, Oregon State University, Corvallis, OR 97331, USA}
\author[0000-0003-1065-9872]{Yu-Ting Chang}
\affiliation{Department of Astronomy, Yale University, New Haven, CT 06511, USA}
\author[0000-0003-2742-3321]{Jeffrey S. Hazboun}
\affiliation{Department of Physics, Oregon State University, Corvallis, OR 97331, USA}
\author[0000-0002-4307-1322]{Chiara M. F. Mingarelli}
\affiliation{Department of Physics, Yale University, New Haven, CT 06511, USA}
\affiliation{Center for Computational Astrophysics, Flatiron Institute, 162 5th Avenue, New York, NY, 10010, USA}
\author[0000-0003-1407-6607]{Joseph Simon}
\altaffiliation{NSF Astronomy and Astrophysics Postdoctoral Fellow}
\affiliation{Department of Astrophysical and Planetary Sciences, University of Colorado, Boulder, CO 80309, USA}
\author[0000-0002-5455-3474]{Matthew T. Miles}
\affiliation{Department of Physics and Astronomy, Vanderbilt University, 2301 Vanderbilt Place, Nashville, TN 37235, USA}
\author[0000-0001-5134-3925]{Gabriella Agazie}
\affiliation{Center for Gravitation, Cosmology and Astrophysics, Department of Physics and Astronomy, University of Wisconsin-Milwaukee,\\ P.O. Box 413, Milwaukee, WI 53201, USA}
\author[0000-0002-8935-9882]{Akash Anumarlapudi}
\affiliation{Department of Physics and Astronomy, University of North Carolina, Chapel Hill, NC 27599, USA}
\author[0000-0003-0638-3340]{Anne M. Archibald}
\affiliation{Newcastle University, NE1 7RU, UK}
\author[0009-0008-6187-8753]{Zaven Arzoumanian}
\affiliation{X-Ray Astrophysics Laboratory, NASA Goddard Space Flight Center, Code 662, Greenbelt, MD 20771, USA}
\author[0000-0003-2745-753X]{Paul T. Baker}
\affiliation{Department of Physics and Astronomy, Widener University, One University Place, Chester, PA 19013, USA}
\author[0000-0003-3053-6538]{Paul R. Brook}
\affiliation{Institute for Gravitational Wave Astronomy and School of Physics and Astronomy, University of Birmingham, Edgbaston, Birmingham B15 2TT, UK}
\author[0000-0002-6039-692X]{H. Thankful Cromartie}
\affiliation{National Research Council Research Associate, National Academy of Sciences, Washington, DC 20001, USA resident at Naval Research Laboratory, Washington, DC 20375, USA}
\author[0000-0002-1529-5169]{Kathryn Crowter}
\affiliation{Department of Physics and Astronomy, University of British Columbia, 6224 Agricultural Road, Vancouver, BC V6T 1Z1, Canada}
\author[0000-0002-2185-1790]{Megan E. DeCesar}
\altaffiliation{Resident at the Naval Research Laboratory}
\affiliation{Department of Physics and Astronomy, George Mason University, Fairfax, VA 22030, resident at the U.S. Naval Research Laboratory, Washington, DC 20375, USA}
\author[0000-0002-6664-965X]{Paul B. Demorest}
\affiliation{National Radio Astronomy Observatory, 1003 Lopezville Rd., Socorro, NM 87801, USA}
\author[0000-0001-8885-6388]{Timothy Dolch}
\affiliation{Department of Physics, Hillsdale College, 33 E. College Street, Hillsdale, MI 49242, USA}
\affiliation{Eureka Scientific, 2452 Delmer Street, Suite 100, Oakland, CA 94602-3017, USA}
\author[0000-0001-7828-7708]{Elizabeth C. Ferrara}
\affiliation{Department of Astronomy, University of Maryland, College Park, MD 20742, USA}
\affiliation{Center for Research and Exploration in Space Science and Technology, NASA/GSFC, Greenbelt, MD 20771}
\affiliation{NASA Goddard Space Flight Center, Greenbelt, MD 20771, USA}
\author[0000-0001-5645-5336]{William Fiore}
\affiliation{Department of Physics and Astronomy, University of British Columbia, 6224 Agricultural Road, Vancouver, BC V6T 1Z1, Canada}
\author[0000-0001-8384-5049]{Emmanuel Fonseca}
\affiliation{Department of Physics and Astronomy, West Virginia University, P.O. Box 6315, Morgantown, WV 26506, USA}
\affiliation{Center for Gravitational Waves and Cosmology, West Virginia University, Chestnut Ridge Research Building, Morgantown, WV 26505, USA}
\author[0000-0001-7624-4616]{Gabriel E. Freedman}
\affiliation{NASA Goddard Space Flight Center, Greenbelt, MD 20771, USA}
\author[0000-0001-6166-9646]{Nate Garver-Daniels}
\affiliation{Department of Physics and Astronomy, West Virginia University, P.O. Box 6315, Morgantown, WV 26506, USA}
\affiliation{Center for Gravitational Waves and Cosmology, West Virginia University, Chestnut Ridge Research Building, Morgantown, WV 26505, USA}
\author[0000-0001-8158-683X]{Peter A. Gentile}
\affiliation{Department of Physics and Astronomy, West Virginia University, P.O. Box 6315, Morgantown, WV 26506, USA}
\affiliation{Center for Gravitational Waves and Cosmology, West Virginia University, Chestnut Ridge Research Building, Morgantown, WV 26505, USA}
\author[0000-0003-4090-9780]{Joseph Glaser}
\affiliation{Department of Physics and Astronomy, West Virginia University, P.O. Box 6315, Morgantown, WV 26506, USA}
\affiliation{Center for Gravitational Waves and Cosmology, West Virginia University, Chestnut Ridge Research Building, Morgantown, WV 26505, USA}
\author[0000-0003-1884-348X]{Deborah C. Good}
\affiliation{Department of Physics and Astronomy, University of Montana, 32 Campus Drive, Missoula, MT 59812}
\author[0000-0003-1082-2342]{Ross J. Jennings}
\altaffiliation{NANOGrav Physics Frontiers Center Postdoctoral Fellow}
\affiliation{Department of Physics and Astronomy, West Virginia University, P.O. Box 6315, Morgantown, WV 26506, USA}
\affiliation{Center for Gravitational Waves and Cosmology, West Virginia University, Chestnut Ridge Research Building, Morgantown, WV 26505, USA}
\author[0000-0001-6607-3710]{Megan L. Jones}
\affiliation{Center for Gravitation, Cosmology and Astrophysics, Department of Physics and Astronomy, University of Wisconsin-Milwaukee,\\ P.O. Box 413, Milwaukee, WI 53201, USA}
\author[0000-0001-6295-2881]{David L. Kaplan}
\affiliation{Center for Gravitation, Cosmology and Astrophysics, Department of Physics and Astronomy, University of Wisconsin-Milwaukee,\\ P.O. Box 413, Milwaukee, WI 53201, USA}
\author[0000-0002-0893-4073]{Matthew Kerr}
\affiliation{Space Science Division, Naval Research Laboratory, Washington, DC 20375-5352, USA}
\author[0000-0003-0721-651X]{Michael T. Lam}
\affiliation{SETI Institute, 339 N Bernardo Ave Suite 200, Mountain View, CA 94043, USA}
\affiliation{School of Physics and Astronomy, Rochester Institute of Technology, Rochester, NY 14623, USA}
\affiliation{Laboratory for Multiwavelength Astrophysics, Rochester Institute of Technology, Rochester, NY 14623, USA}
\author[0000-0003-1301-966X]{Duncan R. Lorimer}
\affiliation{Department of Physics and Astronomy, West Virginia University, P.O. Box 6315, Morgantown, WV 26506, USA}
\affiliation{Center for Gravitational Waves and Cosmology, West Virginia University, Chestnut Ridge Research Building, Morgantown, WV 26505, USA}
\author[0000-0001-5373-5914]{Jing Luo}
\altaffiliation{Deceased}
\affiliation{Department of Astronomy \& Astrophysics, University of Toronto, 50 Saint George Street, Toronto, ON M5S 3H4, Canada}
\author[0000-0001-5229-7430]{Ryan S. Lynch}
\affiliation{Green Bank Observatory, P.O. Box 2, Green Bank, WV 24944, USA}
\author[0000-0001-5481-7559]{Alexander McEwen}
\affiliation{Center for Gravitation, Cosmology and Astrophysics, Department of Physics and Astronomy, University of Wisconsin-Milwaukee,\\ P.O. Box 413, Milwaukee, WI 53201, USA}
\author[0000-0001-7697-7422]{Maura A. McLaughlin}
\affiliation{Department of Physics and Astronomy, West Virginia University, P.O. Box 6315, Morgantown, WV 26506, USA}
\affiliation{Center for Gravitational Waves and Cosmology, West Virginia University, Chestnut Ridge Research Building, Morgantown, WV 26505, USA}
\author[0000-0002-4642-1260]{Natasha McMann}
\affiliation{Department of Physics and Astronomy, Vanderbilt University, 2301 Vanderbilt Place, Nashville, TN 37235, USA}
\author[0000-0001-8845-1225]{Bradley W. Meyers}
\affiliation{Australian SKA Regional Centre (AusSRC), Curtin University, Bentley, WA 6102, Australia}
\affiliation{International Centre for Radio Astronomy Research (ICRAR), Curtin University, Bentley, WA 6102, Australia}
\author[0000-0002-3616-5160]{Cherry Ng}
\affiliation{Dunlap Institute for Astronomy and Astrophysics, University of Toronto, 50 St. George St., Toronto, ON M5S 3H4, Canada}
\author[0000-0002-6709-2566]{David J. Nice}
\affiliation{Department of Physics, Lafayette College, Easton, PA 18042, USA}
\author[0000-0001-5465-2889]{Timothy T. Pennucci}
\affiliation{Institute of Physics and Astronomy, E\"{o}tv\"{o}s Lor\'{a}nd University, P\'{a}zm\'{a}ny P. s. 1/A, 1117 Budapest, Hungary}
\author[0000-0002-8509-5947]{Benetge B. P. Perera}
\affiliation{Arecibo Observatory, HC3 Box 53995, Arecibo, PR 00612, USA}
\author[0000-0002-8826-1285]{Nihan S. Pol}
\affiliation{Department of Physics, Texas Tech University, Box 41051, Lubbock, TX 79409, USA}
\author[0000-0002-2074-4360]{Henri A. Radovan}
\affiliation{Department of Physics, University of Puerto Rico, Mayag\"{u}ez, PR 00681, USA}
\author[0000-0001-5799-9714]{Scott M. Ransom}
\affiliation{National Radio Astronomy Observatory, 520 Edgemont Road, Charlottesville, VA 22903, USA}
\author[0000-0002-5297-5278]{Paul S. Ray}
\affiliation{Space Science Division, Naval Research Laboratory, Washington, DC 20375-5352, USA}
\author[0000-0003-4391-936X]{Ann Schmiedekamp}
\affiliation{Department of Physics, Penn State Abington, Abington, PA 19001, USA}
\author[0000-0002-1283-2184]{Carl Schmiedekamp}
\affiliation{Department of Physics, Penn State Abington, Abington, PA 19001, USA}
\author[0000-0002-7283-1124]{Brent J. Shapiro-Albert}
\affiliation{Department of Physics and Astronomy, West Virginia University, P.O. Box 6315, Morgantown, WV 26506, USA}
\affiliation{Center for Gravitational Waves and Cosmology, West Virginia University, Chestnut Ridge Research Building, Morgantown, WV 26505, USA}
\affiliation{Giant Army, 915A 17th Ave, Seattle WA 98122}
\author[0000-0001-9784-8670]{Ingrid H. Stairs}
\affiliation{Department of Physics and Astronomy, University of British Columbia, 6224 Agricultural Road, Vancouver, BC V6T 1Z1, Canada}
\author[0000-0002-7261-594X]{Kevin Stovall}
\affiliation{National Radio Astronomy Observatory, 1003 Lopezville Rd., Socorro, NM 87801, USA}
\author[0000-0002-2820-0931]{Abhimanyu Susobhanan}
\affiliation{School of Physics, Indian Institute of Science Education and Research Thiruvananthapuram, Maruthamala PO, Thiruvananthapuram, Kerala 695551, India}
\author[0000-0002-1075-3837]{Joseph K. Swiggum}
\altaffiliation{NANOGrav Physics Frontiers Center Postdoctoral Fellow}
\affiliation{Department of Physics, Lafayette College, Easton, PA 18042, USA}
\author[0000-0001-9678-0299]{Haley M. Wahl}
\affiliation{Department of Physics and Astronomy, West Virginia University, P.O. Box 6315, Morgantown, WV 26506, USA}
\affiliation{Center for Gravitational Waves and Cosmology, West Virginia University, Chestnut Ridge Research Building, Morgantown, WV 26505, USA}




\begin{abstract}
\label{sec:abstract}
    Pulsar timing arrays conduct low-frequency gravitational wave searches, which require comprehensive accounting of various noise sources to achieve robust results. Interstellar propagation effects (e.g., dispersion and scattering) are especially complex noise sources, introducing chromatic delays that can reduce sensitivity to gravitational waves and bias their inference if left unmodeled. These delays also strongly depend on the line of sight properties to each individual pulsar. To address this, we present customized chromatic noise models for 67 pulsars in the NANOGrav 15 yr dataset. These models are selected from an expanded suite of Gaussian processes to simultaneously characterize multiple types of chromatic delays and are tailored to each pulsar’s dataset. Alongside probing the interstellar medium, we use these models to infer the solar wind electron density over the course of $\sim 1.5$ solar cycles. We also find evidence for non-dispersive chromatic delays in 21 out of 67 NANOGrav pulsars. After applying our chromatic models, we observe significant impacts on the inference of achromatic noise in 19 out of 67 pulsars, finding in several cases that a previously significant achromatic noise process can be partially or entirely described as chromatic. These results demonstrate that refined noise modeling is essential to enhance the sensitivity and accuracy of low-frequency gravitational wave searches with pulsar timing arrays.
\end{abstract}

\section{Introduction}
\label{sec:intro}

Recent evidence for a nanohertz gravitational-wave background (GWB) reported by pulsar timing array (PTA) collaborations worldwide has opened a new observational window on the universe \citep{ng15gwb, pptadr3:gwb, eptadr2_3:gwb, MPTA_gwb, cptadr1_1:gwb}. PTAs use ultra-precise pulse times of arrival (TOAs) from millisecond pulsars to probe fluctuations in the space-time interval between the Earth and each pulsar induced by gravitational waves (GWs). In practice, these TOAs also include contributions from a hierarchy of non-GW processes, requiring pulsar noise models to capture their uncertain properties. Robust inference therefore hinges on the fidelity of the noise model: if a noise component is omitted or mis-specified, it may partially absorb or leak into other components and bias parameter estimates, including (in the worst-case scenario) common processes such as the GWB \citep{hazboun+2020model, hazboun:2020slice, zic+2022spurious_correlations, DiMarco2025}.

Two broad classes of non-GW contributions are relevant for PTAs: deterministic and stochastic delays. Deterministic delays typically include effects such as pulsar spin-down, astrometry, relativistic effects due to binary companions, and pulse profile evolution. These effects are together captured by deterministic timing models, which effectively remove sensitivity to GW power at specific frequencies in the GW search context \citep{hazboun:2019sc}. Meanwhile, a variety of pulsar-intrinsic and extrinsic processes introduce stochastic timing delays \citep{ng15detchar, cordes2026fundamental_noise}. Intrinsic variations in pulse shape give rise to jitter, which is mildly chromatic and contributes to each pulsar’s white noise budget \citep{lam2019jitter}, while further rotational and magnetospheric irregularities at the pulsar may give rise to slower timing variations, which are often modeled as achromatic red noise. Extrinsic ``propagation'' effects arise as radio pulses traverse the interstellar medium (ISM) and the interplanetary medium. These effects are typically \emph{chromatic} (or radio-frequency dependent) and as such can be resolved from other noise terms using TOA measurements from multiple bands. Among the propagation effects, dispersion measure (DM) variations are typically dominant. DM, the integrated free-electron column density along the line of sight, introduces a frequency-dependent delay that scales as $\Delta t \propto \nu^{-2}$, where $\nu$ is the observing frequency. DM variations are driven by stochastic ISM structure and by time-dependent line-of-sight geometry, which may include linear and quasi-annual components alongside a dominant component from the solar wind (SW; \citealt{Keith+2013, lcc+2016, jones+2017_ng9_dm, Madison+2019}). Scattering, which results from the refraction and diffraction of the pulses through an inhomogeneous medium, introduces additional chromatic noise through a time-variable $\nu$-dependent broadening of the pulse profile. The frequency scaling is typically steeper (to first order, $\Delta t \propto \nu^{-4}$; \citealt{Cordes1998, LorimerKramer2004, Hemberger2008}), but a much larger range of chromatic dependencies may result, depending on the underlying medium and the intrinsic pulse shape \citep{ShannonCordes2017, Geiger+2024}. These chromatic processes are not merely nuisance terms: they shape the PTA's effective low-frequency sensitivity alongside the timing model \citep{ng15detchar} and can covary with achromatic red processes, especially when the data have irregular cadence, evolving frequency coverage or limited multi-band leverage \citep{sosa+2024dmest, Ferranti+2025}.

The need for careful chromatic noise modeling is not new, but it has become increasingly consequential as PTAs enter a regime where sensitivity to a common signal is comparable to the residual systematics from ISM delays. As a common signal continues to emerge in the PTA band, many studies have simultaneously underscored the requirement of meticulous chromatic models for robust spectral characterization of a common process \citep{ng12p5_cnm, DiMarco2025, Ferranti+2025}. Meanwhile, the currently inferred GWB amplitude is in mild tension with standard expectations for a background generated by supermassive black hole binaries, at roughly the $2$--$4.5\sigma$ level~\citep{SatoPolito2024}. While an astrophysical explanation remains plausible~\citep{Mingarelli2026}, it was shown in \citet{Goncharov2025nature} that the apparent excess power is sourced at least in part by mis-characterized pulsar noise.

One key challenge is that many chromatic models can be simultaneously successful in a narrow sense and incomplete in the ways that matter for GW inference. For instance, the standard DM prescription (DMX) used frequently within NANOGrav \citep{ng15data}, which treats DM as piecewise-constant in time bins, has been shown to capture stochastic DM variations and SW effects effectively in many cases \citep{lentati+:2016, ng15detchar, Larsen+2024, Iraci+2024}. However, higher-order chromatic processes typically have no dedicated channel in such a framework. When present, they can manifest as a mixture of DMX parameters, white noise inflation, and apparent achromatic red noise, which is precisely the combination that can bias GW searches \citep{ng12p5_cnm,DiMarco2025}.


Gaussian processes (GPs) offer an effective alternative by allowing multiple time-correlated stochastic processes with distinct chromaticities to be modeled simultaneously. GPs have been applied extensively for this purpose in PTAs (e.g., \citealt{lentati+:2016, goncharov+2021, Chalumeau+2022, inpta_dr1_noise, pptadr3:noise, Larsen+2024, MPTADR2_noise, ng12p5_cnm, inpta-dr2-noise-2025}). GP-based chromatic models offer multiple practical advantages: they typically occupy a smaller effective prior volume than DMX \citep{gitika2025-mpta-tuning}, they enable straightforward evidence comparisons via Bayes factors \citep{lentati+:2016}, and their hyperparameters are interpretable in ways that can be linked back to line-of-sight propagation physics. For instance with the time-domain models used in this work \citep{ng12p5_cnm}, the prior function on GP covariance (or GP kernel), $k$, is related to the structure function, $D$, via $D(\tau)=2[k(0)-k(\tau)]$, where $\tau$ is the lag between observations \citep{ng12p5_cnm,lcc+15,jones+2017_ng9_dm}. 

The choice of GP kernel also requires extensive consideration of the underlying physics and/or analysis of the data itself \citep{AigrainForeman-Mackey2023}. For example, using a simple power-law for the DMGP prior spectrum is rarely sufficient to jointly describe SW and ISM-induced DM variability \citep{susarla+24sw, DiMarco2025, Iraci+2025, hazboun+2022sw}, which motivates separate model components for ISM-driven and SW-driven variations. In general, the fact that every pulsar has a unique line of sight through the ISM and interplanetary medium (IPM) as well as varying sensitivity motivates a per-pulsar approach to chromatic noise modeling as opposed to a uniform prescription applied across the array. Although some recent studies such as \citet{vanHaasteren2025ma} and \citet{DiMarco2025} have advocated in favor of including all justifiable model components by default (i.e. not customizing the noise model), the process of customizing noise models allows a comprehensive exploration of the space of possible models, and yields valuable insights into the data quality, noise characterization, and physical processes at hand which might be overlooked otherwise. Moreover, our customized noise models are computationally cheaper than including all noise processes. Lastly, we still conservatively include red noise and DM processes which will be directly covariant with the GW signals of interest, which is the primary concern raised by \citet{vanHaasteren2025ma}.

In this work we customize the chromatic noise models for all $67$ pulsars in the NANOGrav 15 yr dataset using a state-of-the art model comparison procedure for PTAs. This procedure includes a comprehensive exploration of both Fourier-basis and time-domain GP models for chromatic noise, including a novel application of time-domain GPs for SW variability. The resulting models provide a thorough characterization of detectable chromatic variations along each pulsar's line of sight. We expect the resulting noise models to improve the robustness of GW inference by reducing chromatic leakage into achromatic red parameters.

This paper is organized as follows. \S\ref{sec:dataset} describes the dataset and preprocessing. \S\ref{sec:methods} presents the Bayesian framework and covariance-matrix construction used to evaluate and compare models, with the individual components of each pulsar's noise model, including the GP components and their chromatic scalings, defined in \S\ref{sec:noise_models}. The model-selection strategy and practical workflow are described in \S\ref{sec:model_selection}. We present the favored chromatic models across all pulsars in \S\ref{sec:results}, and we compare these outcomes directly to a baseline DMX treatment in \S\ref{sec:comparisons-with-dmx}. Finally, we discuss implications for PTA analyses and outline next steps in \S\ref{sec:discussion}, and we summarize our conclusions in \S\ref{sec:conclusions}. {The appendices supplement the main text with detailed parameter definitions, signal model definitions, tables of priors and posteriors, explanations of model selection criteria, and auxiliary solar wind sensitivity analyses.}


\begin{table*}[ht!]
    \centering
    \begin{tabularx}{\linewidth}{c | c | l}
        \hline\hline
        \textbf{Category} & \textbf{Term} & \textbf{Definition and section where introduced} \\ \hline
        Acronyms & PTA; TOA; GW & Pulsar timing array; Time of arrival; Gravitational wave (\S\ref{sec:intro}) \\
        (General) & GBT; AO; VLA & Green Bank Telescope; Arecibo Observatory; Very Large Array (\S\ref{sec:dataset}) \\
        & GASP/GUPPI; YUPPI & Different electronic backends used at the GBT; Backend used at the VLA (\S\ref{sec:dataset}) \\
        & ASP/PUPPI & Different electronic backends used at the AO (\S\ref{sec:dataset}) \\
        & GWB; NG15 & Gravitational wave background (\S\ref{sec:intro}), NANOGrav 15 yr dataset (\S\ref{sec:dataset}) \\
        & GP; PSD & Gaussian process; Power Spectral Density (\S\ref{sec:intro}) \\
        & MCMC; INS  & Markov chain Monte Carlo; Importance Nested Sampling (\S\ref{sec:methods}) \\
        \hline
        Acronyms & DM; FC & Dispersion measure; Free chromatic noise with $\Delta t \propto \nu^{-\chi}$ (\S\ref{sec:intro}) \\
        (Noise & CNM; DMX/DMGP & Custom Noise Model; Piecewise-constant/GP-based model for DM estimation (\S\ref{sec:intro}) \\
        processes) & ISM; IPM; SW & Interstellar medium; Interplanetary medium; Solar wind (\S\ref{sec:intro}) \\
        & RN; FS; ADS & Red Noise; Free Spectral Red Noise; Anderson-Darling Statistic (\S\ref{sec:methods}, \S\ref{sec:achromatic-red-noise-comparison}, \S\ref{sec:whitened-residuals}) \\
        & CRN; IRN; WN & Common Red Noise; Intrinsic Red Noise; White Noise (\S\ref{sec:methods}, \S\ref{sec:achromatic-red-noise-comparison}) \\
        & TD; TD\_RF & Time-domain; Time and Radio-frequency-domain, to model $\nu$-dependent DM (\S\ref{sec:methods}) \\
        & SE/QP/RQ & Squared-Exponential/Quasi-periodic/Rational Quadratic kernels (\S\ref{sec:methods}) \\
        & \textsc{Model\_Fourier} & Labels a full pulsar noise model using Fourier basis for all chromatic GPs (\S\ref{sec:model_selection}) \\
        & \textsc{Model\_TD} & As above, using TD interpolation basis for all chromatic GPs (\S\ref{sec:model_selection}) \\
        & \textsc{Model\_Ridge} & As above, using TD interpolation basis w/ Ridge kernels for all chromatic GPs (\S\ref{sec:model_selection}) \\ \hline
        Symbols & $T_{\rm psr}$/$T_{\rm NG15}$ & Total observation timespan of single pulsar/NG15 dataset \\
        (General) & $\Delta t$, $\vec{\delta t}$, $\rho$ & Time delay, Timing residual vector, Free spectral timing residual amplitude \\
        & $f$; $\nu$ & Spectral (fluctuation) frequency; Radio frequency \\
        & $N_f$/$dt$ & \# of frequencies in GP Fourier basis/days between interpolation nodes in TD basis \\
        & $\mathcal{L}$; $\pi$; $\mathcal{P}$; $\mathcal{Z}$; $\mathcal{M}$ & Likelihood; Prior; Posterior; Evidence; Model \\
        & $\mathcal{B}$; $\mathcal{N}$/$\mathcal{U}$; $p$ & Bayes factor; Normal/Uniform distribution; $p$-value \\
        \hline Symbols & $\vec{b}$/$\vec{\eta}$ & GP parameter/hyperparameter vector \\
        (Model & $A$/$\gamma$; $\chi$ & Spectral amplitude/index; Chromatic scaling index of the TOA with radio frequency \\
        parameters) & $n_{E,i}$/$n_E(t)$ & Local electron number density at Earth's average orbit (binned/time-dependent) \\
        & $\sigma$/$\ell$/$\Gamma_p$/$p$ & SE and QP kernel amplitude/lengthscale/QP weight/quasi-period parameters \\
        & $A_{\rm exp}$/$\tau_{\rm exp}$/$t_{0,\rm exp}$ & Amplitude/timescale/initial time of decaying exponential \\
        \hline\hline
    \end{tabularx}
    \caption{Definitions of acronyms and symbols commonly used throughout this work. Other terms used only once or twice are defined in their specific sections. Acronyms may also be a combination of those listed in this table (such as ``SWGP''). {For priors and units associated with the model parameters listed here, see Table~\ref{tab:priors} in Appendix~\ref{appendix:tables}.}}
    \label{tab:acronyms}
    \vspace{-0.5\baselineskip}
\end{table*}



\section{Data Set}
\label{sec:dataset}
A detailed discussion of the data set can be found in \citet{ng15data}, but for convenience, we reiterate the most noteworthy points here. The NANOGrav 15 yr data set (NG15) consists of observations of 67 millisecond pulsars made between July 2004 and August 2020. Note that the total time between the first time of arrival to the last time of arrival is $16.03$ years, but the dataset is referred to as the ``15 yr Dataset'' since no single pulsar has a time span of $16$ years or longer. $21$ new pulsars have been added to the dataset since the $12.5$ year dataset \citep{alam:2020nb,alam:2020wb}, and $2.9$ years worth of data have been added to all the pulsars that are continually observed into this dataset.

Observations of pulsars were made using the $305$-m Arecibo Observatory (AO), the $27$ $25$-m antennae that make up the Very Large Array (VLA), and the $100$-m Greenbank Telescope (GBT). Data were taken at VLA and GBT until $2020$ April $4$, which were the last data taken by the GUPPI backend at GBT, and AO had data taken until $2020$ August $10$, corresponding to the cable breakage and the date that observations ceased at AO. Pulsar observations happen on roughly a monthly cadence, except for a $5$ day cadence for {6 particular pulsars}. From $2015$ to $2020$, AO was observing PSRs J0030+0451, J1640+2224, J1713+0747, J2043+1711, and J2317+1439 at this high cadence, and since 2013, GBT has performed weekly observations of PSRs J1713+0747 and J1909$-$3744 \citep{alam:2020nb,alam:2020wb}. For the entirety of the dataset, pulsars were typically observed at two widely separated frequencies for each epoch. AO observed pulsars at $430$ MHz and $1.4$ GHz, $1.4$ GHz and $2.1$ GHz, or all three frequencies (additionally, ASP-era observations at the AO were taken at 327 MHz and 430 MHz for PSR J2317+1439 exclusively). The GBT observed all pulsars at $820$ MHz and $1.4$ GHz. Lastly, the VLA supplemented AO and GBT pulsars by additionally observing several pulsars at $3$ GHz. \psrplus{J1713}{0747} was observed by all observatories at $1.4$ GHz to provide additional cadence and cross-check instrument systematics. NG15 can be found on Zenodo at \url{doi:10.5281/zenodo.7967584}. 

In this work, we use the data released in NG15 \citep{ng15data} with the exception of a single, additionally excised observation's worth of TOAs for \psrminus{J0613}{0200}, which is discussed in \S\ref{sec:sw_excision}. We start from the timing solutions presented in \citet{ng15data} as well, but we remove the DMX parameters and instead supplement the modeling of dispersion measure variations in the timing model with linear and quadratic terms for DM variations (\texttt{PINT} parameters \texttt{DM1} and \texttt{DM2}), marginalizing over these as well as the mean DM. See \citet{vanhaasteren+2014_gp} for details on the low frequency fit and \S\ref{sec:methods} for details on the marginalized timing model. 



\section{{The Covariance Matrix} \& Bayesian Methods}
\label{sec:methods}

\subsection{\label{sec:likelihood}Likelihood}

The PTA likelihood, $\mathcal{L}$, is expressed as a multivariate Gaussian distribution, which we construct with the \texttt{enterprise }\citep{enterprise} software suite,
\begin{equation}\label{eqn:likelihood2}
     \mathcal{L} 
     =\frac{1}{\sqrt{\text{det}(2\pi\mathbf{C})}}\text{exp}\left(-\frac{1}{2}\vec{r}^T\mathbf{C}^{-1}\vec{r}\right).
\end{equation}
$\vec{r}=\vec{\delta t}-\vec{s}(\vec{\theta})$, where $\vec{\delta t}$ are the timing model residuals and $s$ is a deterministic delay function parameterized by $\vec{\theta}$.
$\mathbf{C}$ is a covariance matrix that houses the stochastic portions of the PTA likelihood, 
\begin{equation}\label{eq:C}
\mathbf{C} =\mathbf{N} + \mathbf{TBT}^T. 
\end{equation}

$\mathbf{N}$ is the white noise covariance matrix composed of TOA measurement uncertainties, $\sigma_{\rm S/N}$, that get inflated by white noise parameters EFAC ($\mathcal{F}$), EQUAD ($Q$), and ECORR ($\mathcal{J}$), which are applied to each TOA uncertainty according to its receiver and back end combination, re/be,
\begin{equation}\label{eq:wn}
N_{ij}
= \mathcal{F}^2\big(\mathrm{re}/\mathrm{be}\big)\big[\sigma_{S/N,i}^2 + Q^2(\mathrm{re}/\mathrm{be})\big]\delta_{ij}
+ \mathcal{J}^2(\mathrm{re}/\mathrm{be})\,\mathbf{\mathcal{U}}_{ij}.
\end{equation}
$\delta_{ij}$ is the Kronecker delta function and the $\mathbf{\mathcal{U}}$ matrix is block diagonal with each subblock containing entries of 1 for TOAs in the same epoch and zeros elsewhere. This imposes an overall block diagonal structure to $\mathbf{N}$ so TOAs from the same observing epoch have a completely correlated error term (ECORR; \citealt{ng11_gwb}).

$\mathbf{T}$ is a stacked matrix of design matrices,  
\begin{equation}
    \label{eq:T_matrices}
    \textbf{T} = [\textbf{M}, \textbf{F}_{\mathrm{RN}}, \textbf{F}_{\mathrm{DM}}, \textbf{F}_{\mathrm{FC}}, \textbf{F}_{\mathrm{SW}}],
\end{equation}
where the subscripts RN, DM, FC, and SW denote different model components (respectively, red noise, dispersion measure, free chromatic, and solar wind) which are described in \S\ref{sec:noise_models}. $\mathbf{T}$ includes $\mathbf{M}$ and $\mathbf{F}$'s where $\mathbf{M}$ is the linearized timing model design matrix, which is constructed from the first derivative of timing delays with respect to the timing model parameters evaluated at the optimal timing model parameters from \citet{ng15data}, and $\mathbf{F}$ is either a Fourier design matrix or a linear interpolation matrix (see \S\ref{sec:noise_models}).

$\mathbf{B}$ is a block diagonal covariance matrix whose subblocks are individual signal covariance matrices subscripted similarly to Eq.\ref{eq:T_matrices},
\begin{equation}
    \label{eq:B_matrix}
    \textbf{B} = \rm diag(\textbf{E}, \mathbf{\phi}_{\mathrm{RN}}, \mathbf{\phi}_{\mathrm{DM}}, \mathbf{\phi}_{\mathrm{FC}}, \mathbf{\phi}_{\mathrm{SW}}),
\end{equation}
where $\mathbf{E}=\rm diag(\infty) $ is set from the improper  prior placed on the timing model perturbations and the $\mathbf{\phi}$ matrices are covariance matrices which are constructed from parameterized power spectral densities in the case of the Fourier basis models and parameterized kernels in the case of the time-domain models. The definitions and selections of the Fourier and time-domain model components are covered in \S\ref{sec:noise_models} and Appendix~\ref{appendix:GP_kernels}.

Finally, $\mathbf{C}$ is more efficiently inverted with the help of the Woodbury matrix identity \citep{Johnson+2024}, 
\begin{equation}
    \label{eq:Cinv}
    \mathbf{C}^{-1}=\mathbf{N}^{-1} - \mathbf{N}^{-1}\; \mathbf{T} \Big(\;\mathbf{B}^{-1} + \mathbf{T}^{T}\mathbf{N}^{-1}\mathbf{T}\; \Big)^{-1}\mathbf{T}^{T}\mathbf{N}^{-1}.
\end{equation}
Note that this is the one-step marginalization version of PTA likelihood which is more efficient than the two-step marginalization for single pulsar noise analyses since the white-noise matrix cannot be cached in its entirety when sampling in the white noise parameters \citep{Johnson+2024}. For more details on the PTA likelihood, the reader is referred to \citet{taylor2021_nhz_GW_astronomy_book, Johnson+2024, ng15detchar}.

To perform Bayesian inference, we write Bayes' Theorem,
\begin{equation}\label{eqn:posterior}
     \mathcal{P}(\vec{\eta},\vec{\theta}|\vec{\delta t},\mathcal{M}) \propto \mathcal{L}(\vec{\delta t}|\vec{\eta},\vec{\theta},\mathcal{M})\pi(\vec{\eta},\vec{\theta}|\mathcal{M}),
\end{equation}
where $\vec{\eta}$ is a vector of noise model hyperparameters, 
$\vec{\theta}$ are the determinsitic signal parameters, $\mathcal{L}(\vec{\delta t}|\vec{\eta},\vec{\theta},\mathcal{M})$ is the likelihood given the model $\mathcal{M}$, $\pi(\vec{\eta},\vec{\theta}|\mathcal{M})$ is the prior over free parameters, and $\mathcal{P}(\vec{\eta},\vec{\theta}|\vec{\delta t},\mathcal{M})$ is the posterior \citep{taylor2021_nhz_GW_astronomy_book}. We use \texttt{PTMCMCSampler} \citep{ptmcmcsampler} to perform parameter estimation, making direct use of this proportionality. 

\subsection{Model Comparison Methods}
\label{sec:selection_methods}

Alongside parameter estimation, we use importance nested sampling (INS) and the Savage-Dickey density ratio for model comparison. The Savage-Dickey density ratio \citep{Dickey1971} is well suited to test whether a particular model component is statistically significant or not if the larger model, $\mathcal{M}_2$ can be written as a nested model in which $\mathcal{M}_2$ reduces to another model, $\mathcal{M}_1$, at a specific region in parameters $\vec{\eta}$. If $\vec{\eta} = \vec{\eta}_0$ denotes the region of parameter space where the two models are equivalent, then taking the ratio between the prior and posterior density at $\vec{\eta}_0$ yields an approximation to the Bayes factor for the larger model over the reduced model. Mathematically this is expressed as,

\begin{align}
    \label{eq:SDratio}    \mathcal{B}^{\mathcal{M}_2}_{\mathcal{M}_1} = \frac{\pi(\vec{\eta} = \vec{\eta}_0|\mathcal{M}_2)}{\mathcal{P}(\vec{\eta} = \vec{\eta}_0|\vec{\delta t},\mathcal{M}_2)},
\end{align}
where $\pi, \mathcal{P}$ are the prior and posterior, respectively.

On the other hand, both nested sampling \citep{skilling2004NS} and INS \citep{importanceNS} directly compute the Bayesian evidence, $\mathcal{Z}_i$, for a model, $\mathcal{M}_i$, which can be used to compare any two models since the Bayes factor between the models is the ratio of their evidences, 
\begin{align}
    \label{eq:z_ratio}    \mathcal{B}^{\mathcal{M}_1}_{{\mathcal{M}}_2}=\frac{\mathcal{Z}_1}{\mathcal{Z}_2}.
\end{align}
The advantages of nested sampling and INS approaches are that they handle complicated posterior geometries like multi-modalities and one can directly compare any two models whether they are nested or not. The drawbacks are that they struggle with high dimensional parameter spaces $(N_{\rm dim}\gtrapprox 60)$, and require separate analyses for each model. 
In this work, we use \texttt{nautilus}, which employs neural networks for efficient bounding for INS \citep{Lange2023}. See \citet{ashton+22NSreview} for a review of nested sampling in physics. 

We explored a number of model comparison techniques in order to pick the favored chromatic model. We tried to use product space sampling \citep{Hee+2016} and likelihood reweighting \citep{Hourihane+2023} to compare models that are identical other than the basis size of the chromatic noise models; however, we found that the chromatic GP hyperparameter posteriors were so disjoint in many cases that neither technique could be used efficiently or accurately. That is to say, the posterior recovery of some of the noise hyperparameters depends on the chromatic basis size, which complicates model comparison. \citet{ng12p5_cnm} observed this effect, showing how ECORR parameters change between different chromatic basis sizes. We also tried to use thermodynamic integration \citep{gelman+98, littenberg+2009} as well as the stepping stone algorithm \citet{steppingstone2018} both of which require the use of parallel tempering in MCMC to calculate a model's evidence. Ultimately, we found that our desired precision in Bayes factors required many temperatures, which proved computationally prohibitive for the number of models and pulsars we compare in this work.

For a broader review of model comparison in PTA data analysis, see \citet{Johnson+2024}.

\subsection{Chromatic noise models}
\label{sec:noise_models}

The various models we test here are implemented primarily as modifications to the PTA covariance matrix \citep{ng15detchar}. We focus here on testing chromatic models, while leaving the original red and white noise models unchanged, similarly to \citet{ng12p5_cnm}. The various chromatic processes we consider include DM variations from the ISM \citep{jones+2017_ng9_dm}, SW-induced DM variations \citep{Madison+2019, hazboun+2022sw}, higher-order chromatic (e.g. scattering or profile) variations \citep{Lentati+2017}, annual sinusoidal chromatic variations due to orbital motion of the line-of-sight, and $\nu$-dependent DM variations \citep{cordes+2016_dm_f}. The majority of these variations are stochastic and described using Gaussian processes (GPs). For each GP we must determine both a set of \emph{basis functions} and a \emph{prior} on the GP parameters in the form of a kernel or a PSD, which directly models the autocorrelations of the process. See \citet{rw06, AigrainForeman-Mackey2023} for more on GPs in general, as well as \citet{vanhaasteren+2014_gp, ng15detchar, ng12p5_cnm} on their application in PTAs. We occasionally supplement these models with deterministic fits for decidedly nonstationary or transient processes (\S\ref{sec:event_models}), while the SW is described by a mixture of stochastic and deterministic components (\S\ref{sec:solar_wind_models}).

Section 3 of \citet{ng12p5_cnm} describes in detail the various GP models we consider for chromatic noise, which we briefly reintroduce here (see also Appendix~\ref{appendix:GP_kernels} for full mathematical descriptions). As in \citet{ng12p5_cnm}, the chromatic models \emph{add} to the standard red and white noise models used in the NANOGrav 15 year dataset \citep{ng15data, ng15detchar}, which we leave unchanged. As mentioned in \S\ref{sec:dataset}, the \texttt{DMX} parameters of the timing model are replaced with a quadratic DM filter. From there, the three different classes of GPs we consider for chromatic noise include:
\begin{itemize}
    \item \emph{Fourier-domain models:} The Fourier basis (\citealt{Lentati+2013}; {Eq.~\ref{eq:Fourier_basis}}) is represented by sine and cosine terms at linearly-spaced frequencies (starting always at the fundamental, $f_0 = 1/T_{\mathrm{NG15}}$, where $T_{\mathrm{NG15}} = {16.03}$ yr), while the prior on the Fourier coefficients is given by a power-law spectrum {with hyperparameters $A$, $\gamma$}, assuming spectral power is uncorrelated between different frequencies (\citealt{ng15detchar}; {Eq.~\ref{eq:powerlaw}}). The number of frequencies to include in the basis is controlled by the $N_f$ hyperparameter, which also determines the high frequency cutoff $f_{\mathrm{max}} = N_f/T_{\mathrm{NG15}}$. 
    \item \emph{Time-domain (TD) models:} The time-domain models use the linear interpolation basis introduced in \citet{ng12p5_cnm}, {Eq.~\eqref{eq:TD_Basis}}. We consider the various kernels introduced therein, which are the diagonal (Ridge) kernel, the square-exponential (SE) kernel, and the quasi-periodic (QP) kernel, {Eq.~\eqref{eq:QP_kernel}}, which model time-correlations of increasing complexity. {The SE kernel introduces hyperparameters $\sigma$, $\ell$ to control the variational amplitude and lengthscale in the SE kernel, and the QP kernel introduces $\Gamma_p$, $p$ hyperparameters to control the behavior of quasi-periodic oscillations.} The spacing between interpolation nodes is controlled by the $dt$ hyperparameter.
    \item \emph{Time and radio-frequency domain (TD\_RF) models:} This multi-dimensional model from \citet{ng12p5_cnm}, {Eq.~\eqref{eq:RF_band_basis}}, allows DM variations to decorrelate over radio-frequency through a 2D interpolation basis in both time and radio-frequency. The kernel is factorized as the product of a rational-quadratic (RQ) kernel in radio frequency and the TD kernel (Ridge, SE, or QP), {Eq.~\eqref{eq:RQ_kernel}, with hyperparameters $\alpha_{\rm wgt}$, $\ell_2$ to control the amplitude and scale of decorrelations across radio frequency}.
\end{itemize}

To account for chromatic noise, all bases are scaled by a factor $(\nu/1400$ MHz$)^{-\chi}$ unless otherwise noted {(Eq.~\ref{eq:chromatic_basis})}. Setting $\chi = 0$ reduces the noise to an achromatic process. For DM variations, $\chi = 2$. For higher-order chromatic variations, which are most likely (but not necessarily) induced by interstellar scattering variations, we let $\chi$ vary as a free parameter. Throughout this work we refer to this component as a ``Free Chromatic'' (FC) noise. To constrain its behavior, we set $\chi$ to use a uniform prior with a Gaussian tail $\chi \sim $ \textsc{UniformGaussUpperTail}($\chi_{\rm low}=2.5,\chi_{\rm high}=7,\sigma_\chi=1$), such that $\chi \sim \mathcal{U}(2.5,7)$ if $\chi < 7$ and $\chi \sim \mathcal{N}(7,1)$ if $\chi > 7$. Explicitly, the prior probability density function (PDF) is defined
\begin{align}
    &\pi(\chi;\chi_{\rm low},\chi_{\rm high},\sigma_\chi) \nonumber \\
    &= \tfrac{1}{\chi_{\rm high} - \chi_{\rm low} + \sqrt{\frac{\pi}{2}}\sigma_\chi} \begin{cases}
        0 & \chi < \chi_{\rm low}, \\
        1 & \chi_{\rm low} < \chi < \chi_{\rm high}, \\
        e^{-\frac{(\chi-\chi_{\rm high})^2}{2\sigma_\chi^2}} & \chi > \chi_{\rm high}.
    \end{cases}
    \label{eq:chi_prior}
\end{align}
The lower bound $\chi_{\rm low} = 2.5$ is placed to decouple the process from DM variations. Meanwhile, the Gaussian tail above $\chi_{\rm high} = 7$ is motivated as these large $\chi$ values have previously been measured \citep{MPTADR2_noise}, but at the time when we began this work there were a lack of mechanisms proposed to explain chromatic noise above $\chi \sim 6.4$ \citep{ShannonCordes2017}. Nonetheless, it has been shown that the FC index $\chi$ is not necessarily the same as the scaling index on the scattering delay, since the TOA may depend nonlinearly upon the scattering delay \citep{Geiger+2024}. A recent study by \citet{Kulkarni+2025} has also shown that larger $\chi$ values may result as a by-product of model misspecification, e.g. due to non-Gaussianity of scattering delays. 

\begin{table*}
    \begin{center}
        \caption{Types of GP bases/models we consider for chromatic noise. Further detailed descriptions of the basis types, kernels, basis size parameters, and the chromatic processes they are modeling can be found in \S\ref{sec:noise_models}, and further details on our SW model can be found in \S\ref{sec:solar_wind_models}. Equations defining different types of GPs are reviewed in Appendix~\ref{appendix:GP_kernels}.}
        \begin{tabular}{c|cccc}
            \hline\hline Basis Type & Kernels \citep{ng12p5_cnm} & Basis Size Parameter & Chromaticity (Basis scaling) \\
            \hline Fourier & Power Law PSD & $N_f \in [50,100,150,200]$ & DM ($\nu^{-2}$), FC ($\nu^{-\chi}$), SW ($\nu^{-2}$; \S\ref{sec:solar_wind_models}) \\
            TD & SE, QP, Ridge & $dt \in [3,7,15,20,30]$ days & DM ($\nu^{-2}$), FC ($\nu^{-\chi}$), SW ($\nu^{-2}$; \S\ref{sec:solar_wind_models}) \\
            TD\_RF & TD Kernel $\times$ RQ($\nu$) & $dt \in [3,7,15,20,30]$ days & DM ($\nu^{-2}$) \\
            Triangular (\S\ref{sec:solar_wind_models}) & Ridge & $N_{\mathrm{conj}}$ & SW ($\nu^{-2}$; \S\ref{sec:solar_wind_models}) \\
            \hline \hline
        \end{tabular}
        \label{tab:noise_models_table}
    \end{center}
\end{table*}

Alongside the prior on the chromatic index $\chi$, we implement a few additional modifications on these models from \citet{ng12p5_cnm}. Firstly, the number of GP coefficients for each model are determined on a per-pulsar basis following our model selection procedure (cf. \S\ref{sec:model_selection}). Secondly, we marginalize over the phase of a chromatic annual variations model, using a Fourier basis with a single free-spectral bin at $f = 1/\mathrm{yr}$ and amplitude parameter $\rho_{1\mathrm{yr}} \sim \log_{10}\mathcal{U}(10^{-10},10^{-2})\;\rm{s}$ {(cf. Eq.~\ref{eq:annual_basis},~\ref{eq:annual_prior})}, as opposed to a two-parameter deterministic model. Thirdly, in order to avoid incurring a bias on the TOA covariance matrix in certain regions of the kernel parameter space where the kernel variance $\sigma$ is large, we reduce a numerical regularization term along the diagonal of the TD kernels from $(\sigma/500)^2$ to $(\sigma/50000)^2$. We additionally reduce the prior on TD kernel parameters to $\sigma \sim \log_{10}\mathcal{U}(10^{-10},10^{-4.5})$ s and $\ell \sim \log_{10}\mathcal{U}(10,T_{\mathrm{psr}})$ day, where $T_{\mathrm{psr}}$ is the observation timespan of the pulsar (see Table~\ref{tab:priors} for a full list of priors).

The remainder of model settings are the same as those in \citet{ng12p5_cnm}. See Table~\ref{tab:noise_models_table} for a summary of the various types of GP models we consider for chromatic noise, organized by their GP basis and the various options we consider for each, which are selected for each pulsar following the procedures in \S\ref{sec:model_selection}.


\subsection{Solar wind}
\label{sec:solar_wind_models}
Our SW models account for changes in dispersive delays due to the highly time-variable nature of the ionized interplanetary medium. We largely follow the approach used in \citet{hazboun+2022sw} which consists of both a fixed, deterministic component and a stochastic component, the latter of which effectively adds perturbations on top of the base deterministic model. This deterministic plus stochastic perturbations schema was originally introduced into the PTA literature in \citet{hazboun+2022sw}, utilized in \citet{pptadr3:noise, MPTADR2_noise}, and expanded on in \citet{susarla+24sw, Nitu+2024}. 

\subsubsection{Piecewise global deterministic solar wind}
\label{sec:global_deterministic_solar_wind}
The deterministic model we use is adopted from the global schema introduced in \citet{hazboun+2022sw} and used in \citet{ng12p5_cnm}. It is a piecewise constant inference of the solar electron density at $1 \rm AU$ in $180$-day bins across all $67$ pulsars.

The SW DM contribution can written as
\begin{align}
    \rm{DM}_\odot &= n_E\frac{\pi-\theta_i}{r_\oplus\sin\theta_i}(1 \rm AU)^2,
    \label{eq:sw_dm}
\end{align}
where $n_E$ is the solar electron density at Earth, $\theta_i$ is the solar impact angle at the $ith$ observation, and $r_\oplus$ is the distance between the observatory and the Sun. This assumes a spherically symmetric solar electron density which decreases proportional to $1/r^2$ from the Sun. The dispersive time delay due to the IPM relative to an infinite-frequency observation is then
\begin{align}
    \Delta t_{\rm DM_\odot} &= \frac{e^2}{2\pi m_e c^2}\frac{\rm{DM}_\odot}{\nu^2}.
    \label{eq:sw_dt}
\end{align}

A common approach to modeling dispersive delays in the IPM is to fit a single $n_E$ value for each pulsar for the duration of the data set \citep{tempo2timing}. However, this model assumes that solar electron density is constant in time. In our deterministic model, we fit $33$ different values for $n_E$ which we will denote as $n_{E_i}$ where $i$ indexes the $n_E$ bin. {This is similar to the SWX model in \texttt{PINT} except we fit these parameters in the noise model rather than the timing model.} We then proceed to infer common values for these among different pulsars via a factorized posterior method as was done in \citet{ng12p5_cnm}.

A higher order SW term was first introduced into PTA literature by \citet{You+2007}. In this work, though, we test for the higher order term from \citet{hazboun+2022sw} which is also found in \citet{ng12p5_cnm}. We report here that we did not find it to be overall significant, so we do not include it in the final models.

\subsubsection{Stochastic solar wind}
\label{sec:stochastic_solar_wind} 
We apply GP models on top of the global deterministic model to capture the more highly time-variable nature of the SW, as well as the idea that different pulsars will each scan a unique line-of-sight through the IPM. This type of model is commonly referred to as ``SWGP''. It should be noted that the stochastic component of the model is not shared among the pulsars as they do not share individual realizations of a stochastic process in PTA GP models; thus, the overall SW model is not a truly global model. In this work, we use two classes of GP models to capture the deviations from the base deterministic SW model: a piecewise GP on the solar electron density weighted towards observations near conjunctions \citep{Nitu+2024} and a more commonly adopted GP on the solar electron density which was introduced in \citet{susarla+24sw,hazboun+2022sw}.

The former GP model consists of a piecewise-yearly inference of the variance in solar electron density from the underlying deterministic model. It utilizes a triangular-basis function to up-weight observations near conjunction and down-weight observations farther from conjunction {(Eq.~\ref{eq:Nitu_basis})}. This schema comes with minimal computational costs since the covariance matrix is diagonal and is rank $\mathcal{O}(1)$ as it only depends on the number of solar conjunctions rather than the number of observations in the dataset. However, this class of GPs is less flexible because it effectively enforces a piecewise constant perturbation of solar electron density at each conjunction and assumes no SW variability on a sub-annual timescale. 
We place a uniform hyperprior of $\mathcal{U}(-4,2)$ {cm$^{-3}$} on {$\log_{10}\sigma_{n_E}$}, which represents the Gaussian standard deviation of solar electron density perturbations from the deterministic model across each year, centered at conjunction. For more details on this model see \citet{Nitu+2024}.

We also expand on the type of SWGP models introduced in \citet{hazboun+2022sw} and studied in \citet{susarla+24sw}, which are implemented by multiplying the achromatic basis (either the Fourier basis or the linear interpolation basis described in \S\ref{sec:noise_models}) by a chromatic, geometric factor, $\Delta t_{\rm DM\odot}(n_E=1)$ which is the SW delay for a particular line of sight for a solar electron density at the Earth of $1$ $e$ $\rm{cm}^{-3}$ {(cf. Eq.~\ref{eq:SW_basis})}. The weights for this basis then infer the changes in solar electron density such that the GP encodes the chromatic delay experienced at each TOA due to a time-varying but spherically-symmetric SW. In this work, we extend this type of SWGP model to now include the TD basis from \citet{ng12p5_cnm} in addition to the Fourier basis previously used. For the SWGP priors, we choose either a ridge prior with hyperprior $\log_{10}\sigma_{n_E} \in \mathcal{U}(-4, 3)$ with a linear interpolation basis or a power-law prior with hyperpriors $\gamma_{n_E}\in\mathcal{U}(-6, 5)$ and $\log_{10}A_{n_E}\in\mathcal{U}(-12,0)$ with a Fourier basis. Note that the negative prior range in $\gamma_{n_E}$ constitutes a blue spectrum.

It should be noted that these SWGP models are not supplemented by a polynomial fit of solar electron density at low frequencies as is done for the achromatic and DM GP models in \citet{vanhaasteren+2014_gp} and this work. Rather, we use a binned, deterministic model to absorb low frequency power. Future work could explore a polynomial fit to this low frequency power.

\subsection{Events}
\label{sec:event_models} 

Some pulsars have been observed to experience sudden changes in their pulse profiles in time and radio frequency, which may manifest as chromatic delays in the timing residuals, and are well-described as stationary GPs \citep{ng15detchar}. The most infamous case is PSR J1713+0747, which has experienced three such events, two of which are included the NANOGrav $15$ yr dataset (the most dramatic event is beyond the timespan of the $15$ yr dataset and its mitigation is an area of active research, see e.g., \citealt{jennings+2022, mandow_J1713_2025}). Here we apply the typically used ``exponential dip'' model for these events, i.e., the product of a Heaviside step function with a decaying exponential function (\citealt{lam+2018, hazboun:2020slice, ng12p5_cnm}; {Eq.~\ref{eq:det_dip}}), which has been shown to strongly improve PSR J1713+0747's noise characterization in NANOGrav datasets \citep{hazboun:2020slice, ng12p5_cnm, Larsen+2024}. Evidence for similar dip events has been found in \citet{Shannon+2016, goncharov+2021}, while \citet{ng12p5_cnm} found evidence for cusp events, and \citet{pptadr3:noise, MPTADR2_noise} found evidence for additional Gaussian events {(Eq.~\ref{eq:gauss_event})}. Alongside J1713+0747's events, we test for the presence of such events in PSRs J0613$-$0200 \citep{ng12p5_cnm}, J1600$-$3053 \citep{pptadr3:noise}, J1643$-$1224 \citep{Shannon+2016, pptadr3:noise}, and J2145$-$0750 \citep{goncharov+2021, pptadr3:noise}. Alongside the GP bases and priors, definitions of the event signals we search for can be found in Appendix~\ref{appendix:GP_kernels}.

Searches for new events could be undertaken in the profile domain \citep{Brook+2018} or in the timing residual domain using a generic deterministic signal search \citep{MPTADR2_noise} or a model-agnostic method such as a wavelet basis \citep{lentati+:2016}. While we do not explore these methods here, the GPs we use in this work can occasionally diagnose short-timescale events, which can be diagnosed with conditional GP posterior recovery (cf. \S\ref{sec:consistency_checks}). 



\section{Model Selection Process}
\label{sec:model_selection}
\begin{figure}[ht]
    \centering
    \includegraphics[width=0.45\textwidth]{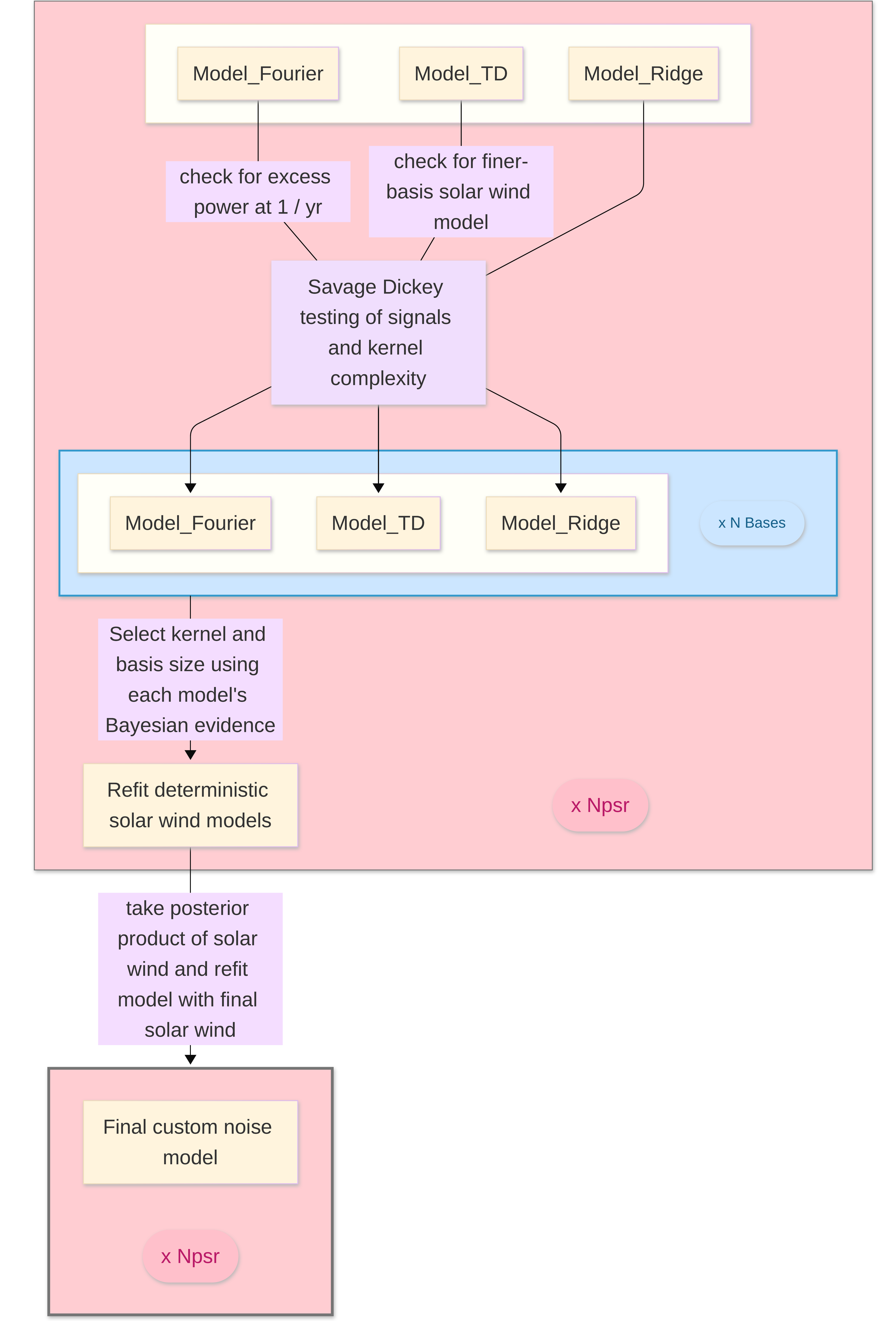}
    \caption{Model selection flow chart. The beige boxes represent steps along our model selection process while the purple boxes label the action arrows of the model selection process. The pink plates indicate that the analysis was performed over all $67$ pulsars while the blue plate shows that the analyses were additionally performed over each basis size. }
    \label{fig:flowchart}
\end{figure}

In this section we describe our approach to select optimal chromatic models for each pulsar among the options discussed in \S\ref{sec:methods}. Our approach builds on previous Bayesian model selection procedures used in past PTA analyses (e.g., \citealt{lentati+:2016, goncharov+2021, Chalumeau+2022, eptadr2_2:noise, pptadr3:noise, MPTADR2_noise, ng12p5_cnm}). This approach balances between over and underfitting, yields insights into multiple model components simultaneously (e.g., the choice of DM model may depend on whether or not a FC component is included), and also harmonizes with most GW search frameworks in PTAs. As in \citet{ng12p5_cnm}, we leave achromatic red and white noise models unchanged in each pulsar, considering only the selection of chromatic components.

The sheer range of possible models we consider (discussed in \S\ref{sec:methods}, Table~\ref{tab:noise_models_table}) and computational limitations in comparing between them requires a multi-step noise model selection process. To simplify the selection process, we assume that all chromatic components of the model can be described by the same basis, e.g., if we include DM and FC variations, both components will use a Fourier basis with the same number of frequencies, as opposed to using an Fourier basis for DM and a TD basis for FC, or vice versa. 

Fig.~\ref{fig:flowchart} shows a visual overview of our implementation. Every pulsar is moved through each stage of the model selection process in parallel with each other. Schematically, we start from a few initial models which include all chromatic processes we consider, remove extraneous components for which the data do not yield enough evidence, select the best fitting basis size of the GP models, and finally refit the global SW parameters. The next subsections discuss each phase of the model selection in further detail. Throughout this process, we also apply various posterior-predictive consistency checks, as discussed in \S\ref{sec:consistency_checks}, which may further inform our selection process for specific pulsars.

\subsection{Initial Models}
\label{sec:initial_models}

We start by testing three different models, each listed at the top of Fig.~\ref{fig:flowchart}, for each pulsar. These models are each iterated on until one is selected for each pulsar:
\begin{itemize}
    \item \textsc{Model\_Fourier}: A fully Fourier-domain model using a power law PSD for DM and FC variations (5 chromatic hyperparameters), on top of achromatic red and white noise. The Fourier basis is the most common implementation of chromatic models in the PTA community (e.g., \citealt{lentati+:2016, Chalumeau+2022, pptadr3:noise, Larsen+2024, ipta3p+2024}). The preference for a FC component and the value of $N_f$ is subject to change after starting from this model. We set $N_f=100$ to start.
    \item \textsc{Model\_TD}: A fully time-domain model initially using the QP\_RF kernel for DM, and a QP kernel for FC variations (11 chromatic hyperparameters), on top of (Fourier-basis) achromatic red and white noise. This follows the implementation of chromatic models in \citealt{ng12p5_cnm} and may be best suited to pulsars with complicated, non-powerlaw chromatic variations. The kernel type, preference for a FC component, preference for $\nu$-dependent DM, and the value of $dt$ in days (starting using $dt=15$ days) are subject to change after starting from this model.
    \item \textsc{Model\_Ridge}: A fully time-domain model using only the Ridge kernel for all components (3 chromatic hyperparameters), on top of (Fourier-basis) achromatic red and white noise. This is equivalent to \textsc{Model\_TD} using the simplest kernel (i.e., if $\ell = \Gamma_p = \alpha_{\mathrm{wgt}} = 0$) and is most appropriate for pulsars with little to no chromatic structure. The preference for a FC component and the value of $dt$ in days (starting using $dt=15$ days) are subject to change after starting from this model.
\end{itemize}
On top of these, all initial noise models include the global, time-binned SW component with fiducial $n_E(t)$ values obtained from an initial fit using a subset of the 15yr pulsars. As pulsars with ecliptic latitude less than $35^\circ$ are especially sensitive to the SW \citep{susarla+24sw}, for these pulsars we also consider further GP perturbations in $n_E(t)$ to account for any measurable errors in the global SW component. Two additional pulsars, B1937+21 and J1640+2224 with ecliptic latitudes of $42.3^\circ$ and $44.1^\circ$, are also tested for a SWGP component (cf. Appendix~\ref{appendix:SW_SNR}). As a baseline, we use the triangular-basis SWGP component to account for these perturbations in $n_E(t)$ \citep{Nitu+2024}. However, neither the global SW component nor the triangular basis GP account for variations in $n_E(t)$ on timescales shorter than 6-months, so for select pulsars with apparent \emph{sub-annual} SW variations (determined initially from posterior analyses, see \S\ref{sec:consistency_checks}), we also consider models that use either a $N_f=100$ Fourier basis SW component with a power-law PSD (\textsc{Model\_Fourier}) or the $dt=15$ TD basis SW component with a ridge kernel (\textsc{Model\_TD}, \textsc{Model\_Ridge}), on top of the global model. For select pulsars, we also include transient signals in their models (\S\ref{sec:event_models}), and for others we test for the presence of annual chromatic variations in their models following posterior analyses (\S\ref{sec:consistency_checks}).




\subsection{Kernel Significance Testing}
\label{sec:significance_testing}

Each initial model's parameter estimation analysis is performed using \texttt{PTMCMCSampler} \citep{ptmcmcsampler}. The purpose of these initial analyses is to start with the fairly comprehensive versions of each class of model described in \S\ref{sec:initial_models} and understand how much model complexity is required in the first place. For example, several pulsars do not even show any significant DM or chromatic variations, so any further consideration of the 6 parameter QP\_RF kernel in \textsc{Model\_TD} would unnecessarily slow down subsequent analyses with insignificant parameters. During this stage, we can also verify some of our results qualitatively by comparing the results of the three initial models, and diagnose issues prior to a more extensive model comparison. The signal significance tests described here are also repeated after later stages of the model selection.

To test the significance of each component of the model quantitatively, we use the Savage-Dickey Bayes Factor. As a baseline, we always include a basic DM GP, and we assume pulsars near the ecliptic will always require an additional SWGP. This means for the two models \textsc{Model\_Fourier} and \textsc{Model\_Ridge} we only need to test the significance of the FC component, using the $\log_{10}A_{\mathrm{FC}}$ and $\log_{10}\sigma_{\mathrm{FC}}$ posteriors, respectively. During this initial stage, we use a conservative threshold $\mathcal{B}^{\mathrm{FC}}_\oslash > 10$, above which FC variations are kept in the model. 
We use the region $(\log_{10}\sigma_{\mathrm{FC}} < -9,\; \chi < 4)$ to compute the Savage-Dickey Bayes Factor for FC variations, as we find FC variations with large $\chi$ may be significant in the lowest band even for small $\log_{10}\sigma_{\rm FC}$. Alongside FC variations, for \textsc{Model\_TD} the selection of DM kernel among various options (SE, QP, SE\_RF, QP\_RF) is also required. Due to the nested properties of these kernels, these selections can also be made with the Savage-Dickey Bayes Factor, again using a threshold of $\mathcal{B} > 10$ to select a more complicated DM kernel over a simpler one during this initial stage. See Appendix~\ref{appendix:TDSDBF} for further details on the Savage-Dickey Bayes Factor implementation for \textsc{Model\_TD} kernel selection. Before the final models are selected, thresholds are increased to $\mathcal{B} > 100$ to keep FC variations and the higher-order terms in the TD kernels.

We also consider the significance of additional signals, namely annual processes and events (\S\ref{sec:event_models}), using the same initial criteria $\mathcal{B} > 10$, and $\mathcal{B} > 100$ before final models, to determine if the extra signals merit inclusion in the model.

Caveats to this analysis include firstly that these Savage-Dickey Bayes Factors are dependent on the lower cutoff of the log-uniform prior. The Bayes Factor threshold is similarly arbitrary, where a high Bayes Factor threshold imposes an Occam penalty and hedges against overfitting but may overzealously eliminate subthreshold models that provide slightly better fits to the data. Nonetheless this practice of culling models based on Bayes Factors is consistent with prior work (e.g., \citealt{lentati+:2016, goncharov+2021}), and importantly we only consider removing high-order components of the chromatic models, so GW results are relatively unlikely to be impacted by these choices. In future work we may explore alternative approaches such as model averaging \citep{vanHaasteren2025ma}.

\vspace{\baselineskip}

\subsection{Basis Size Selection}

After making our initial assessments of kernel complexity for each pulsar, we test for which \emph{basis size}, or number of GP coefficients, is preferred, following e.g., \citet{Chalumeau+2022, ng12p5_cnm}. For each model, we test the finite number of basis sizes listed in Table~\ref{tab:noise_models_table}, leaving the basis size consistent across the DM, FC, and (where applicable) SW components. While Table~\ref{tab:noise_models_table} is certainly non-exhaustive, the key is that we consider sufficiently many options to allow our models to explore a variety of timescales. The largest Fourier basis, $N_f = 200$, notably covers only up to a time lag of $T_{\mathrm{NG15}}/N_f \approx 30$ days, while the TD GP models start at this timescale and extend to shorter timescales. Extending the Fourier basis to shorter timescales could enhance spectral leakage or aliasing errors due to the uneven data cadence of PTA datasets \citep{Crisostomi+2025}, which is one reason why we use the TD models to explore this space. Additionally, we do not employ the variable $N_f$ model presented in \citet{eptadr2_2:noise} for \textsc{Model\_FD} as we do not yet have an analogous model for the TD GPs allowing variable $dt$.

As discussed in \S\ref{sec:selection_methods}, we use INS as our primary method to compute the Bayesian evidence $\mathcal{Z}$ for each model, which may be compared against each other to compute any number of Bayes Factors. We specifically use the \texttt{nautilus} code, which leverages INS and neural network estimation of bounding distributions to sample accurately and efficiently \citep{Lange2023}.

For this model selection we additionally consider the likelihood evaluation time, $t_{\mathcal{L}}$, which scales approximately as the cube of the number of basis elements in the pulsar's noise model. We account for this because if $t_{\mathcal{L}}$ is too large, the model may not be practical for GW searches. To introduce an Occam penalty to account for this, we select the basis size which maximizes the ratio $\mathcal{Z}/t_{\mathcal{L}}$. This is primarily useful for pulsars that are more or less agnostic to the choice of basis size. While a caveat to using $t_{\mathcal{L}}$ is that it is hardware-dependent, the ratio between $t_{\mathcal{L}}$ for different models is more likely to remain consistent across different computing systems. When computing $t_{\mathcal{L}}$ for the model selection, we cache steps of the covariance matrix inversion involving the white noise and timing model components, as those are typically cached during GW searches \citep{Johnson+2024}. Unfortunately, for PSR J1713+0747 using \textsc{Model\_TD} with basis sizes $dt=7$ and $dt=3$ days, $t_{\mathcal{L}}$ grew too large for INS to complete, so these two sub-models could not be considered.

Considering a smaller or larger basis size may also result in previously significant parameters no longer being significant. To account for this, we apply the procedures from \S\ref{sec:significance_testing} to see if any model components can be removed and update the Bayesian evidence accordingly, $\mathcal{Z} \to \mathcal{Z}\times\mathcal{B}^\oslash$ for the final model selection, where $\mathcal{B}^\oslash$ is the Bayes Factor for the model with the extra signal removed (inverse of the Savage-Dickey density ratio).

As a cross-check for the INS method, we also used product space sampling to compare between different basis sizes, performing one MCMC for each of the 3 model types (\textsc{Model\_Fourier}, \textsc{Model\_TD}, \textsc{Model\_Ridge}). While product space sampling does not always compute the Bayes Factors between models reliably due to large differences between model posteriors, we find the final choice of selected basis is consistently in agreement with the results from INS.

\begin{figure}
    \centering
    \includegraphics[width=\linewidth]{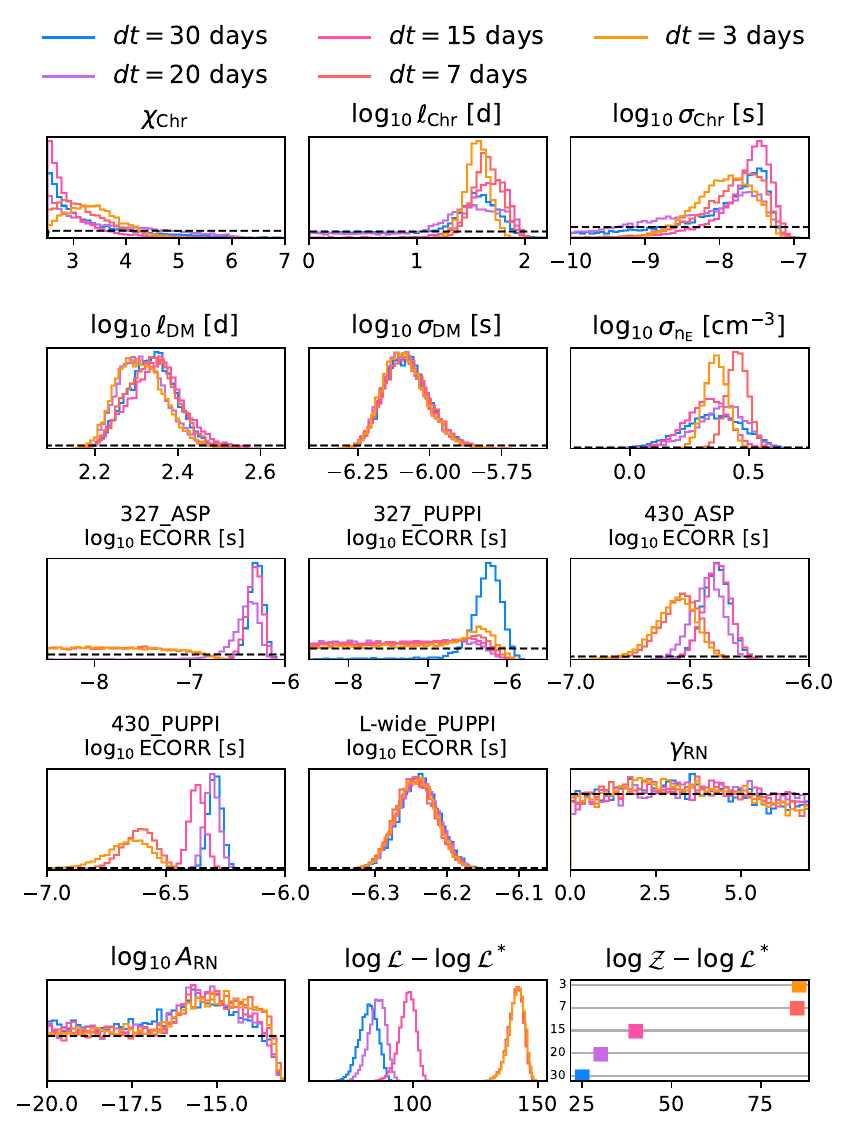}
    \caption{Posteriors given different basis sizes for PSR J2317+1439 using \textsc{Model\_TD}. Different basis sizes are distinguished by color and marked in the legend. The dashed black lines indicate the prior values over the plotted parameter ranges (cf. Table~\ref{tab:priors} for the full prior distributions). The bottom right 2 panels also show marginalized log likelihood $\log\mathcal{L}$ and log evidence $\log\mathcal{Z}$ values for each model, from which a constant factor $\log\mathcal{L}^* = 174600$ has been subtracted. $dt = 7$ days is adopted in the final model.}
    \label{fig:J2317_basis_selection}
\end{figure}

An example case showing the importance of proper basis size selection for chromatic noise using NG15 is shown in Fig.~\ref{fig:J2317_basis_selection} for PSR J2317+1439 using \textsc{Model\_TD} with SE kernel DM, SE kernel FC, and Ridge kernel SWGP components included. Some of the parameter posteriors (particularly ECORR) are dramatically affected by the basis size, which here is controlled by the spacing between interpolation nodes $dt$. 
The bottom right two panels of Fig.~\ref{fig:J2317_basis_selection} compare the marginalized log likelihood distributions and log evidences $\log\mathcal{Z}$ between each basis size. For PSR J2317+1439, $\log\mathcal{Z}$ is significantly larger for the smallest basis sizes, $dt = 7$ and $dt = 3$ days, than it is for the larger basis sizes. However, the preference for $dt = 3$ days is not significantly larger than $dt = 7$ days; as such the penalty to the likelihood evaluation time $t_\mathcal{L}$ incurred from the larger basis results means $dt = 7$ days is selected for the final model. \citet{ng12p5_cnm} includes additional demonstrations showing the impact of chromatic basis size on ECORR.

\subsection{Global Binned Solar Wind Refit}
With our model selections finalized, we refit the binned deterministic SW model (\S\ref{sec:global_deterministic_solar_wind}) using a factorized likelihood approach. We follow the same procedure used in both \citet{hazboun+2022sw,ng12p5_cnm}. It should be noted however, that in this work, we refit the binned model alongside our SWGP models when applicable. The binned model takes care of the long time scale changes in solar electron density while SWGP takes care of the short time scale changes for pulsars which were found to be sensitive to these changes. 

We carry out the refit first by using \texttt{PTMCMCSampler} to sample in the usual parameters and hyperparameters (white noise, chromatic noise, achromatic noise, and timing model perturbations) and $33$ additional SW parameters, which are a piecewise, constant fit of the $n_{E}$ time series with parameters $n_{E_i}$, where $i$ indexes the bin number. We then use \texttt{kalepy}~\citep{kalepy} to create a 1-dimensional kernel density estimate over each SW parameter's posterior and multiply the posterior densities together across all $67$ pulsars. Different pulsars have differently constrained solar-wind parameters, but the multiplication across pulsars ensures that they are broadly consistent with each other.

The resulting binned $n_{E}$ time series is presented in Figure~\ref{fig:sw_ne_series}. {We take the medians of our new, common $n_{E_i}$ posteriors, and apply them as fixed deterministic delays for each pulsar as part of the final round of parameter estimation.} 

\subsection{Consistency checks}
\label{sec:consistency_checks}

We last detail a few consistency checks we applied along the way which proved to be important to our model selection procedure. Firstly, 
we frequently observed intermediate analyses would recover a FC index $\chi$ posterior PDF which rails against the lower prior bound, $\chi = 2.5$. Since we expect chromatic noise below this value to be entirely degenerate with DM variations, this railing suggests an insufficient DM variations model. After further investigation, in the majority of cases the issue was ameliorated by using a more comprehensive SWGP model (\S\ref{sec:stochastic_solar_wind}). A similar railing on an initial upper bound of $\chi = 7$ informed our choice to extend the upper range of the prior with a Gaussian tail, Eq.~\eqref{eq:chi_prior}.

Several more consistency checks are enabled by GP parameter reconstruction. While typically marginalized over, the coefficients of the GPs, {denoted $\vec{b}$}, may be sampled {from the conditional posterior $\mathcal{P}(\vec{b}|\vec{\delta t},\vec{\eta})$ as
\begin{equation}
    \label{eq:conditional_GP}
    \vec{b}|\vec{\delta t},\vec{\eta} \sim \mathcal{N}(\hat{b},\mathbf{\Sigma}_b),
\end{equation}
}with mean $\hat{b} = \mathbf{\Sigma}_b\mathbf{T}^T\mathbf{N}^{-1}\vec{\delta t}$ and covariance $\mathbf{\Sigma}_b = (\mathbf{B}^{-1} + \mathbf{T}^T\mathbf{N}^{-1}\mathbf{T})^{-1}$. This can be done after an initial estimation of hyperparameters $\vec{\eta}$ to perform various posterior predictive checks on the model itself \citep{Meyers+2023}. {Note that ECORR is moved from the $\mathbf{N}$ matrix into Eq.~\ref{eq:T_matrices} as a reduced-rank GP for this analysis, as in \citet{ng9_gwb}.}

One way we apply this method is to test for the presence of annual sinusoidal DM or FC variations at an early stage. Specifically, we use the initial results of \textsc{Model\_Fourier} to determine if any significant noise excursions from the power law prior near a frequency of $1$/yr may be present. 
To do this, we first approximate the prior by computing the power law spectra given the MAP values of $A$ and $\gamma$ from the posterior $\vec{\eta}$. We next draw coefficients $\vec{b}$ from the posterior $\mathcal{P}(\vec{b}|\vec{\delta t},\vec{\eta})$ by first drawing 500 values of the noise hyperparameters $\vec{\eta}$ using the posteriors $\mathcal{P}(\vec{\eta}|\vec{\delta t})$ and then for each sample, drawing from the conditional distribution, Eq.~\eqref{eq:conditional_GP}. For each value in the distribution, we then isolate the individual Fourier coefficients $\vec{a}_{\rm DM/FC}$ from $\vec{b}$ and compute the RMS sum of the coefficients $\vec{\rho}$ at each frequency. Finally, if both 95\% of the coefficients $\vec{\rho}$ lie above the power law prior at $f = 1$/yr and the Bayes Factor of the noise process is greater than 3, we test the pulsar for a separate annual process in subsequent rounds of the model selection.

Eq.~\eqref{eq:conditional_GP} also allows reconstruction of posterior noise time series as one may simply contract the coefficients $\vec{b}$ with the design matrix $\mathbf{T}$. For example, \citet{Larsen+2024} used this method to analyze noise sources in NG15 pulsars, identifying covariant signals that would ``compete'' with each other to fit the data (an effect we call \emph{contramodeling}, since the models appear to contradict each other). \citet{NG15_PPC} also used this method to construct waveforms of the GWB, decoupled from intrinsic noise and timing model contributions. Here we apply this method to determine the sign of the delays induced by the SW, which scale as the electron density $n_E(t)$ along the line of sight. Intrinsically, $n_E(t)$ is always positive, but there is no constraint in the SWGP models we use that force it to remain so. As such, the reconstruction method may be used to diagnose any epochs where SWGP is overfitting the data, which may indicate the presence of some additional noise source, or potential TOA outliers (see \S\ref{sec:sw_excision} for 2 identified cases). We also used time series reconstruction to diagnose a particularly egregious case of chromatic index railing in PSR J1909$-$3744 and understand the improvements made by use of a fine-grained, TD SWGP (cf. Fig.~\ref{fig:swgp_realizations} in \S\ref{sec:swgp-results}). 

\section{Results: Final Chromatic Models}
\label{sec:results}

\begin{figure*}
    \includegraphics[width=\linewidth]{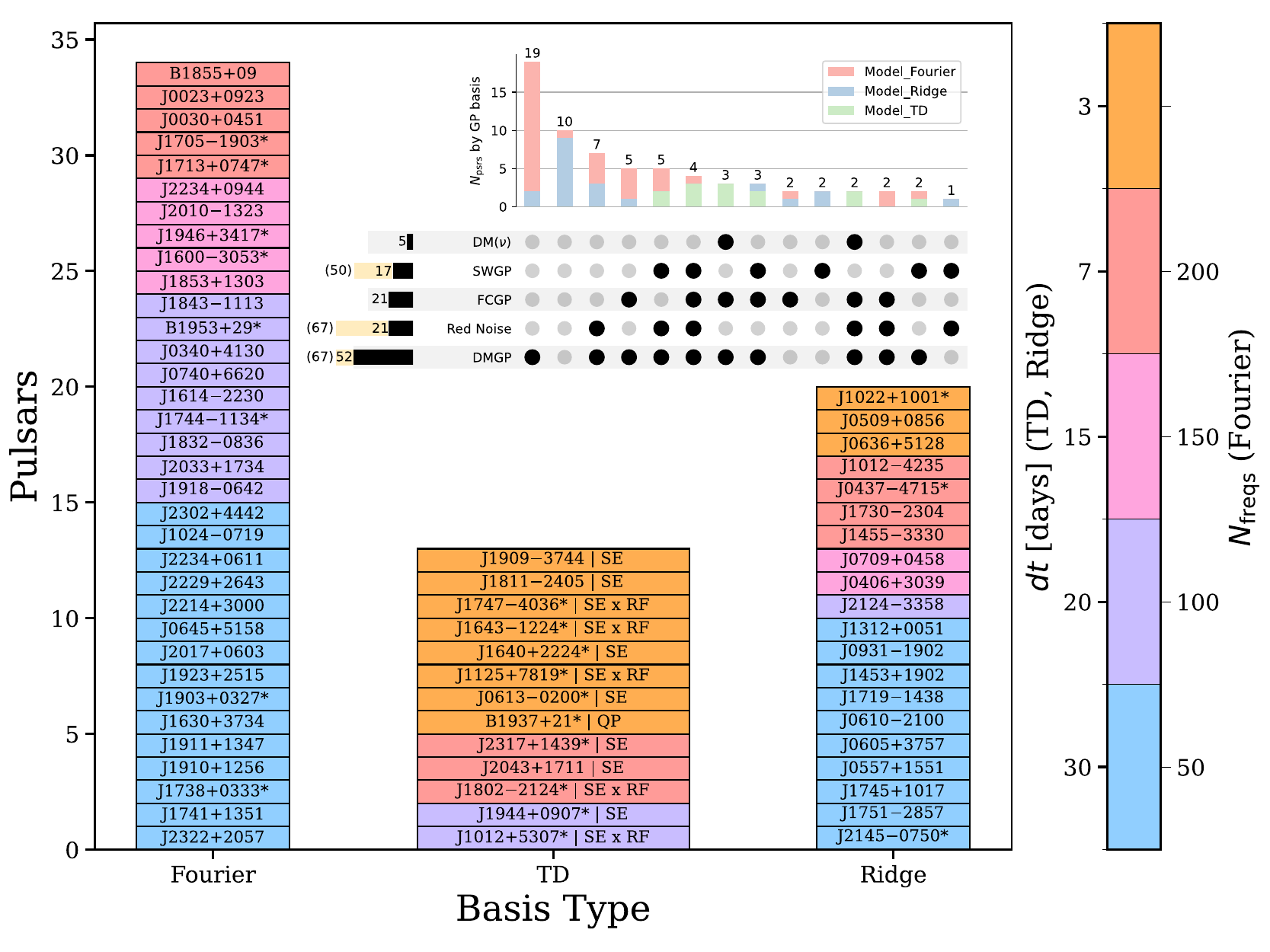}
    \caption{The final models after our model selection process are represented in this bar chart. The pulsars are split into three columns corresponding to the favored basis type. The favored basis size (either $dt$ in days or $N_f$) is color coded where the highest $dt$ shares a color with the smallest $N_f$ since both represent a smaller basis size. A * indicates that the favored model includes a free chromatic (FC) model component. Specific time domain kernels are listed after the pulsar in the TD column. \emph{Inset bar chart:} Horizontal bars show the number of pulsars with significant red, DM, FC, SW, and $\nu$-dependent DM noise (parenthetical values indicate the total number of pulsars using the noise process even if it was not significant). The number of pulsars with different combinations of these noise types are indicated by the vertical bars, with vertical bar colors indicating the type of GP basis used for all chromatic GPs.}
    \label{fig:final_models_bar_chart}
\end{figure*}

Our model selection results can be summarized by the types of noise processes which were found to be significant and the GP bases which were favored for each pulsar. We find that after the completion of our model selection procedure, 21 pulsars show evidence for FC noise and 5 show evidence for $\nu$-dependent DM variations. While red and DM noise are included in all pulsars, DM noise is only significant in 52 pulsars and RN is only significant in 21 pulsars as determined by a Savage-Dickey Bayes Factor larger than 10. In addition to a fixed, deterministic, piecewise global model for SW electron density variations, pulsar-specific GP perturbations to the SW electron density are significant in 17 pulsars. An additional 33 pulsars with ecliptic latitudes less than 35$^\circ$ also include SWGP since the signal is still expected to be present at sub-threshold levels. For tables of parameters corresponding to each noise type, see Appendix~\ref{appendix:tables}.


Fig.~\ref{fig:final_models_bar_chart} summarizes which of the models described at the top of \S\ref{sec:initial_models} are preferred for the chromatic noise in each pulsar. Out of the 67 pulsars in the dataset, 34 preferred \textsc{Model\_Fourier}, 13 pulsars preferred the more complicated \textsc{Model\_TD}, and 20 pulsars preferred \textsc{Model\_Ridge} (a simpler sub-model of \textsc{Model\_TD}). In other words, half the pulsars favored some versions of the TD models explored in \citet{ng12p5_cnm}, and the other half favored the more commonplace Fourier basis models. A spread of different basis sizes are also preferred depending on the pulsar, which may depend on various factors including the set of detected chromatic processes, the observation timespan, and the effective observation cadence. Fig.~\ref{fig:final_models_bar_chart} shows in particular that the shortest-timescale basis size $dt = 3$ days is most often preferred when \textsc{Model\_TD} is favored. 

The inset panel of Fig.~\ref{fig:final_models_bar_chart} additionally shows how the preferred basis type correlates with the types of noise detected in each pulsar, as well as the number of pulsars where different combinations of noise were found to be significant (excluding annual and transient signals). One of the most striking aspects is that \textsc{Model\_TD} is only ever favored when either $\nu$-dependent DM or SWGP components are required. Naturally, pulsars with significant $\nu$-dependent DM favor \textsc{Model\_TD} by construction, as we did not consider a similar $\nu$-dependent DM process in \textsc{Model\_Fourier}. Meanwhile, the coincidence of \textsc{Model\_TD} with SWGP detections suggests that the TD interpolation basis is particularly well suited for modeling the time-variable SW. 
Pulsars that exhibit chromatic (DM and/or FC) noise but do not show SWGP or $\nu$-dependent DM noise overwhelmingly favor \textsc{Model\_Fourier}. After accounting for pulsars with significant annual or non-stationary events, this suggests the ISM-induced chromatic noise in the majority of MSPs is highly consistent with a power law spectrum. Unsurprisingly, when there are no detections of noise at all, the Ridge kernel (with just a single DM hyperparameter) is the the most consistently favored. 



\subsection{ISM-induced DM variations}

The DM noise (or DMGP) component of the model largely captures ISM-induced DM variations as the components induced by SW variability are isolated in a separate channel. Tables~\ref{tab:FD_params} and \ref{tab:TD_params} of Appendix~\ref{appendix:tables} give the posterior medians (or upper limits) for DMGP parameters, depending whether the Fourier or linear-interpolation basis was favored.

\begin{figure}[ht!]
    \centering
    \includegraphics[width=0.45\textwidth]{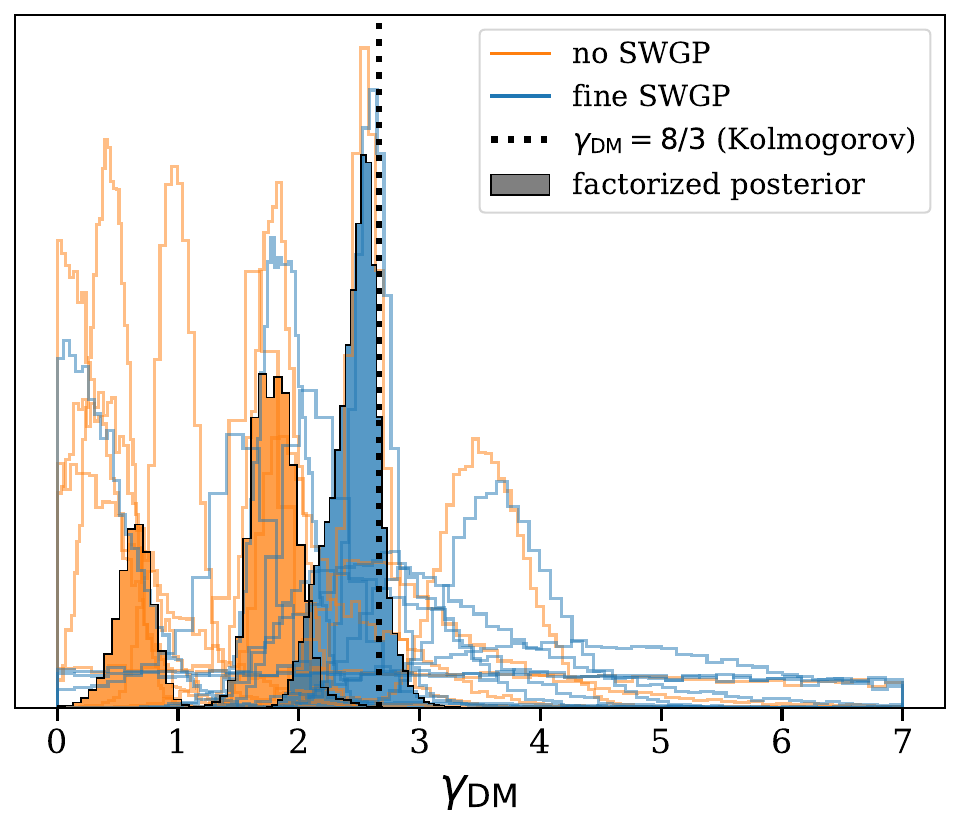}
    \caption{The blue outlines show posteriors for the spectral indices in the Fourier basis models when a SWGP is included while the orange outlines show posteriors when a SWGP is not included. The shaded regions are the normalized posterior products across all $12$ pulsars both cases. (Note that the no SWGP case produces bimodal posterior product.) Both models have the fixed, deterministic SW included. A vertical line demarcates $\gamma_{\rm DM}=8/3$, which is the anticipated value for a Kolmogorov-like ISM. Note that the posteriors shown here are not necessarily those of the final, favored pulsar model.}
    \label{fig:sw-dmgp-gamma-comp}
\end{figure}

Decoupling the SW component is perhaps the biggest challenge for modeling DM variations. To illustrate how the SW and DM models influence each other, Fig.~\ref{fig:sw-dmgp-gamma-comp} compares posteriors on $\gamma_{\rm DM}$ with and without a fine-grained SWGP included in the model, for all pulsars which favored a fine-grained SWGP. For the purposes of this investigation, the preferred version of \textsc{Model\_Fourier} is used for each pulsar in Fig.~\ref{fig:sw-dmgp-gamma-comp}, rather than each pulsar's \emph{final} preferred model which is not necessarily \textsc{Model\_Fourier}. Without a SWGP included, many of the $\gamma_{\rm DM}$ values are measured to be very shallow ($\gamma_{\rm DM} < 2$). With SWGP included, those DM spectral indices tend to higher values and posteriors are clustered primarily near the value predicted for DM variations from the ISM assuming a Kolmogorov-like medium, $\gamma_{\rm DM} = 8/3$ \citep{Keith+2013}.

We quantify this population shift through the shaded regions of blue and orange in Fig.~\ref{fig:sw-dmgp-gamma-comp} which are the $\gamma_{\rm DM}$ posterior products across pulsars for the fine-grained SWGP and no SWGP cases respectively. Omitting a SWGP model yields a factorized posterior with a shallower spectrum than when SWGP is included. Furthermore, the former distribution is multi-modal, suggesting two subpopulations of pulsars where the level of bias induced by the SW depends on the relative strength of SW- vs ISM-induced DM variations in each pulsar's noise budget, which itself depends on e.g. pulsar distance and ecliptic latitude (see Appendix~\ref{appendix:SW_SNR}). For example, PSR J0030+0451 is the most sensitive pulsar to SW in the NG15 dataset (cf. Fig.~\ref{fig:SW_SNR}), with relatively weak ISM-induced DM variations. Consequently, this pulsar's $\gamma_{\rm DM}$ is very biased by the solar wind without a SWGP. Meanwhile, PSR B1937+21 is also sensitive to SW variations but has a very large DM value and exhibits strong ISM DM variations. As such, the SW does not bias its DMGP hyperparameters as significantly as PSR J0030+0451. In summary, the penalty for not including a SWGP component increases for pulsars with higher SW sensitivity and weaker (lower amplitude) ISM-induced DM variations. Lastly, we note that the only posterior with a low spectral index value after applying SWGP in Fig.~\ref{fig:sw-dmgp-gamma-comp} corresponds to PSR J1022+1001, which ultimately favors \textsc{Model\_Ridge} over \textsc{Model\_Fourier}. Overall, these results suggest our models can successfully decouple the majority of ISM and SW-induced DM variations within TOA measurement uncertainties. Future work should explore hierarchical models such as those found in \citet{vanHaasteren2025hm} to further decouple SW and ISM DM.

\begin{figure}[ht!]
    \centering
    \includegraphics[width=\linewidth]{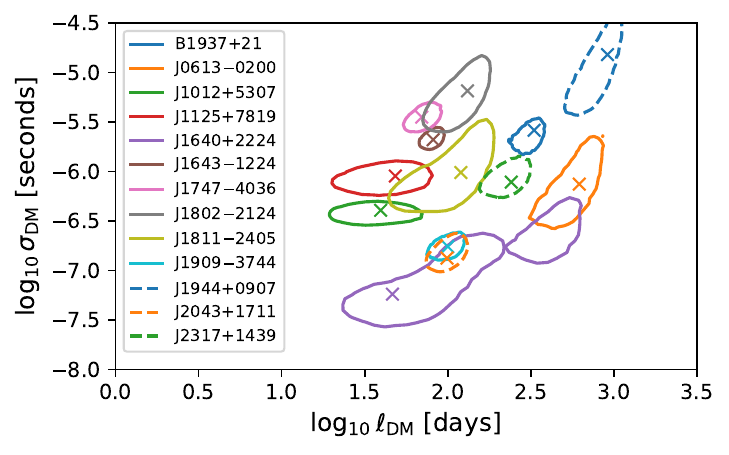}
    \caption{Posterior 95\% credible regions on the DM noise parameters $\log_{10}\sigma_{\rm DM}$, $\log_{10}\ell_{\rm DM}$ of the squared-exponential GP kernel, {cf. Eq.~\eqref{eq:QP_kernel}}, for pulsars that favored \textsc{Model\_TD}. PSR J1640+2224 registers a multimodal distribution. Crosses indicate the location of each pulsar's maximum a posteriori sample.}
    \label{fig:dmgp-td-params}
\end{figure}

Alongside Table~\ref{tab:TD_params}, posteriors on the TD DMGP parameters $\log_{10}\sigma_{\rm DM}$ and $\log_{10}\ell_{\rm DM}$ are shown in Fig.~\ref{fig:dmgp-td-params}, for pulsars that favored \textsc{Model\_TD}. The posterior DMGP amplitude and timescale parameters likely reflect the spread in per-pulsar distances and velocities of the line-of-sight. Timescale parameters run the gamut between 30-1000 days. PSR J1640+2224 notably registers a multimodal posterior distribution in its final model. This implies PSR J1640+2224 may actually prefer 2 DMGPs, as \citealt{ng12p5_cnm} found for PSR B1937+21. Assessment of PSR J1640+2224's FC model component (which may in fact be a manifestation of DM variations) further supports this assessment, as discussed in \S\ref{sec:FC_noise}. A possible interpretation for this could be if the line of sight samples two distinct regions of the ISM with different associated scale parameters.

Referencing Fig.~\ref{fig:final_models_bar_chart}, only seven pulsars with significant DM variations favor \textsc{Model\_Ridge}. As shown in Table~\ref{tab:TD_params}, these are all relatively short timespan pulsars $T_{\rm psr} < 7$ years, with the exception of PSR J1455$-$3330 with $T_{\rm psr} = 15.6$ years. Given its long timespan, the DM variations in this pulsar are exceptionally quiet \citep{Lam+2025}, suggesting chromatic noise in this pulsar should have minimal impact on the measurement of low frequency GWs.

\subsubsection{Radio-frequency/band dependent DM variations}
\label{sec:nu-dependent_DM_results}

\begin{figure}
    \includegraphics[width=\linewidth]{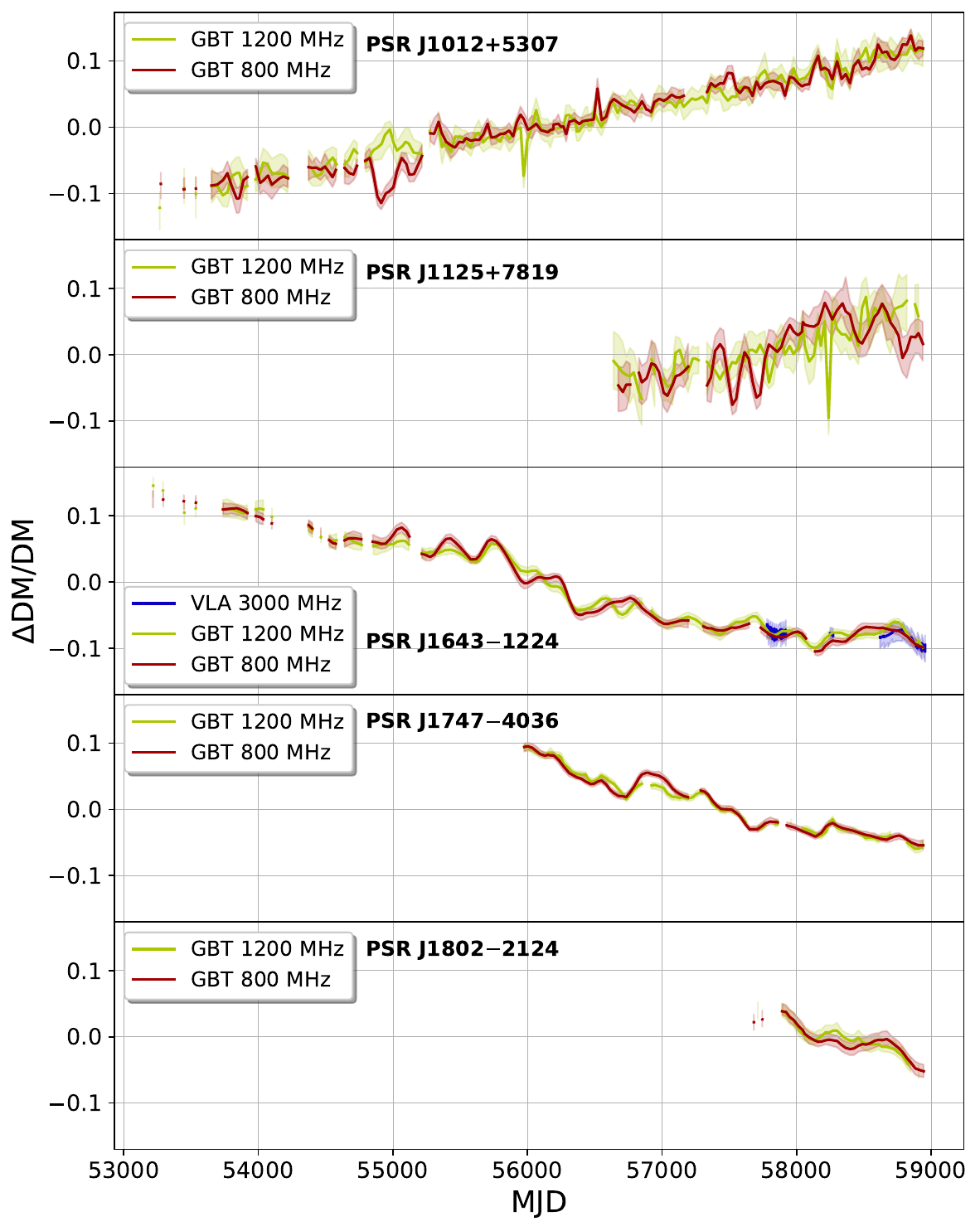}
    \caption{Inferred band-dependent DM variations in 5 pulsars. Delays are computed using Eq.~\eqref{eq:conditional_GP} (including the quadratic term from the timing model), converted to units of DM, and split up by band. Solid lines are median DM values and shaded regions enclose 68\% of realizations in each band. Gaps are placed in the time series where the gap in between TOAs is greater than 60 days. Variations are each divided by the base DM of the pulsar to place all time series on equal scale.}
    \label{fig:RF_band_DM}
\end{figure}

We measure significant band-dependent DM variations in five pulsars (J1012+5307, J1125+7819, J1643$-$1224, J1747$-$4036, J1802$-$2124). These pulsars are each observed at the GBT in two bands centered at 800 MHz and 1200 MHz, and PSR J1643$-$1224 is additionally observed at the VLA at 3000 MHz. The significance of the band-dependence is assessed via a large Savage-Dickey Bayes Factor from the $\log_{10}\alpha_{\rm wgt}$ parameter, which weights the strength of decorrelation across the bands \citep{ng12p5_cnm}, with posterior parameter values given in Table~\ref{tab:TD_params} of Appendix~\ref{appendix:tables}. PSRs J1643$-$1224, J1747$-$4036, and J1802$-$2124 each have relatively large DMs of 62.3, 153.0, and 149.6 pc cm$^{-3}$, respectively. While our $\nu$-dependent DM model is intended to account for multipath propagation, a more plausible explanation may be a deficiency in our FC model for scattering, as $\nu$-dependent DM is not expected to arise at $\nu > 800$ MHz given current sensitivity levels \citep{cordes+2016_dm_f}. Another possible ISM mechanism which could explain the observed decorrelations would be scattering through an anisotropic medium \citep{Kulkarni+2025}. PSRs J1012+5307 and J1125+7819 have relatively low DM values of 8.9 and 11.2 pc cm$^{-3}$, so it is even more surprising that they register a band-dependent DM, and an ISM origin would appear less plausible. To find an explanation for PSR J1012+5307's signal, one should also account for its very shallow spectral RN, which is persistent even across datasets from different PTAs \citep{Chalumeau+2022, ipta3p+2024, Larsen+2024}.

Fig.~\ref{fig:RF_band_DM} further shows the posterior band-dependent DM variations for these five pulsars, which are drawn using Eq.~\eqref{eq:conditional_GP}. The posteriors reveal that the DM variations may remain correlated for long stretches of the observation span, only decorrelating at specific times. For the high-DM pulsars: J1643$-$1224's DM decorrelates at several times between MJDs 55000 and 57000 and J1747$-$4036 experiences only an ``event'' near MJD 57000. In both cases the DM varies more dramatically in the 800 MHz band. PSR J1802$-$2124 has the lowest Bayes Factor $\mathcal{B}^{\rm SE\_RF}_{\rm SE} \cong 200$ for $\nu$-dependent DM out of these five pulsars, and the DM decorrelation is only marginally apparent across its short observation timespan. As for the low-DM pulsars, more rapid DM variations are observed, with several decorrelations and sudden jumps across the timespan of both pulsars (particularly near MJD 55000 in PSR J1012+5307 and MJD 58200 in PSR J1225+7819), which are also apparent in DMX time series \citep{ng15data}. The presence of stronger variations in the upper 1200 MHz band in these two pulsars heightens our skepticism of an ISM origin for the decorrelations. Further work studying scintillation and pulse properties of these pulsars may help inform their future modeling.

\begin{figure*}[ht!]
    \centering
    \includegraphics[width=.95\textwidth]{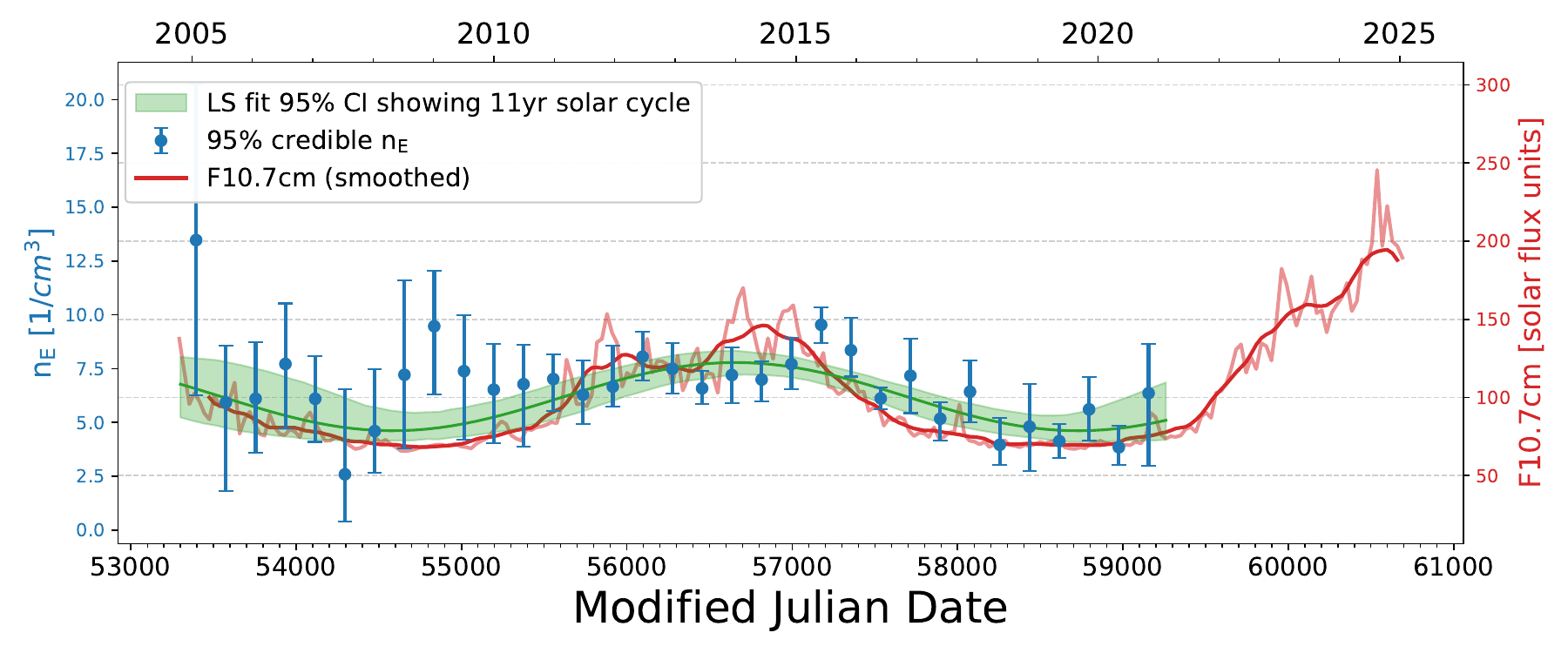}
    \caption{The medians and $95\%$ credible interval for the factorized posterior on the binned solar wind electron density ($n_E$) parameters are shown in blue. A least square's fit of a sine function to these points are shown in green. Monthly solar flux density data obtained from NOAA are plotted in red alongside a smoothed version of those data. The data have been rescaled to better visually compare trends alongside the $n_E$ data.}
    \label{fig:sw_ne_series}
\end{figure*}

\subsubsection{Annual and Quasi-Annual DM}
\label{sec:annual-dm-results}



Annual and quasi-annual DM variations may arise as the Earth's local motion around the Sun impacts the Earth-pulsar line of sight (cf. e.g., \citealt{jones+2017_ng9_dm, Madison+2019, ng12p5_cnm}). We find only two pulsars (B1937+21 and J0613$-$0200) show evidence for annual DM signals in their final models. While we did not directly test PSR B1937+21 for an annual DM signal, we find it is the only pulsar whose final model favors a QP kernel using \textsc{Model\_TD} (with a very large Bayes Factor $\log_{10}\mathcal{B}^{\rm QP}_{\rm SE} \sim 13$), and a quasi-period parameter recovered precisely at $\log_{10}(p_{\rm DM}/\rm{yr}) = 0.00^{+0.01}_{+0.01}$, meaning a component of its DM behaves effectively as a quasi-sinusoidal annual process. It is unclear if this suggests an annual sinusoidal process should be used in the future, or if the \emph{quasi}-annual process is genuinely preferred for PSR B1937+21. As for PSR J0613$-$0200, its final model actually includes an annual FC process (suggestive of annual scattering, see \S\ref{sec:annual-FC-variations}), but the recovered FC index is also consistent with a dispersive origin $\chi \sim 2$. Annual DM variations have previously been reported for PSR J0613$-$0200 by \citet{Keith+2013, jones+2017_ng9_dm, goncharov+2021}.


\subsection{Solar wind}
\label{sec:solar-wind-results}
\label{sec:deterministic-sw-results}
Fig.~\ref{fig:sw_ne_series} presents the results of our $180$ day piecewise-fit to the solar electron density described in \S\ref{sec:global_deterministic_solar_wind}. Note that this does not include the SWGP perturbations on top of this. The credible intervals on the first several bins of the dataset tend to be higher since there were fewer pulsars and also fewer observations. 
Here we compare our measurements of $n_E$ to the F$10.7$cm data, which consist of monthly-averaged radio flux-density measurements from the Sun \citep{NOAA_F107}. These data provide a useful metric for solar activity \citep{tapping2013-f10.7cm}. The solar flux data have been scaled so that the amplitudes of the fluctuation in the data roughly match those of our $n_E$ time series and the offsets are aligned. The similarities in trends in the datasets suggest that our deterministic SW model is successfully capturing the long time scale variations in solar activity. To further show this, we fit a simple sine wave, to the $n_E$ time series using least squares and report a period of $11.3\pm1.3$ years, which is consistent with the solar cycle. We extended the flux data to $2025$, which is the approximate end date of the NANOGrav $20$ year dataset. The increasing trend in the F$10.7$cm flux suggests this data set will contend with higher delays due to a more variable solar electron density.

\subsubsection{SWGP}
\label{sec:swgp-results}


Tables~\ref{tab:SW_params1} and \ref{tab:SW_params2} in Appendix~\ref{appendix:tables} display posteriors on the SWGP hyperparameters as well as the Bayes factors for including a SWGP component (on top of the fixed, deterministic, global SW component) in each pulsar's final model. As previously shown in Fig.~\ref{fig:sw-dmgp-gamma-comp}, the inclusion of SWGP has a striking impact on DMGP parameter recovery overall. The tables show that out of the 17 pulsars with a significant SWGP component, there are only 4 pulsars that favor a Fourier basis SWGP (Table~\ref{tab:SW_params1}) and 2 that favor the annual perturbations (triangular-basis) model from \citet{Nitu+2024}. The remaining 11 prefer to use the TD linear-interpolation basis with a Ridge kernel. The following case study reveals much about the apparent preference for the TD basis across the dataset.

\begin{figure*}[ht!]
    \centering

    \includegraphics[width=\textwidth]{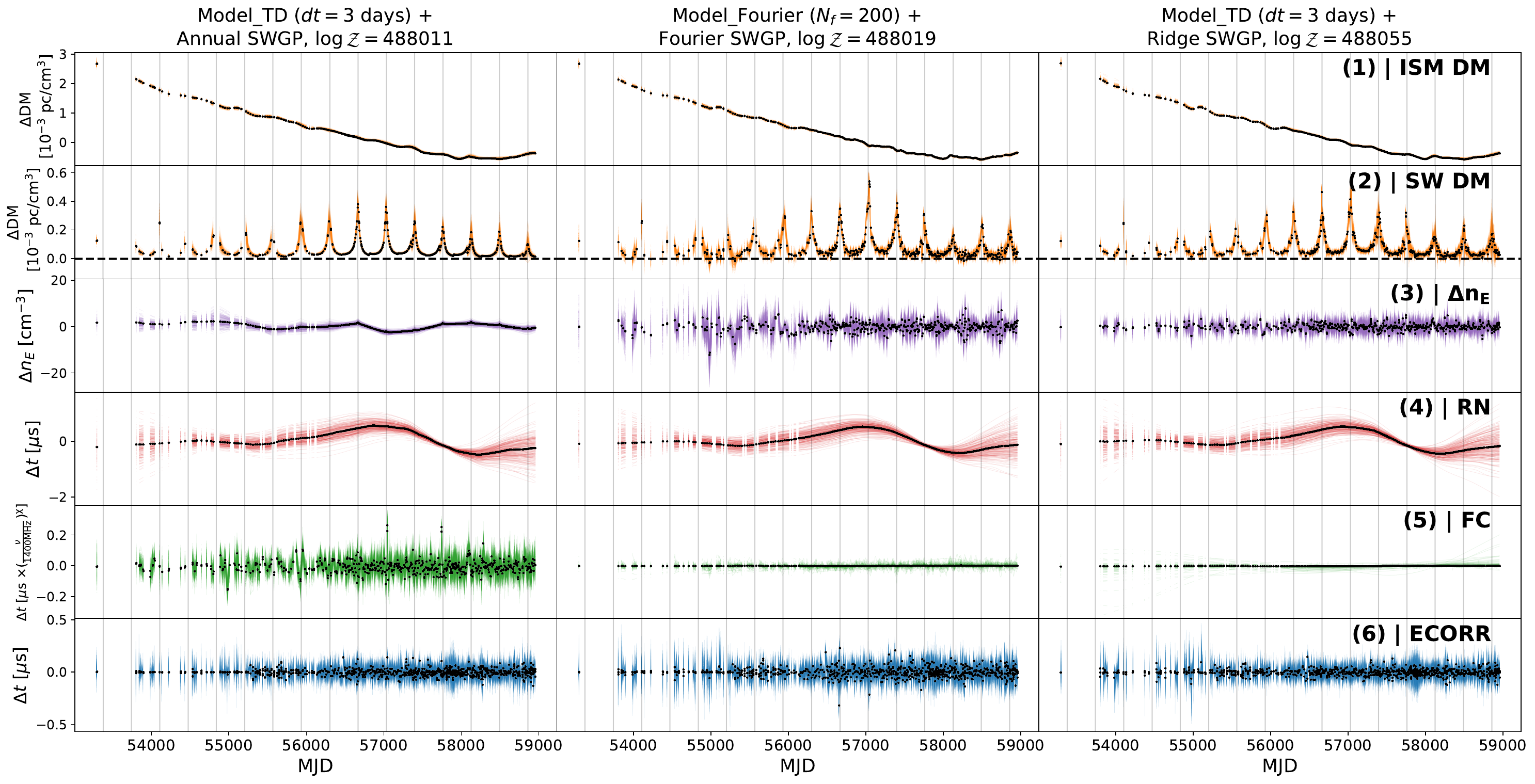}
    \caption{Gaussian process realizations for delays included in PSR J1909$-$3744's model under different versions of SWGP. Left-hand panels correspond to a preliminary favored model (\textsc{Model\_TD}, $dt=3$ days) under annual SW perturbations only; middle panels correspond to the favored \emph{Fourier basis} model (\textsc{Model\_Fourier}, $N_f=200$) with SWGP sharing the Fourier basis; right-hand panels correspond to the final favored model (\textsc{Model\_TD}, $dt=3$ days) with SWGP sharing the linear-interpolation basis. By row, we have: (1) DM variations excluding SW signals (DMGP+DM1+DM2), (2) DM variations including only SW signals (deterministic SW+SWGP), (3) perturbations to $n_E$ introduced by SWGP, and time delays due to: (4) achromatic red noise, (5) free chromatic (FC) noise (at $\nu = 1400$ MHz), and (6) ECORR. Using only annual SW perturbations (left), high-frequency SW effects are spuriously absorbed into the FC noise model, while the Fourier basis SWGP variations (middle) suffer from aliasing between the epochs of closest approach between the Sun and the pulsar on the sky (vertical lines). The favored model (right) recovers no additional FC noise and the $n_E(t)$ series has no noticeably unphysical structures.}
    \label{fig:swgp_realizations}
\end{figure*}

The application of the fine-grained TD GPs from \citet{ng12p5_cnm} for SW electron density variations is a novel advance in this work. We were originally alerted to the need for better SW modeling by examining GP realizations of FC delays in PSR J1909$-$3744 using an intermediate model which recovered $\chi \sim 2.5$, suggesting the FC term was accounting for unmodeled dispersive effects (\S\ref{sec:consistency_checks}). As shown in the left panels in Fig.~\ref{fig:swgp_realizations}, before including a fine-grained SWGP, the vertical lines indicating annual Sun-pulsar conjunctions reveal a SW model deficiency, as both the ECORR and FC contributions to the residuals have spikes near solar conjunction. \citet{inpta-dr2-noise-2025} see a similar ostensible FC noise using InPTA DR2, which they also attribute to the SW after residual inspection and TOA cuts based on solar elongation.

A partial solution to the problem is to use a \emph{high-frequency Fourier basis} SWGP, where corresponding noise realizations for PSR J1909$-$3744 under this model are shown in the middle column of Fig.~\ref{fig:swgp_realizations}. This model can already better account for short-timescale perturbations in $n_E(t)$ near the conjunctions, resulting in no preference for the FC noise in the bottom panel and the total model is favored with a Bayes Factor $\log_{10}\mathcal{B} = 3.47$ over the model with only annual SW perturbations on the left-hand side. However, lingering issues include increased presence of ECORR residual contributions near the conjunctions, and a few epochs of the total SW DM time series which drop below zero, implying an unphysical negative electron density. We also notice in the modulations to $n_E(t)$ in the purple middlemost panel three apparent phenonemena: 1) higher amplitude variations over the first half of the dataset, 2) an annual envelope on the $n_E(t)$ uncertainty between conjunctions, and 3) constrained, short period modulations to $n_E(t)$ within the envelope, in between conjunctions. {The annual envelope indicates that we should not be sensitive to SW effects far from conjunction; as such, the correspondence of constrained, short period modulations between conjunctions is highly unexpected.} 
{We believe these short period modulations} are a manifestation of aliasing unique to the use of a Fourier basis, given the sampling rate of the SWGP is significantly above the effective Nyquist frequency of the SW, $f_{\rm Ny} \sim 1/$yr. 

Upon adding a TD SWGP with a white noise kernel and a finer ($dt = 3$ days) basis (right panel Fig.~\ref{fig:swgp_realizations}), we see these features disappear from the ECORR realizations, the FC model is still no longer favored, and the $n_E(t)$ series is well-behaved outside of the nearest Sun-pulsar conjunctions. The TD model is strongly favored over the Fourier-basis model with $\log_{10}\mathcal{B}^{\rm TD}_{\rm Fourier} = 11.3$. In other words, the TD SWGP can account for short-timescale variations in $n_E(t)$ near the conjunctions, without poor extrapolation of the $n_E(t)$ series away from conjunction. These effects likely explain why \textsc{Model\_TD} and \textsc{Model\_Ridge} are more often preferred in pulsars with a significant SWGP signal while \textsc{Model\_Fourier} is more often preferred otherwise, as shown in Fig.~\ref{fig:final_models_bar_chart}.


\subsection{Free chromatic noise}
\label{sec:FC_noise}




\begin{figure}
    \includegraphics[width=\linewidth]{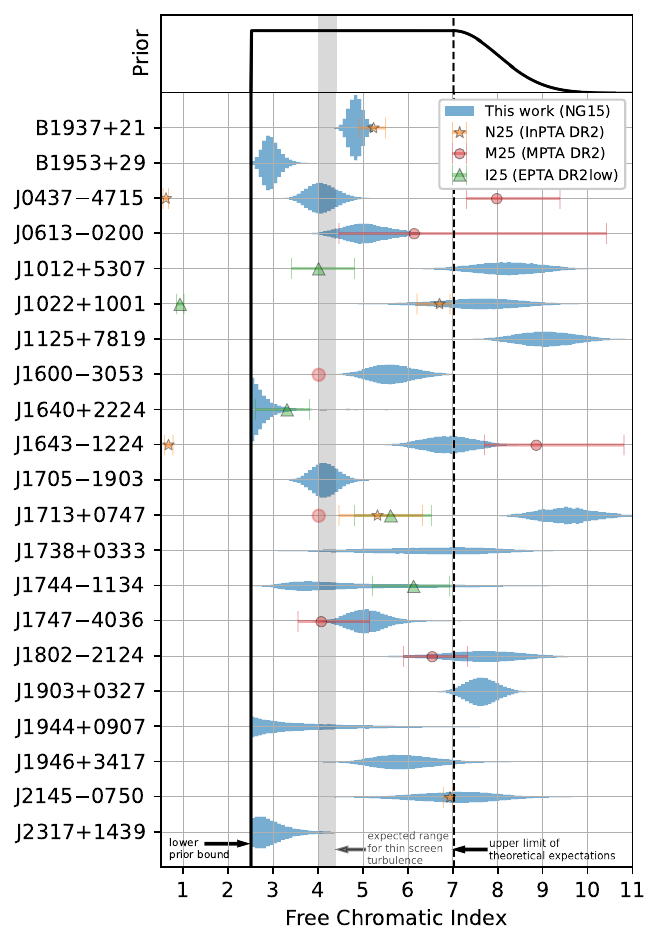}
    \caption{Posteriors on the free chromatic (FC) index $\chi$ are shown in blue for all 21 pulsars whose final noise model favored inclusion of the FC component. We observe a significant spread in the posteriors, with few favoring the expected values of $\chi = 4$ or $\chi = 4.4$ hypothesized for thin screen turbulence. The value $\chi = 7$ marks where the upper region of the prior tapers off following Eq.~\eqref{eq:chi_prior}, while the lower end of the prior is cut off at $\chi = 2.5$ to reduce model degeneracy with DM variations. For comparison, we also show previously reported FC index values from MPTA DR2 (red circles; \citealt{MPTADR2_noise}), EPTA DR2low (green triangles; \citealt{Iraci+2025}), and InPTA DR2 (yellow stars; \citealt{inpta-dr2-noise-2025}).}
    \vspace{-0.5\baselineskip}
    \label{fig:chrom_idxs}
\end{figure}

After the end of the model selection process, we find 21 pulsars still favor the presence of a FC component (Fig.~\ref{fig:final_models_bar_chart}). For six pulsars (J1125+7819, J1705$-$1903, J1738+0333, J1944+0907, J1946+3417, J2317+1439), this is the first reported measurement of significant non-dispersive chromatic noise in PTA timing residuals that we know of (although PSR J1705$-$1903 is already known for its orbital phase dependent noise, \citealt{MPTADR2_noise}). The presence of FC noise is expected for pulsars like PSR J1903+0327, which is the highest DM pulsar in NG15 (DM$=297.6$ pc/cm$^3$) and is known for its large measured scattering delays \citep{Geiger+2024, ng15detchar}. The lack of a significant FC noise is also of interest for some pulsars. For example, PSR J1909$-$3744 previously registered a FC process \citep{Larsen+2024, ng12p5_cnm}, but here we no longer detect FC noise after use of a fine-grained SWGP (cf. \S\ref{sec:swgp-results}; \citealt{inpta-dr2-noise-2025} also arrived at the same conclusion for PSR J1909$-$3744). 4 additional pulsars (B1855+09, J0030+0451, J0645+5158, and J1614$-$2230) which had significant non-dispersive chromatic noise in \citet{ng12p5_cnm} no longer have significant FC noise using NG15.

Fig.~\ref{fig:chrom_idxs} shows the FC index posteriors, $\chi$, for these 21 pulsars, and Table~\ref{tab:FC_params} tabulates all final FC parameters. 
We find consistency with the turbulent thin-screen scattering hypothesis $\chi=4$ or $\chi = 4.4$ only for PSRs J0437$-$4715, J1705$-$1903, and J1744$-$1134 (and to a lesser degree, PSRs J0613$-$0200 and J1944+0907). Meanwhile, we find extremely large FC indices in six pulsars (J1012+5307, J1022+1001, J1125+7819, J1713+0747, J1802$-$2124, J1903+0327), where the median values lie above $\chi > 7$ where our prior begins to exponentially suppress our posteriors. In this regime, it is likely that a relatively small number of TOAs in the very lowest subbands are dominating the inference. \citet{Geiger+2024} recently predicts a very shallow ($\chi \sim 0.6$) chromatic index for PSR J1903+0327's FC noise, since the scattering delays are large in comparison to the intrinsic pulse width. As our priors on $\chi$ do not probe this regime, the large measured value for $\chi$ in PSR J1903+0327 may be a result of mismodeling. Furthermore, \citet{Kulkarni+2025} recently showed that a scattering process may be detected with much larger $\chi$ than expected if the intrinsic variations are non-Gaussian. As such, these processes we recover with large values for $\chi$ may suffer from mismodeling but still plausibly originate from interstellar scattering.

We also find some pulsars prefer relatively small values $\chi < 3$, which suggests some leftover degeneracy between FC and DM variations, despite our lower prior cutoff. In particular, PSR J1640+2224 displays multiple modes in its DM and FC posteriors (cf. Fig.~\ref{fig:dmgp-td-params}), which coupled with its small FC index posterior $\chi = 2.8^{+1.1}_{-0.2}$, indicate the FC model is standing in as a second component of the DM variations model. 


In Fig.~\ref{fig:chrom_idxs}, we also compare our FC index results with values measured in three independent datasets: MPTA DR2 \citep{MPTADR2_noise}, InPTA DR2 \citep{inpta-dr2-noise-2025}, and EPTA DR2low \citep{Iraci+2025}. These datasets probe slightly different timespans and radio-frequency bands: MPTA DR2 covers a 4.5 year timespan with observations at L-band, InPTA DR2 covers a 7.2 year timespan with dual L-band and 300-500 MHz observations, and EPTA DR2low features varying timespans per pulsar while extending radio frequency coverage down below 300 MHz. The FC index is consistent with MPTA DR2 for 4 pulsars (J0613$-$0200, J1643$-$1224, J1747$-$4036, J1802$-$2124), consistent with InPTA DR2 for 3 pulsars (B1937+21, J1022+1001, J2145$-$0750), and consistent with EPTA DR2low for 2 pulsars (J1640+2224, J1744$-$1134). As for inconsistencies, we find our $\chi$ value for PSR J1713+0747 is a notable outlier, being significantly higher than the values estimated from all three independent studies. Our $\chi$ values for PSRs J0437$-$4715 and J1600$-$3053 are also inconsistent with MPTA DR2, and our $\chi$ value for PSR J1012+5307 is very inconsistent with EPTA DR2low. \citet{inpta-dr2-noise-2025} also found the FC index value to be lower than our prior range for 2 pulsars, and \citet{Iraci+2025} for 1 pulsar. The reasons for these inconsistencies are almost certainly related to modeling, e.g. if the noise models are not synchronized across PTAs, or the FC model is misspecified in some way. If the latter is true, this could suggest evidence for deviations from stationarity or Gaussianity in the FC process, for a time-variable FC index, and/or for a FC index that changes across scales. More advanced noise model selection with future IPTA datasets may {shed light on the origin of these discrepancies between datasets}. However, if the origin of any FC delays are related to intrinsic profile variability as suggested in \citet{MPTADR2_noise}, then differences in profile templates used across PTAs may introduce difficulties for characterizing the noise in future IPTA datasets.

\subsubsection{Annual Chromatic Variations}
\label{sec:annual-FC-variations}

\begin{figure}
    \centering
    \includegraphics[width=\linewidth]{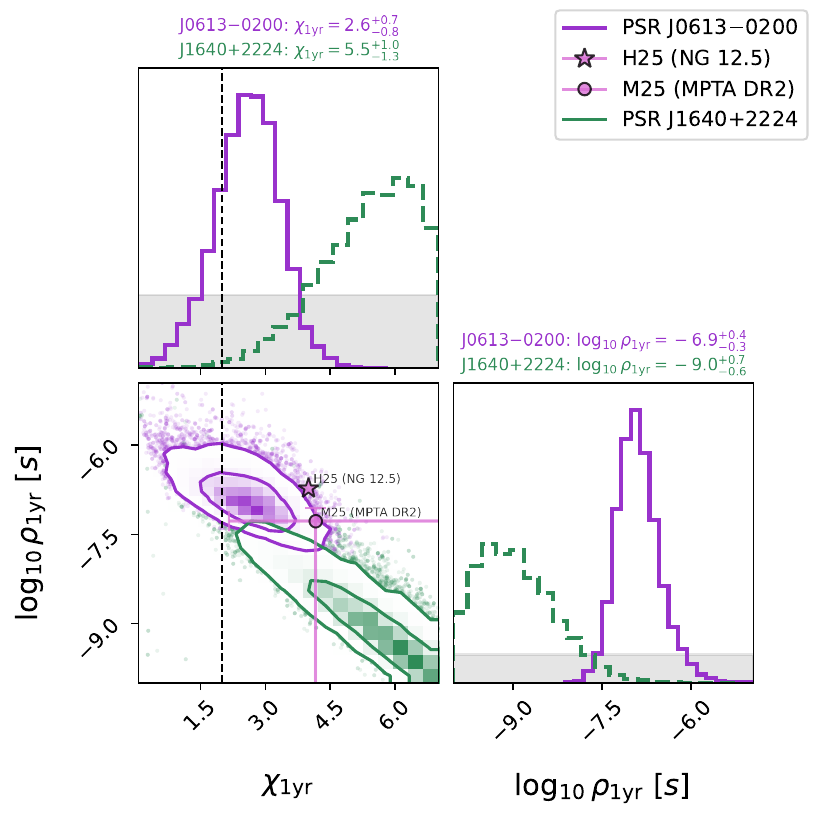}
    \caption{Posteriors on the free chromatic annual parameters $\chi_{1\rm{yr}}$, $\log_{10}\rho_{1\rm{yr}}$ for PSRs J0613$-$0200 (solid purple) and J1640+2224 (dashed green). The posterior distributions for PSR J0613$-$0200 are consistent with a dispersive origin ($\chi_{1\rm{yr}} = 2$, dashed), whereas those for PSR J1640+2224 are not. Shaded regions show the uniform priors. Reported values in the panel headings are the parameter medians and 68\% credible intervals from the posteriors. For comparison, annual chromatic parameters for PSR J0613$-$0200 as reported by \citet{ng12p5_cnm} and \citet{MPTADR2_noise} are indicated by the purple star and circle respectively.}
    \label{fig:annual_variations}
\end{figure}

We find evidence for annual chromatic variations in two pulsars (J0613$-$0200 and J1640+2224). The posterior parameter distributions recovered from the annual FC component under their final noise models are shown in Fig.~\ref{fig:annual_variations}. For PSR J0613$-$0200, the FC index $\chi_{1\rm{yr}} = 2.6^{+0.7}_{-0.8}$ is consistent with a dispersive origin as well as previously reported parameter values from \citet{MPTADR2_noise, ng12p5_cnm} (cf. \S\ref{sec:annual-dm-results}). It is also possible that scattering contributes to this signal based on previous observations of annual scintillation arc variability \citep{Main+2020, Liu+2023scint}. For PSR J1640+2224, we find $\chi_{1\rm{yr}} = 5.5^{+1.0}_{-1.3}$, which is inconsistent with a dispersive process. 


We also test for annual variations in PSR J1614$-$2230 since \citet{ng12p5_cnm} used an annual chromatic model for this pulsar, but the model is not favored in NG15 with $\mathcal{B}^{\rm{FC}_{\rm{1yr}}}_\oslash \sim 0.6$.


\subsection{Deterministic Signals}
\label{sec:det_signal_results}


We include decaying exponential signals in the final noise models for three pulsars. Table~\ref{tab:det_params} in Appendix~\ref{appendix:tables} gives the posterior parameter values for each event. PSR J1713+0747 strongly favors two event signals \citep{lam+2018}, with nearly identical posteriors as \citet{Larsen+2024} previously obtained, also using NG15. As discussed in \citet{Larsen+2024}, the two events have strong impacts on PSR J1713+0747's achromatic RN spectrum, we which see later on in \S\ref{sec:achromatic-red-noise-comparison}. We confirm the presence of a decaying exponential event in PSR J1643$-$1224 in the NG15 dataset, which has previously been studied in PPTA datasets and also manifest as a change in the pulse profile \citep{Shannon+2016, goncharov+2021}. We also recover the event's inverted chromaticity with a posterior $\chi_{\rm exp} = -1.6^{+0.3}_{-0.3}$, which is slightly steeper than the value $\chi_{\rm exp} = -1.0^{+0.1}_{-0.1}$ found by \citet{goncharov+2021}. We also find evidence of an event in PSR J2145$-$0750, which was first identified by \citet{goncharov+2021} with an achromatic spectrum, and showed no associated profile variation. We find using NG15 that this event slightly favors an inverted spectrum, $\chi_{\rm exp} = -0.6^{+0.4}_{-0.4}$. Like PSR J1713+0747, inclusion of this event has a strong impact on this pulsar's achromatic RN spectrum (\S\ref{sec:achromatic-red-noise-comparison}).

Some other pulsars did not warrant sufficient evidence to keep an event signal in their final noise model. For example, a possible event in PSR J0613$-$0200 from \citet{ng12p5_cnm} we identified as an outlier observation rather than a longterm process that requires modeling (cf. \S\ref{sec:sw_excision}). Additionally, we found marginal evidence for a Gaussian event in PSR J1600$-$3053 first identified by \citet{pptadr3:noise}. However, for NG15 we obtained a Savage-Dickey Bayes Factor of only $\mathcal{B}^{\rm Gauss}_\oslash \sim 46$, which we do not consider sufficiently large enough to keep the signal in the final model. When excluded, the signal instead manifests as a slight bump in PSR J1600$-$3053's DM time series near MJD 57572. We interpret that the NANOGrav data cannot quite distinguish this signal as a deviation from DM variations following a power law spectrum.

\subsection{Auxiliary results}
Here we discuss results of our model selection process pertaining to individual pulsars that do not fall neatly into the above discussions.

\subsubsection{Red and DM noise degeneracy}
\label{sec:noise_covariance}

\begin{figure}
    \centering
    \includegraphics[width=\linewidth]{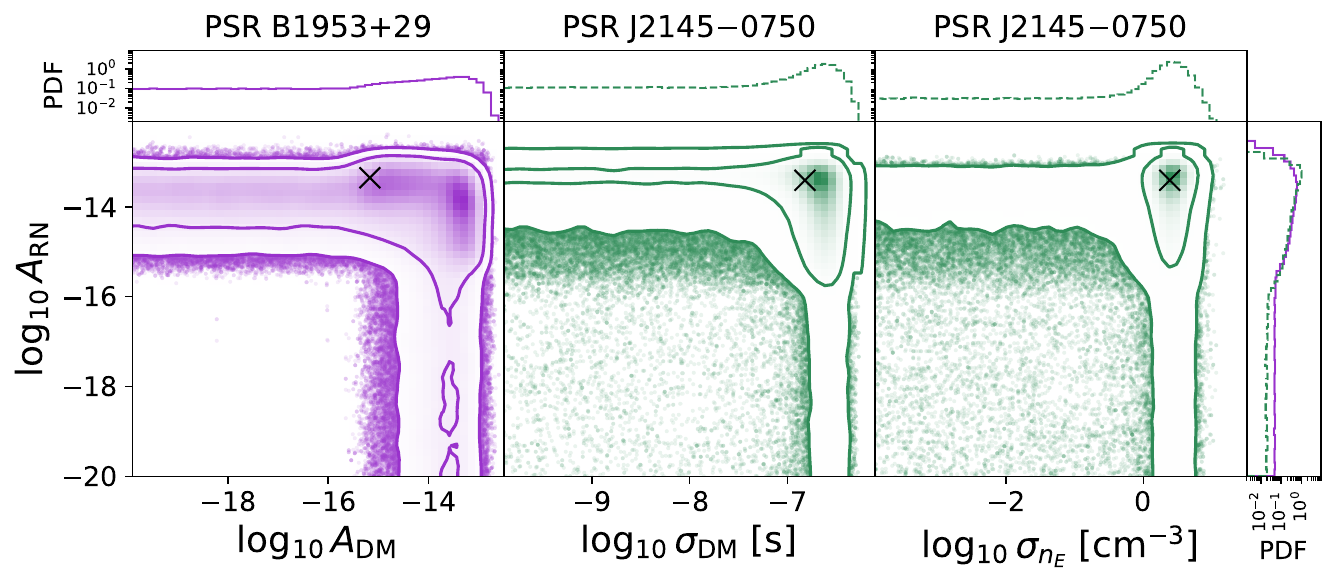}
    \caption{Degeneracies between red noise (RN) and DM model components are apparent only for PSRs B1953+29 (purple) and J2145+0750 (green) via covariances between parameters. For PSR J2145+0750, the covariance is three-way between RN, DMGP, and SWGP. Crosses indicate the location of the maximum a posteriori sample.}
    \label{fig:RN_DM_covariance}
\end{figure}

Covariances between achromatic and chromatic noise parameters are known occur if there are insufficient TOAs across time and radio frequency (relative to the intrinsic noise amplitudes) to resolve the noise processes from each other. The observation of this degeneracy is cause for concern as it has been shown to reduce sensitivity to the GWB \citep{hazboun:2020slice, Ferranti+2025, MPTA_gwb, DR2Lite_2025}. Out of our $67$ custom pulsar models, only two pulsars (B1953+29 and J2145$-$0750) show any noticeable degeneracy between red and DM noise processes. The covariant parameters are shown in Fig.~\ref{fig:RN_DM_covariance}. Interestingly, both pulsars register significant FC noise, constituting the only two pulsars with FC noise that do not have significant DM noise (cf. Fig.~\ref{fig:final_models_bar_chart}). On one hand, this observation suggests that the degenerate red and DM noise in both pulsars is most likely just unresolved DM noise. However, deeper investigation for PSR B1953+29 suggests that the degeneracy is related to the sparse TOAs with the AO ASP backend, collected primarily at L-band and S-band rather than 430 MHz just during the first 3 years of observation \citep{ng15data, sosa+2024dmest}. This is similar to the example of inhomogeneous frequency coverage-induced degeneracies highlighted by \citet{Ferranti+2025} for the full version of EPTA DR2.

For PSR J2145$-$0750, the issue is more complicated. Firstly, its achromatic RN is dominantly impacted by a deterministic transient event included in its model (cf. \S{\ref{sec:det_signal_results}, \S\ref{sec:achromatic-red-noise-comparison}}). After its inclusion, there remains a three-way degeneracy between achromatic RN, DM noise, and stochastic SW variations. The Bayes Factors for each individual process are $\mathcal{B}^{\rm{RN}}_\oslash = 6.0$, $\mathcal{B}^{\rm{DM}}_\oslash = 1.7$, and $\mathcal{B}^{\rm{SWGP}}_\oslash = 5.8$, with marginally sub-threshold SW and RN Bayes Factors. Application of Eq.~\eqref{eq:nested_BF} to compute the simultaneous significance of DM and SW variations yields $\mathcal{B}^{\rm{DM}+\rm{SWGP}}_\oslash \sim 107$, revealing there are indeed prominent DM variations on top of a FC noise, but the data alone cannot isolate the source as pulsar-intrinsic perturbations to $n_E(t)$ or to a generic DM noise induced by the ISM. 

\subsubsection{Unphysical outliers identified by SWGP}
\label{sec:sw_excision}

Modeling perturbations to $n_E(t)$ as a GP has one limitation that it allows the SW electron density model to return unphysical negative values. As the SW electron density is intrinsically positive, the data are unlikely to support negative values of $n_E(t)$ unless additional noise leaks into the SWGP. As such, identification of these negative SW epochs is useful to diagnose unmitigated noise from the ISM or potential outliers related to RFI.

While post-processing some of our noise models at an intermediate stage of the selection process, we observed a particularly egregious case in PSR J0613$-$0200, where a very large and significant drop in $n_E(t)$ down to negative values took place near MJD 58000. Both \citet{Larsen+2024} and \citet{ng12p5_cnm}, who do not use a time-variable SW model, instead notice a sudden drop in the chromatic model at the time of this observation. Applying a deterministic dip model isolated the timing residual offset to the observation at MJD 57923, and follow-up analysis revealed a large offset in the timing residuals between GBT 800 and 1200 MHz receivers which was inconsistent with a DM trend. Further investigation of the observation indicated likely presence of unmitigated RFI in the 800 MHz band. As such, we decided to excise the TOAs from the 800 MHz band at MJD 57923 as outliers, after which the unphysical trend in $n_E(t)$ was no longer present. This observation was not identified as an outlier in \citet{ng15data} as the offset was instead absorbed as a sudden drop in DMX. This example highlights the need for further development of outlier analyses in tandem with alternative chromatic modeling for future analyses.

Similarly to PSR J0613$-$0200, we also found a very significant, unphysical drop in the SW electron density, $n_E(t)$, associated with an observation at MJD 58482 in PSR J1944+0907. However, unlike PSR J0613$-$0200, we find the timing residuals within the observation (between AO 430 MHz and L-band) to be consistent with a DM trend. As such, this observation could have an astrophysical origin if there was a small underdensity in the ISM which has been misattributed to the SW by our models. As the origin of this outlier is unclear, we do not excise the observation in our analyses. However, this pulsar deserves further scrutiny in the future to understand if the observation could be affected by RFI.



\section{Results: Comparisons with base chromatic noise model (DMX)}
\label{sec:comparisons-with-dmx}

Here we detail how changing the treatment of chromatic noise using our individually-tailored pulsar noise models impacts the \emph{achromatic} noise properties of the pulsars. The standard treatment uses the \textsc{DMX} model parameters in \texttt{PINT} generically across all pulsars to account for all chromatic contributions as a piecewise-constant DM fit at each observation epoch \citep{ng15data, ng15detchar}. Additionally, in \S\ref{sec:chromgp_only_compare} we see how single pulsar noise would instead be impacted using a more generic GP model which has not undergone the detailed model selection process of \S\ref{sec:methods} but does include non-dispersive chromatic noise.

\subsection{Achromatic Red Noise Comparison}
\label{sec:achromatic-red-noise-comparison}

\begin{figure*}
    \centering
    \includegraphics[width=0.85\linewidth]{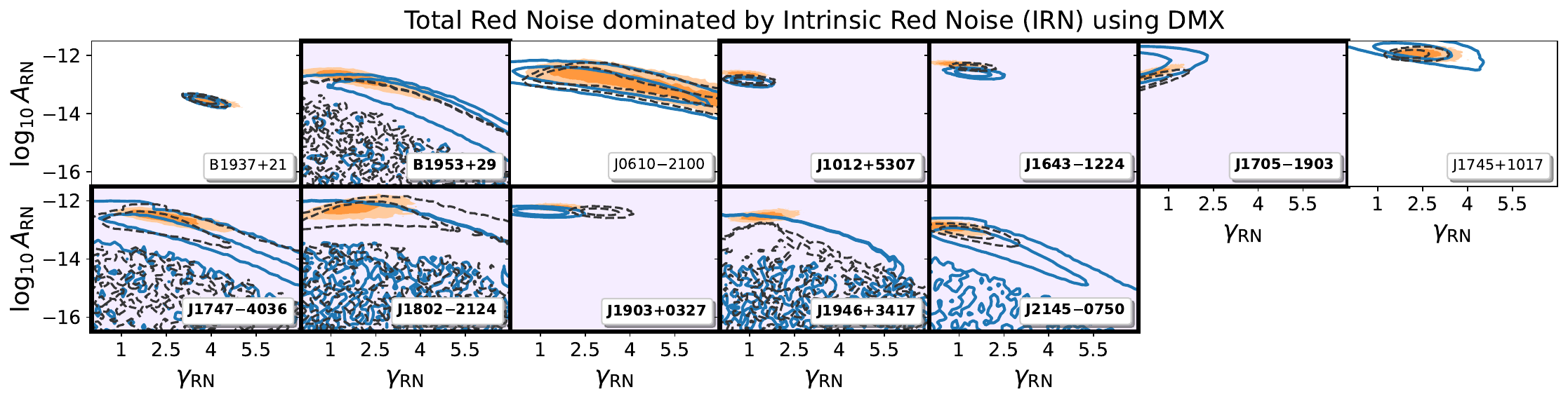}
    \centering
    \includegraphics[width=0.85\linewidth]{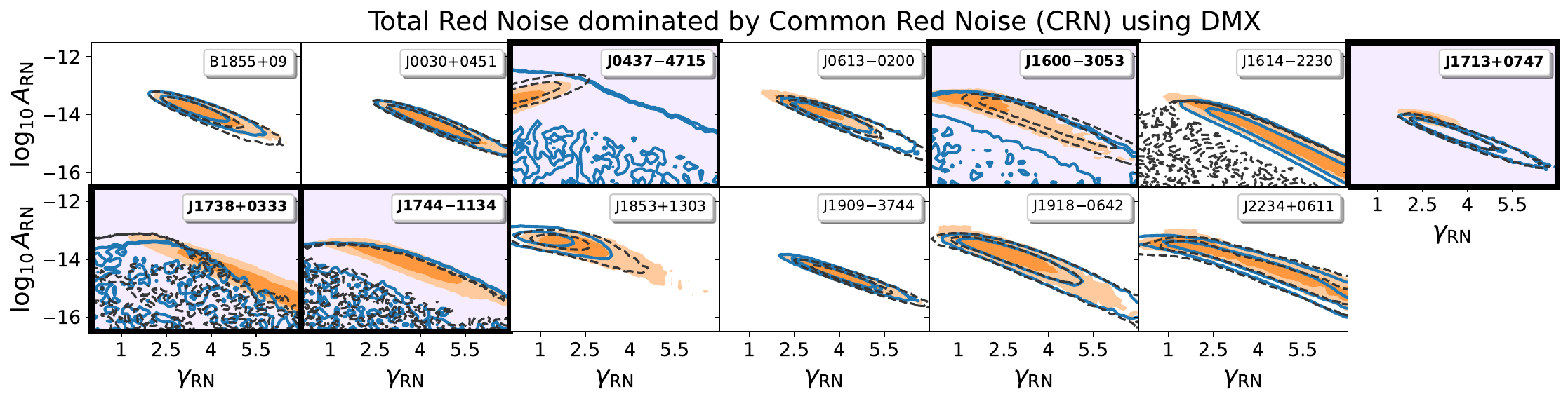}
    \centering
    \includegraphics[width=0.85\linewidth]{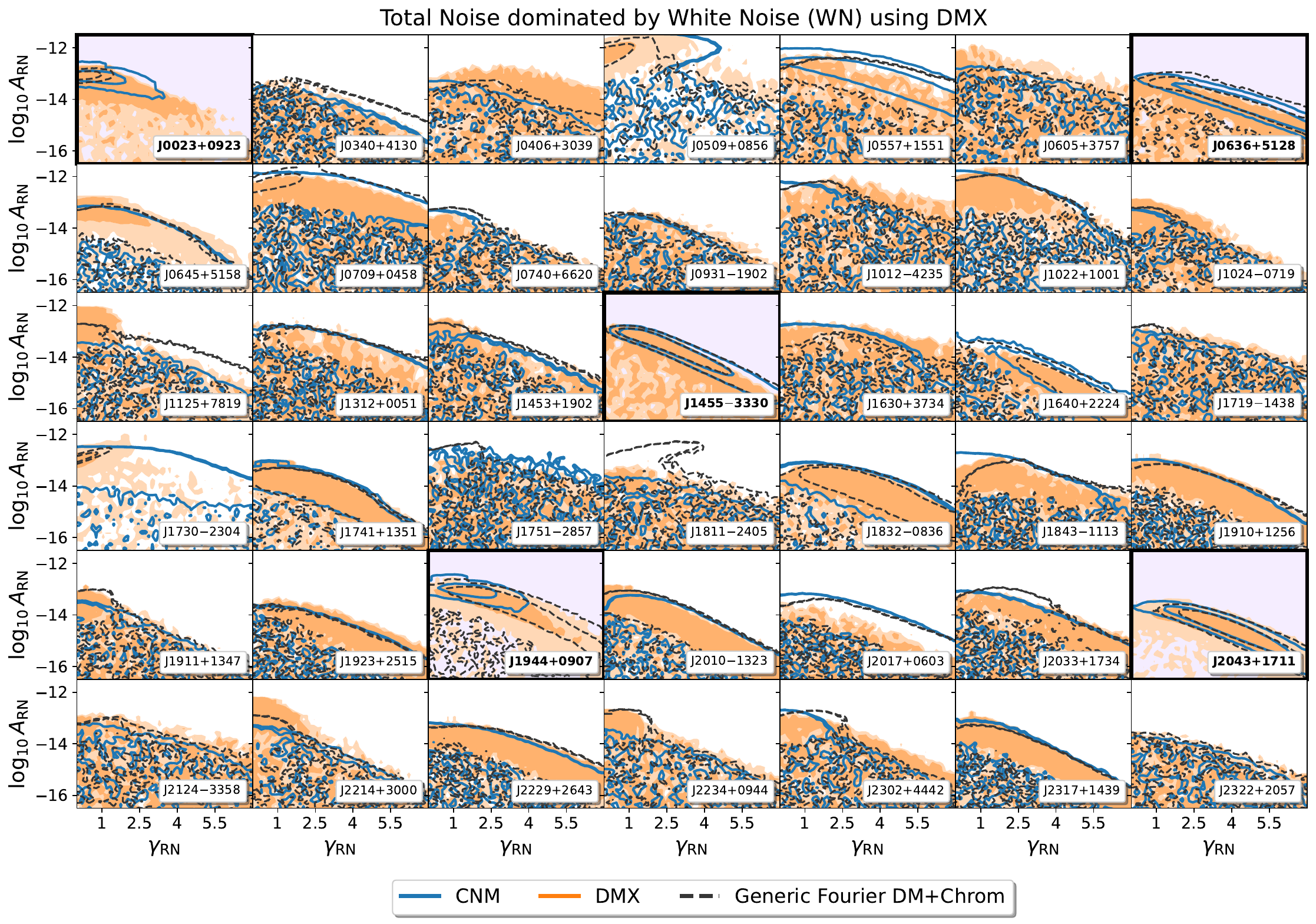}
    \caption{Comparison of posterior achromatic red noise (RN) hyperparameters ($\log_{10}A_{\rm RN},\gamma_{\rm RN}$) under our custom chromatic noise models (CNM; blue), the standard DMX model (orange shaded), and a generic ``uncustomized'' power law spectral DM GP + chromatic GP scaling as $\Delta t \propto \nu^{-4}$, with no time-variable SW components ({grey} dashed; see \S\ref{sec:chromgp_only_compare}). Pulsars are grouped depending whether they are dominated by an intrinsic red noise (IRN), a common red noise (CRN), or a white noise (WN) process under the DMX model, as identified in \citet{ng15detchar}. Highlighted subplots identify pulsars where either the achromatic RN hyperparameters or the Savage-Dickey Bayes Factors have changed significantly after applying the CNM. Among the CRN-dominated pulsars, PSRs J1713+0747 and J0437$-$4715 are the most significantly impacted. 9 out of 12 IRN-dominated pulsars are significantly impacted by the CNMs, and 5 of the previous WN-dominated pulsars newly have significant RN using their CNMs.}
    \label{fig:RN_params}
\end{figure*}

The impacts on achromatic RN are perhaps the most consequential for GW analyses, since the GWB should manifest as a component of the total RN in single pulsar noise analyses. \citet{ng15detchar} previously identified two populations of NG15 pulsars with significant RN, one with significant \emph{intrinsic} red noise (IRN) which lies 1-2 orders of magnitude above the \emph{common} red noise (CRN), and one with RN similar to the CRN and no additional significant IRN. We label these populations the ``IRN-dominated'' and ``CRN-dominated'' pulsars. 
Pulsars with no significantly measured RN are instead dominated by white noise (WN-dominated). We emphasize that mismodeled chromatic noise may comprise a component of the total noise, introducing biases affecting the IRN, CRN, WN, or combination thereof, for pulsars in any of these categories. By analyzing how the RN changes under our custom chromatic models, we can infer how mismodeled chromatic noise may impact the analysis.

RN parameters are summarized in Fig.~\ref{fig:RN_params}, with exact values under the custom nosie models (CNMs) in Tables~\ref{tab:FD_params}, \ref{tab:TD_params}. Across the board, we find that applying our CNMs tends to either reduce the significance of RN, shift the hyperparameters to lower amplitude and higher spectral index, or have negligible effect on the RN. There are also exceptions and interesting edge cases. $19$ pulsars in total display \emph{significant changes} in RN, as determined either by a large change in parameters ($\Delta\theta > 1\sigma$) or Savage-Dickey Bayes Factor ($|\Delta\log_{10}\mathcal{B}^{\rm RN}_\oslash| > 1$). To continue the discussion, the pulsar RN parameters in Fig.~\ref{fig:RN_params} are broadly grouped based on the previous categorization (IRN-dominated, CRN-dominated, WN-dominated).

\subsubsection{IRN-dominated | 12 pulsars}

The majority of these pulsars, shown at the top of Fig.~\ref{fig:RN_params} display significant changes to their RN parameters under the CNMs, indicating that mismodeled chromatic noise indeed underpins the origins of their noise. All 12 of these pulsars except for PSRs J0610$-$2100 and J1745+1017 favor an FC component to their noise model (intended to mitigate scattering) with 5 additionally favoring band-dependent DM variations and/or chromatic events in their models.

For 5 out of 12 pulsars (B1953+29, J1747$-$4036, J1802$-$2124, J1946+3417, J2145$-$0750), the achromatic RN is no longer significant ($\mathcal{B}^{\rm RN}_\oslash < 10$), although in no cases do the new posteriors completely rule out the previously significant IRN process. Nonetheless, use of the CNMs implies these pulsars are more likely to eventually contribute to or constrain a CRN process in the full PTA. The RN hyperparameters for PSRs B1953+29 and J2145$-$0750 notably show covariances with the DM hyperparameters (cf. \S\ref{sec:noise_covariance}), indicating the RN will be highly sensitive to alternative priors (e.g. hierarchical instead of log-uniform) on the DM hyperparameters in NG15.

Another of the 4 IRN-dominated pulsars (J1012+5307, J1643$-$1224, J1705$-$1903, J1903$+$0327) have a significant, low spectral index IRN which reduces in amplitude but still remains after applying our CNMs. The noise in each of these four pulsars is especially intriguing in its own right (cf. e.g., \citealt{Chalumeau+2022, Lentati+2017, Morello+2019, Geiger+2024}). PSR J1705$-$1903 in particular is in tight binary and eclipses for $\sim30\%$ of its orbit \citet{ng15data}, while PSR J1643$-$1224 lies behind an \textsc{Hii} region \citep{Ocker+2024}. While our CNMs appear to have mitigated a portion of their IRN, it is likely that our models are still insufficient to properly model all the chromatic noise in these pulsars. We suggest more sophisticated TOA models or mitigation on the level of pulse profiles will be required for complete characterization.

3 pulsars remain where IRN persists and does not significantly change based on the chromatic model. PSR B1937+21, an isolated millisecond pulsar, is well-known for its unusually dominant achromatic RN whose origin has been purposed to be due to an asteroid belt \citep{shannon+2013asteroidbelt} or rotational instabilities from torques in the magnetosphere \citep{kramer+2006}. The RN in this pulsar remains despite this pulsar favoring among the most complicated chromatic models. Meanwhile, PSRs J0610$-$0200 and J1745+1017 are spider pulsars and could be impacted by unmodeled orbital irregularities. Interestingly, the model for J1745+1017 favors a fine-grained solar wind despite theoretically being very insensitive to the SW (Fig.~\ref{fig:SW_SNR}), which suggests that this pulsar's model requires a unique variety of ISM model not covered by our suite of GPs.

\subsubsection{CRN-dominated | 13 pulsars}

As shown in the middle rows of Fig.~\ref{fig:RN_params}, 5 out of these 13 show significant changes to their total RN under our CNMs, with the remaining 9 experiencing only minor shifts in their posteriors. The pulsar with arguably the most important change is PSR J1713+0747, where RN is detected with a significantly lower amplitude ($\log_{10}A_{\rm RN} = -14.7^{+0.3}_{-0.4}$) and higher spectral index ($\gamma_{\rm RN} = 3.7^{+1.2}_{-0.8}$) than before. As this is the only pulsar in this category with such a significant change, we assess that PSR J1713+0747's CNM is expected to have the most outsized impact on a GWB analysis. For more detailed discussion on the impact of including model components for PSR J1713+0747's chromatic events in the NG15 dataset, see \citet{Larsen+2024}, and \citet{hazboun:2020slice, ng12p5_cnm} for previous NANOGrav datasets.

As for the remaining pulsars, RN is no longer detected in PSRs J0437$-$4715, J1600$-$3053, J1738+0333, and J1744$-$3744 using the CNM. As all 4 favor a FC noise component, we interpret that these pulsars do not yet have enough sensitivity to detect both the CRN and a FC noise, and using their CNMs will conservatively downweight their impact in GW analyses. PSR J1600$-$3053, which is especially known for the presence of scattering variations, has similarly lost detection significance for a CRN in previous work \citep{Chalumeau+2022, ng12p5_cnm}. Using the CNMs, we no longer register the very shallow spectral RN in PSR J0437$-$4715, and find that a portion of the previous posterior in PSR J1738+0333 is now disfavored. Although the changes are not significant, PSRs B1855+09 and J1853+1303 do register slight increases to their RN amplitudes under the CNMs. These changes imply either that pulsars could drive the CRN to larger amplitude, or that a more dominant intrinsic component of their RN has emerged (due either to the increased sensitivity afforded by, or some insufficiency of, their CNMs) which may reduce their impact on GWB analysis.

\subsubsection{WN-dominated | 42 pulsars}

For these pulsars, no significant RN process is detected under the standard DMX chromatic model, as shown on the bottom of Fig.~\ref{fig:RN_params}. After applying our CNMs, we now find a significant RN process in 5 pulsars (J0023+0923, J0636+5128, J1455$-$3330, J1944+0907, and J2043+1711). We attribute the change in PSRs J1455$-$3330 and J0636+5128 to the more simplified ridge-model for DM variations increasing sensitivity to RN (see \citealt{Lam+2025, ng12p5_cnm} for details on effects of simplified DM modeling in PSR J1455$-$3330). PSR J2043+1711 favors a slightly more complicated SE-kernel DMGP and Ridge-kernel SWGP, nonetheless its RN is still more significant using our CNMs and highly in line with the CRN. PSR J0023+0923 meanwhile registers a much more significant, shallow spectral RN ($\gamma_{\rm RN} = 1.1^{+0.9}_{-0.7}$) under the CNM. PSR J0023+0923 is especially challenging as it is both a very short binary-period pulsar and is near the solar ecliptic plane on the sky, so its newly significant RN may arise either because we have misspecified the SW model or we are seeing unmodeled orbital and/or stellar wind effects from its binary companion. PSR J1944+0907 similarly registers newly shallow-spectral RN ($\gamma_{\rm RN} = 1.7^{+0.8}_{-0.7}$). While this RN could have an intrinsic origin, we find it most likely that PSR J1944+0907 has hit an edge case where our consideration of DM kernels was not comprehensive enough, as the favored SW model yields unphysical negative values for the SW electron density $n_E$ at certain epochs (cf. \S\ref{sec:sw_excision}). In contrast, the DMX model places no assumptions on the autocorrelations or origins of the DM variations and can easily account for such outliers.

While the remaining 37 pulsars still do not register any significant RN after applying our CNMs, we can still place upper limit constraints on their RN, and GWs by extension. Overall, 24 out of 37 pulsars have a reduced upper limit $A_{\rm RN}^{95\%}$ after applying the custom chromatic model, whereas only 13 have a larger value of $A_{\rm RN}^{95\%}$. This shows that applying CNMs nets an overall sensitivity gain among the pulsars with insignificant RN. A few interesting cases are gleaned from Fig.~\ref{fig:RN_params}. For PSR J1125+7819, the CNM (which includes FC noise and band-dependent DM variations) now rules out a low spectral-index mode of the posterior. The CNM yields a much broader RN posterior for the highly ecliptic PSR J1730$-$2304, most likely resulting from our new treatment of the SW. The largest increase in RN upper limit occurs for PSR J2017+0603, whose CNM favors only a simple but well-constrained 50 frequency Fourier basis DMGP. Interestingly, two pulsars (J1640+2224, J2317+1739) in this category register no changes to their RN posteriors, despite favoring complicated chromatic models due to their highly informative and low frequency AO data.

\subsubsection{Comparison to a generic chromatic model}
\label{sec:chromgp_only_compare}

Alongside our custom chromatic models and the standard DMX model, we also apply a third noise model on each pulsar which includes a generic DMGP and non-dispersive chromatic noise process. This is based on the chromatic model from \citet{Larsen+2024}, which applied a Fourier-basis DM GP with $N_f = 100$, a Fourier-basis chromatic GP with $\chi = 4$ and $N_f = 150$ intended to model scattering, and a time-independent fit for $n_E$, alongside the same timing model settings as our CNMs. The two chromatic events are also included for PSR J1713+0747. 
Differences between our CNMs and this generic Fourier-basis model (henceforth ``uncustomized model'') may therefore result from 1) use of a TD basis, 2) inclusion of a more sophisticated SW model, 3) use of a FC index, or 4) use of a coarser or finer GP basis size. Recall the uncustomized model in general has a much lower Bayes Factor than the CNMs as even for pulsars with no measurable SW, the basis size for DM and FC variations is not optimized.

The posterior RN parameters in each pulsar under the uncustomized model are shown in Fig.~\ref{fig:RN_params}, alongside the results for our CNMs and DMX. Starting with the IRN-dominated pulsars, we find the pulsars that lost a detection of RN using the CNMs also lost it using the uncustomized models, strengthening the evidence that the originally significant RN is a result of mismodeled scattering, with the exception of PSR J2145$-$0750 which only lost RN significance using our CNMs due to the inclusion of a chromatic event. The RN posteriors are noticably shifted using the uncustomized model for PSRs J1903+0327, J1643$-$1224, and J1705$-$1903, suggesting the exact form of the non-dispersive noise model has large impact on noise (mis)characterization for these pulsars. Strangely, the RN upper limit for PSR J1946+3417 using the uncustomized model is noticeably more constrained than using the CNM, despite the two models being nearly identical outside of using $N_f = 150$ for the DM model and a deterministic SW model using the CNM.

As for the CRN-dominated pulsars, the RN posteriors using the uncustomized model largely track the CNM results, particularly for PSR J1713+0747 where the outsized impact of the change is due to its chromatic events, which are included in the uncustomized model. The uncustomized model most strongly deviates from the CNM results for PSR J0437$-$4715, a short timespan pulsar which favors \textsc{Model\_Ridge} for chromatic noise using the CNMs. Overall, the comparison is suggestive that most of the impacts of using the CNMs on the inference of a CRN process may be achieved with an uncustomized chromatic model, particularly if PSR J1713+0747's events are included. Further work is currently underway using simulated datasets to more deeply understand the expected sensitivity gains from using this model \citep{Chang+2026}.

Some of the RN posterior changes among the WN-dominated pulsars using the uncustomized model are also revealing. For example, not including a time-variable SW model in the uncustomized model yields very shallow and constrained RN spectra for PSRs J0023+0923 and J1730$-$2304, while the CNMs result in broader posteriors for both cases. A few pulsars (e.g., J0340+4130, J1125+7819, J1811$-$2405, J2234+0944) have noticeably larger upper limits on their RN using the uncustomized model instead of the CNMs. For PSR J1125+7819, the CNM likely yields a lower RN upper limit due to inclusion of the band-dependent DM component. In other cases, the RN uncertainties may be larger using the uncustomized model due to inclusion of scattering term where it is not favored in the CNM. The uncustomized model also produces much broader constraints on RN for PSR J1944+0907, which only registers a constrained RN using the CNM. This again suggests J1944+0907 may have hit an edge case in our model selection process where the SW model is overfitting to high-frequency noise fluctuations at the expense of other components of the model.

\subsubsection{Free spectral analysis}

\begin{figure}
    \centering
    \includegraphics[width=\linewidth]{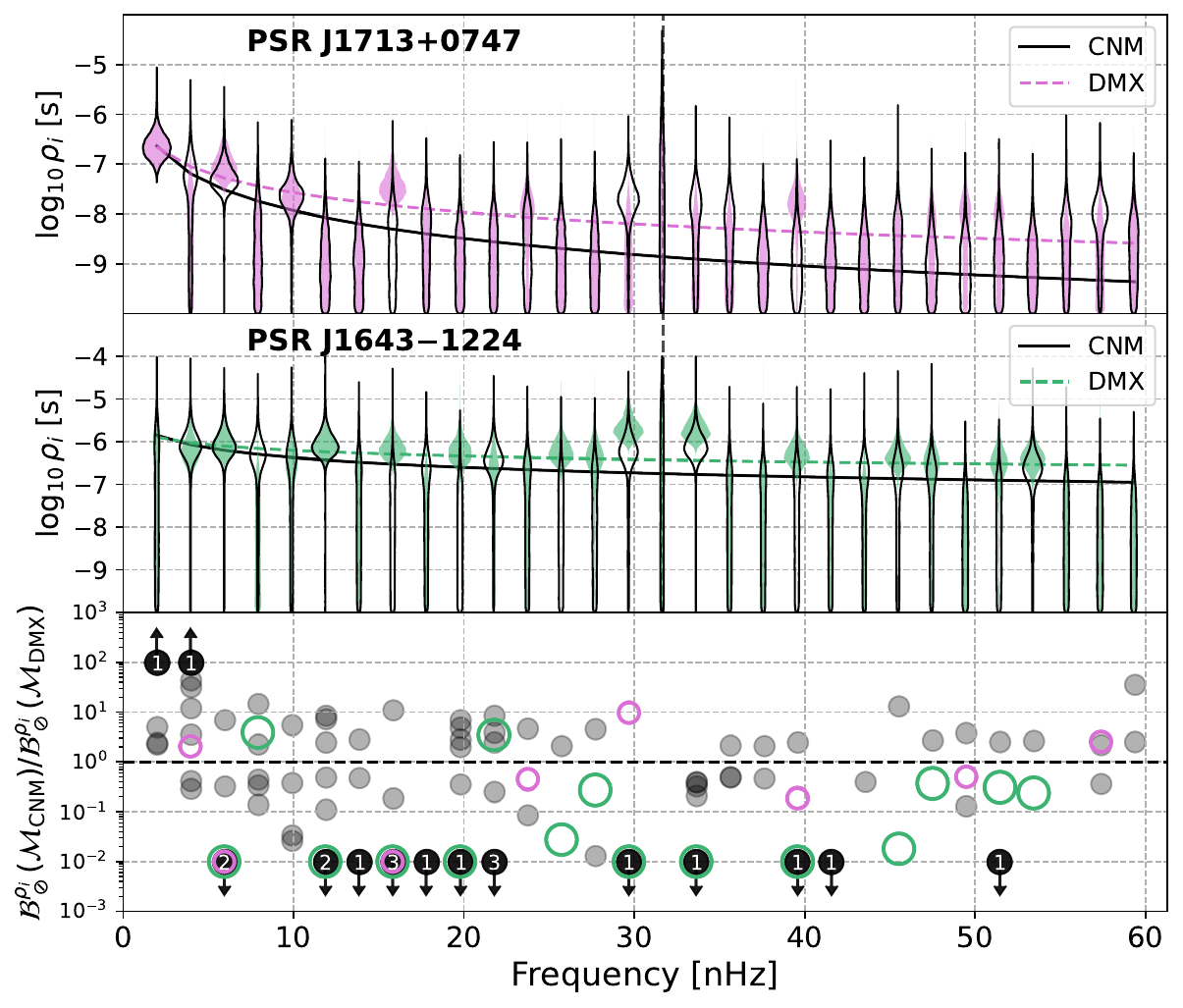}
    \caption{Free spectral red noise (RN) analysis under our custom chromatic noise models (CNMs) as compared with the standard DMX model. \emph{Top two panels:} The log of the timing delay spectrum (where violins represent posterior PDFs, curves are median power law spectra) for two representative pulsars (J1713+0747 and J1643$-$1224) under our CNM (outlined black) and under DMX (filled colors). \emph{Bottom panel:} The ratio of the per-frequency Savage-Dickey Bayes Factor, Eq.~\eqref{eq:FSSDBF}, evaluated under CNM vs the DMX at each frequency for PSR J1713+0747 (pink), J1643$-$1224 (green) and all remaining pulsars (black), wherever this ratio is greater than 2 or less than 1/2. {Black circles with numbers inside indicate how many pulsars gain or lose a detection of power at the given frequency bin, as determined by the ratio being greater than 100 or less than 0.01.} Results suggest an overall reduction in detections of power at higher frequency, with some new detections of power at lower frequency using the CNMs. No significant power is measured at $f = 31.7\text{ nHz} = 1/\rm{yr}$ due to the fit for astrometric timing model parameters.}
    \label{fig:freespec_BFs}
\end{figure}

We further examine the total achromatic noise of our pulsars below the cutoff frequency of our RN models $f < 60$ nHz using a free spectrum (FS) model, {Eq.~\ref{eq:free_spectrum}}, which relaxes the assumption of a power law PSD by instead assuming independent timing residual amplitude at each frequency, $\rho_i \equiv \rho(f_i)$. A FS may also be used to assess statistical detection significance of power in each pulsar at a per-frequency level using the Savage-Dickey Bayes Factor, which may be computed from the tails of the posteriors on $\log_{10}\rho_i$ as
\begin{align}
    \mathcal{B}^{\rho_i}_\oslash = \frac{\pi(\log_{10}\rho_i)}{\mathcal{P}(\log_{10}\rho_i < -9|\vec{\delta t})}.
    \label{eq:FSSDBF}
\end{align}
We then compare the achromatic FS in each pulsar using our CNMs and the standard DMX model for chromatic noise to understand how CNMs impact the noise characterization on a per-frequency level.

The top two panels of Fig.~\ref{fig:freespec_BFs} shows the FS under both our CNMs and under DMX for two interesting case pulsars where RN is detected under both models but $A_{\rm RN}$ significantly dropped after applying our CNMs. In the top panel, we have the FS for PSR J1713+0747, whose noise is primarily dominated by the CRN. We observe the most significant reductions in power at 6, 16, and 40 nHz, although there are also some increases in power at 4, 30, and 58 nHz, and minor changes elsewhere. Comparing the median best fit power law curves to the spectra under each model suggests the decrease in excess power at 16 nHz is largely responsible for the change in power law parameters for PSR J1713+0747, and will have implications on free spectral analyses of the GWB. Meanwhile, the middle panel shows the FS for PSR J1643$-$1224, one of the IRN-dominated pulsars, experiences reductions in power at many frequencies (particularly 16, 20, 26, 28, 30, 34, 40, 46 nHz) from the CNM, but not yet enough to drive the power law down to the level of the CRN.

The bottom panel of Fig.~\ref{fig:freespec_BFs} visualizes changes to the per-frequency FS Bayes Factor, Eq.~\eqref{eq:FSSDBF}, across all pulsars, where each scatter point is the ratio of the Bayes Factor estimated under the CNM vs DMX for a particular pulsar/bin. 
The interpretation of this ratio is by what factor the detection significance of noise in a particular frequency bin for a particular pulsar increases or decreases under the CNM. 
Overall, the results reveal that the CNMs tend to reduce per-frequency detections of power at high frequencies, and increased detections are mostly prevalent at low frequencies. In the high-frequency range (above 10 nHz), we observe 16 \emph{decreases} in per-frequency Bayes Factor by a factor of 100 or more across all pulsars. These previously significant detections of power were likely spurious due to the lack of comprehensive chromatic modeling. There are also some marginal increases in detection significance in this range, the highest being an increase to the Bayes Factor by a factor of $\sim 40$ at $f\sim60$ nHz in one pulsar. In the low-frequency range (below 10 nHz), 2 \emph{increases} in Bayes Factor by a factor of 100 or more are observed at 2 and 4 nHz respectively, alongside 2 \emph{decreases} in Bayes Factor by a factor of 100 or more at 6 nHz. Many of the significant changes in Bayes Factor take place for PSRs J1713+0747 and J1643$-$1224, which are marked by pink and green open circles respectively in the bottom panel of Fig.~\ref{fig:freespec_BFs}, which can be compared to the changes in posterior power violins in the upper panels.

\subsection{White Noise Comparison}
\label{sec:white-noise-comparison}
{The white noise parameter definitions and applications are defined in Eq. ~\ref{eq:wn}. EFAC, EQUAD, and ECORR can be characterized as  a scaling factor, addition in quadrature, and epoch correlated addition respectively. They are inferred and applied to residuals across unique receiver and backend combinations.} Comparing changes in white noise between different models is a daunting task. There are a variety of reasons for which white noise parameters and levels might change between different models; some of which indicate an improvement and modeling and some of which indicate worse modeling. With this caveat in mind, we proceed to compare white noise parameters between the standard noise analysis (DMX) and our CNMs.

\subsubsection{Measurement Noise -- EFAC and EQUAD}

The EFAC posteriors are very consistent across all $215$ EFAC noise parameters for this data set with $3$ exceptions. {The EFAC parameters for B1937+21 at $800$ MHz with the GUPPI backend, J1903+0327 at S band with PUPPI, and J1903+0327 at L-band with PUPPI all have $95\%$ credible intervals which do not overlap between the standard model and the CNMs.} In all cases, the custom noise EFAC is lower than the standard noise and noteably PSRs B1937+21's EFAC is unusually high ($\sim3.9$ in the case of standard noise). PSRs B1937+21 and J1903+0327 show signs of being two of the most highly scattered pulsars in the $15$-year dataset \citep{Geiger+2024, turner2024cycspec}. Both of these pulsar's favored models include an FC model component, so it is reasonable to surmise that the dedicated scattering model is reducing the large EFAC values in the standard model.

The EQUAD parameters are also completely consistent whether one uses DMX or the CNMs. {Interesting to note though is that while the EQUAD in J1713+0747 at 800 MHz with GUPPI is consistent, it slightly less significant with the CNMs. This can be explained by the deterministic model included for this pulsar's second chromatic event better modeling the event than just the DMX model alone.}

\subsubsection{ECORR}
\begin{figure*}
    \centering
    \includegraphics[width=\linewidth]{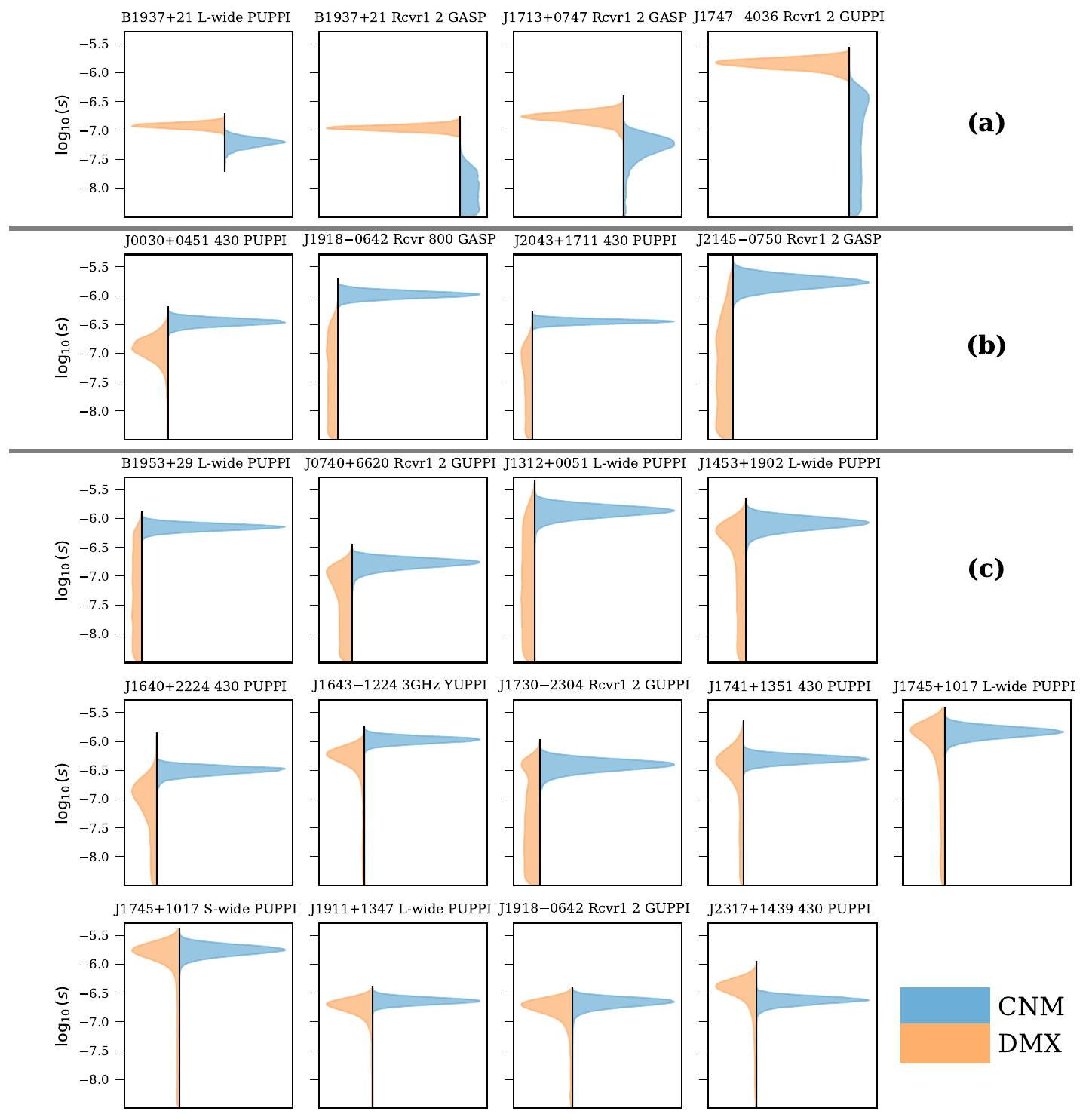}
    \caption{Changes in ECORR significance. Violin plots compare ECORR in the standard noise to our custom noise are plotted for all of the cases in which ECORR changes from very significant to insignificant between DMX and our custom chromatic noise model (CNM). We group these into $3$ categories: \textbf{(a)} ECORR goes from very significant with DMX to  insignificant with the CNM,
    \textbf{(b)} ECORR goes from insignificant with DMX to very significant with CNM \textit{and} the $90\%$ credible intervals do not overlap, and \textbf{(c)} ECORR goes from insignificant with DMX to very significant with CNM \textit{and} the $90\%$ credible intervals overlap.}
    \label{fig:ecorr_violins}
\end{figure*}

The physical motivation for inclusion of an ECORR model stems from the need to model stochastic variations in the shape and phase of pulses, known as pulse jitter \citep{LorimerKramer2004, lcc+16b}. However, it has been shown that ECORR is often measured to be much greater than pulse jitter alone \citep{Shannon+2016, lam2019jitter, gitika2025-mpta-tuning}. The interepoch uncorrelated nature of ECORR combined with NANOGrav's different-day observations with different receivers gives it the flexibility to absorb noise which is otherwise unmodeled, thus making ECORR the ``chromatic gutter'' of a PTA noise budget. As a result, comparing ECORR to jitter estimates proves to be a grounding tool in a chromatic noise analysis. \citet{ng12p5_cnm} examines the changes in select ECORR parameters between the DMX model and the Gaussian process chromatic noise models presented for the NANOGrav $12.5$-yr dataset. Fig.~\ref{fig:ecorr_violins} presents a similar comparison of the ECORR parameters, highlighting parameters which change from very significant ($\mathcal{B^J_\varnothing}>100$) to insignificant ($\mathcal{B^J_\varnothing}<10$) or vice-versa. In this work, we group these parameters into $3$ categories:
\begin{itemize}
    \item[\textbf{(a)}] ECORR goes from very significant with DMX to insignificant with CNM
    \item[\textbf{(b)}] {ECORR goes from insignificant with DMX to very significant with CNM \textit{and} the $90\%$ credible intervals do not overlap}
    \item[\textbf{(c)}] {ECORR goes from insignificant with DMX to very significant with CNM \textit{and} the $90\%$ credible intervals overlap.}
\end{itemize}

Group \textbf{(a)} can be interpreted as an overall improvements in the noise modeling when we replace DMX with the CNM and the ECORR decreases. Specifically, the reduction in B1937+21's $2$ ECORR parameters can be attributed to the inclusion of the FC variations and the fine-grained SWGP. J1713+0747's ECORR reductions can be attributed to both the included FC variations model as well as the modeling of the later chromatic event. Lastly, the reduction in J1747$-$4036's L-band ECORR is attributed the use of the band dependent DM model used for J1747$-$4036, which shows DM decorrelating around $57000$ MJD between bands (see Fig.~\ref{fig:RF_band_DM}).

Group \textbf{(b)} can be interpreted as cases where the favored CNM model is still lacking in some capacity. All of the pulsars in this category with the exception of J1918$-$0642 have realizations of ECORR which spike near solar conjunctions despite all having a SWGP in the model. The extreme flexibility of DMX on the other hand allows it to capture these outliers in the SW DM provided that the DMX bins are narrow enough near solar conjunction.
Pulsar J0030+0451 favors a \textsc{Model\_Fourier} with a fine-grained solar wind, but it appears that the fine-grained solar wind is still not sufficient for this very ecliptic pulsar. 
PSR J2043+1711 is an interesting edge case in our model selection since a finer time domain basis was slightly favored based on Bayesian evidence alone, but our selection process also incorporates likelihood evaluation time as a factor, which ultimately favored a coarser basis (Sec.~\ref{sec:model_selection}). A finer basis here reduces ECORR and improves the model at the expense of compute time. This pulsar in particular would benefit from an interpolation basis which more finely gridded near solar conjunction and sparsely gridded off of conjunction, which is aim of future work. \citet{ng12p5_cnm} found that for this pulsar in particular, ECORR levels kept decreasing in both PUPPI bands as $dt$ was decreased even down to $dt=1$day, which is lower than this work considers. PSR J2145$-$0750 favors \textsc{Model\_Ridge} with a yearly SWGP. The symmetric-about-conjunction nature of this SW model limits its ability to well model solar wind outliers which are near but off of solar conjunction. PSR J1918$-$0642 does not appear to suffer from an insufficient solar wind model but appears to lack some other chromatic (but probably non-dispersive) component in its model particularly in the GASP era when the frequency coverage was much poorer. Residual plots in \citet{ng15data} suggest that some non-dispersive chromatic component is missing from the model. The MPTA collaboration reported significance for a chromatic Gaussian event in this pulsar in \citep{MPTADR2_noise}. Though this event is outside of the $15$-year dataset, it lends credence to the need for additional non-dispersive chromatic noise in this pulsar. However, our model selection process did not favor FC variations for this pulsar enough to warrant its inclusion.
Overall, this category of changes in ECORR suggests that further improvements in SW modeling will continue to improve chromatic-noise mitigation and reduce very large ECORR values.

\begin{figure}
    \centering
    \includegraphics[width=\linewidth]{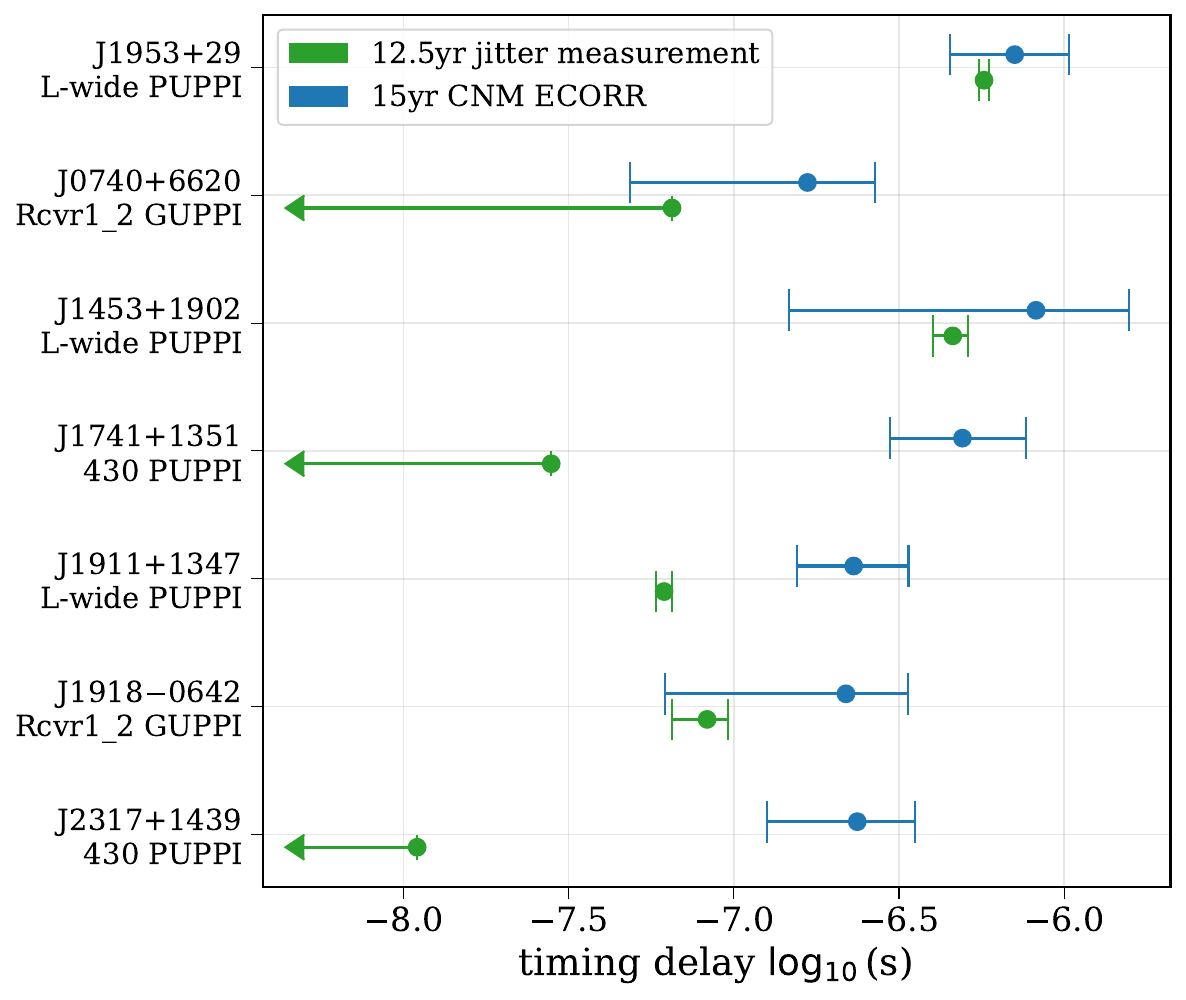}
    \caption{Comparing jitter measurements and ECORR. The blue (above) intervals capture the $99.7\%$ credible interval for the ECORR posteriors we recover with CNMs. The green (below) intervals are the estimated jitter values from Table~$4$ (Model B) in \citet{lam2019jitter} derived from the NANOGrav $12.5$-yr dataset.}
    \label{fig:ecorr_jitter_comp}
\end{figure}

Group \textbf{(c)} consists of ECORR parameters which are consistent between DMX and CNM but are more significantly measured using CNM. As noted previously, chromatic noise is infamous for leaking into the ECORR noise channel. Consequently, improved chromatic noise modeling ought to resolve tension between ECORR measurements jitter estimates. Furthermore, using a GP model instead of DMX, we reduce the prior volume, making us more sensitive to different processes. (See for instance the comparison of the DMX and CNM transmission functions in Section~\ref{sec:sensitivity-comparison}.) These reasons suggest that it is plausible we are more sensitive jitter noise detections in certain pulsars with CNM as opposed to DMX. To explore this idea, we compare the per-band jitter estimates (for 30 minute integration times) reported in \citet{lam2019jitter} for the NANOGrav $12.5$yr data set to the ECORR measurements with our favored noise models in this work. Fig.~\ref{fig:ecorr_jitter_comp} shows the $99.7\%$ credible interval for the pulsars in Group (c) which have available jitter estimates for that band. Overall, there is mixed agreement between jitter estimates and ECORR in Fig.~\ref{fig:ecorr_jitter_comp}. $4$ of the $7$ pulsars shown have ECORR measurements which are both very significant ($\mathcal{B^J_\varnothing}>100$) and consistent with ECORR. One could interpret these as an indication that the chromatic models in these pulsars are sufficient as there are no indications of excess chromatic noise leaking into the ECORR noise channel. Of the $3$ pulsars where ECORR is not consistent with the jitter estimates, $2$ of them only have upper limits on jitter and the corresponding bands are low frequency ($430$ Mhz). These cases could be interpreted as the chromatic model (or perhaps some other piece of the model) lacking in some regard since the ECORR credible interval is in disagreement with the jitter measurement/upper limit. We stress that these interpretations are just a hypothesis, and updated jitter estimates and further investigations are required to make stronger claims about increased sensitivity to jitter. However, these results provide a hopeful indication that in some cases ECORR-jitter tensions may be relieved through more in-depth chromatic modeling. As PTA sensitivity increases, radio bandwidths widen significantly, and advancements are made in chromatic noise modeling, future work may allow for the inclusion of direct measurements of jitter rather than fitting for an ECORR parameter, thereby furthering PTA sensitivity to GWs.

Taken as a whole, our CNM models show modest to little improvements in EFAC and EQUAD levels compared to DMX. A comparison of the ECORR shows large improvements from CNM to DMX in some cases alongside indications that some CNM models still lack the flexibility needed to account for ECORR in others. Thankfully, realizations of the ECORR, which show spikes near solar conjunctions, point future work in the direction of  more customized interpolation near solar conjunctions. Some consistency of jitter estimates and ECORR measurements provide encouragement for continued efforts in chromatic modeling.

\subsection{Sensitivity Comparison}
\label{sec:sensitivity-comparison}
\citet{hazboun:2019sc} lays out the foundation for a realistic pulsar timing sensitivity curve formalism using \texttt{hasasia}. This formalism was initially limited to DMX models but was extended to include DMGP in \citet{ipta3p+2024} and recently other Gaussian process noise in \citet{gitika2025-mpta-tuning}. GP noise processes are included in the covariance matrix, {$\mathbf{C} = \mathbf{N} + \mathbf{F}\bm{\phi} \mathbf{F}^T$, similar to Eq. \ref{eq:C} except in this formalism, the linearized timing model is not included in the covariance matrix but instead is used to construct a transmission function which describes the power lost in the timing model fit \citep{hazboun:2019sc}.}
Ordinarily, one includes the GWB as a source of noise when constructing a sensitivity curve. However, in order to compare the DMX and CNM sensitivity on an equal footing, we only use the RN of each from their respective single pulsar noise runs. Note that the power of the GWB is still contained in the data, primarily in the achromatic RN spectra. This approach affords us the ability to compare different model's sensitivity; however, we caution that the interpretation of these curves is not the same as traditional sensitivity curves.

\begin{figure}
    \centering
    \includegraphics[width=0.98\linewidth]{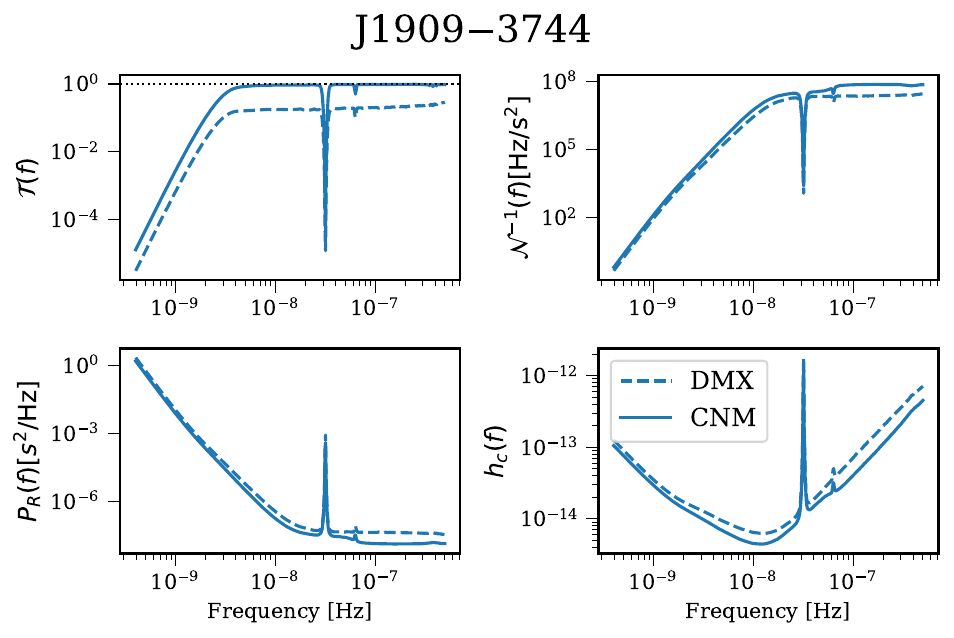}
    \caption{Sensitivity comparison of DMX (dashed lines) and custom noise model (CNM; solid lines) for pulsar J1909$-$3744. Top left: The timing model transmission function.
    Top right: the noise-weighted inversion transmission function ($\mathcal{N}^{-1}$).
    Lower left: residual power spectral density, $P_R(f)$.
    Lower right: characteristic strain power spectral density, $h_c(f)$. Note the dip/spike in the curves is at $\sim3\times10^{-8}$~Hz or $1/$yr, which arises from components of the timing model fit.}
    \label{fig:sensitivity_quad_fig}
\end{figure}

Fig.~\ref{fig:sensitivity_quad_fig} shows $4$ different functions related to pulsar timing sensitivity curves for pulsar J1909$-$3744. This figure compares the CNM and DMX models across all four panels. Since DMX is fit in the timing model, more power is transmitted through the CNM timing model transmission function because the chromatic signals are included in the covariance matrix rather than the marginalized, linearized timing model. This comparison illustrates how DMX devours power indiscriminately across the spectrum (nearly $10\%$ in this case). A comparison of the noise-weighted inverse transmission function, $\mathcal{N}^{-1}(f)$, is in the top right panel; this function encodes both the noise in the covariance matrix and the effects of the timing model fit. Accordingly, this represents a fair comparison of the two models as the DMX fit is encapsulated in the transmission function and the CNM power resides in the noise-weighting. Comparing the DMX and CNM $\mathcal{N}^{-1}(f)$'s, we see that more power is transmitted through the CNM, especially so at higher frequencies (beyond the smaller dip at $2/$yr). The lower left panel presents the residual power spectral density, $P_R(f)$ respectively. Note that $P_R(f)\propto1/\mathcal{N}^{-1}(f)$. The flat line at high frequencies in residual power spectral density constitutes the white noise floor for the pulsar. These curves allow us to directly compare the white noise floor between the DMX and CNM models, revealing that for this pulsar, the white noise floor is $17\%$ lower with CNM. While this does not affect background measurements, continuous wave searches stand to gain much at these higher frequencies. Lastly, the lower right panel touts the increases in sensitivity from DMX to CNM evident across all frequencies in the characteristic strain power spectral density, $h_c(f)$ plotted.

\begin{figure}
    \centering
    \includegraphics[width=0.98\linewidth]{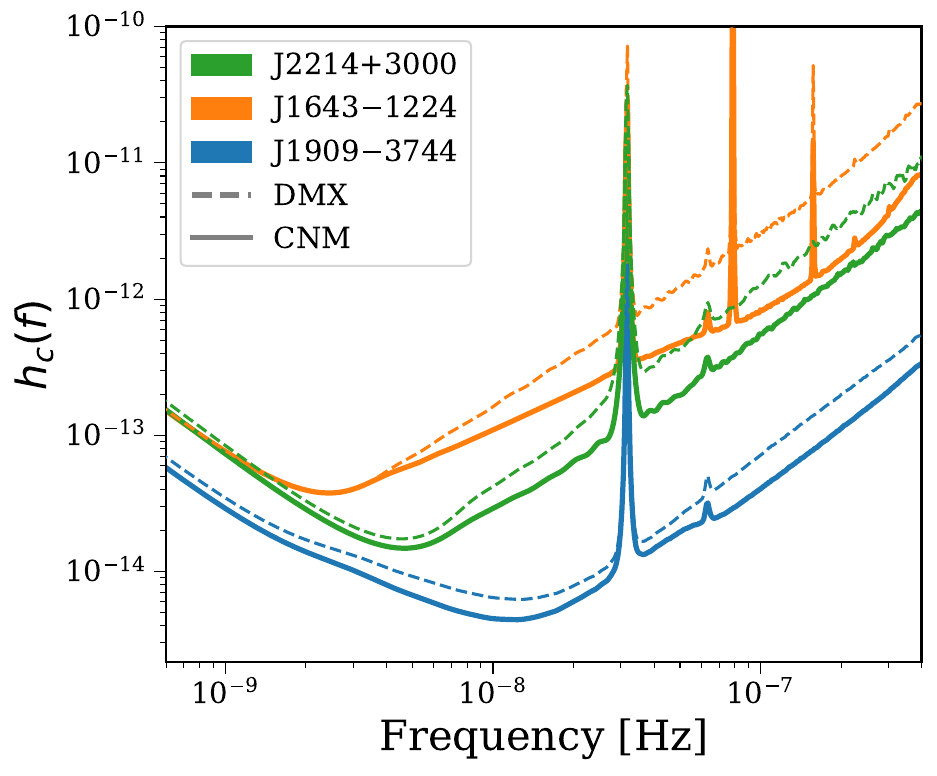}
    \caption{
    The characteristic strain sensitivity, $h_c$, is plotted for 3 different pulsars for both the DMX model (dashed) and the CNM (solid). The spike in insensitivity at $\sim3\times10^{-8}$~Hz is at $1/$yr. A smaller insensitivity appears at $2/$yr and additional insensitivities appear for J1643$-$1224 at its binary frequency and higher harmonics.}
    \label{fig:sensitivity_comp}
\end{figure}

Fig.~\ref{fig:sensitivity_comp} compares the characteristic strain sensitivity of 3 pulsars between DMX and CNM. These $3$ pulsars represent each of the categories laid out in Section~\ref{sec:achromatic-red-noise-comparison}: IRN, CRN, and WN dominated. In each case, we see the pulsars gain sensitivity especially at mid to high frequencies, suggesting that we will be more sensitive to gravitational waves using these CNMs than with the DMX models.

Thus, transmission functions and sensitivity curves provide yet another avenue of comparison for CNM and DMX models. One key advantage to comparison through a sensitivity curve is that the transmission function formalism can take into account the prior volume a model occupies. For this representative set of pulsars, we see how much our sensitivity improves from DMX to CNM.

\subsection{Whitened residuals}
\label{sec:whitened-residuals}

\begin{figure}
    \centering
    \includegraphics[width=\linewidth]{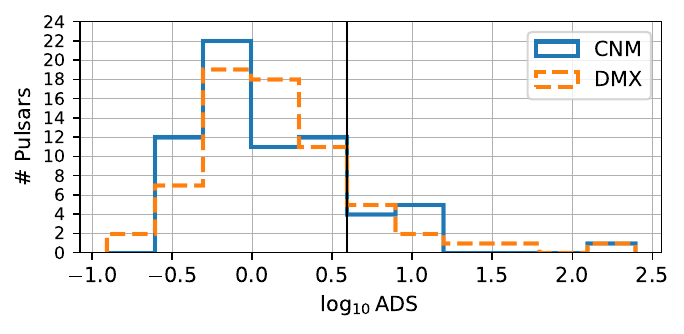}
    \caption{Histogram of the Anderson-Darling statistic (ADS) across 67 pulsars, as computed on the whitened timing residuals under our CNMs (blue) and under DMX (orange dashed). The dashed black line indicates the threshold above which there is a $<1\%$ chance the whitened, noise-subtracted residuals under the model are consistent with a unit normal. Under either model, only 10 out of 67 pulsars lie above this threshold. This shows both our CNMs and DMX provide good fits to the timing residuals.}
    \label{fig:ADS}
\end{figure}

\begin{table}
    \begin{center}
        \label{tab:ADS}
        \caption{ADS under both models CNM and DMX for 14 pulsars where the statistic lies above a threshold $\rm{ADS}>3.9$, indicating $p < 0.01$ that the whitened, noise-subtracted timing residuals are drawn from a zero mean, unit variance normal distribution. Bolded values mark where the $\rm{ADS} > 3.9$ for only one model.}
        \begin{tabular}{c|cc}
            \hline\hline Pulsar & ADS (CNM) & ADS (DMX) \\
            \hline B1937+21 & 12.6 & 31.6 \\
            J0406+3039 & 0.8 & \textbf{5.2} \\
            J0437$-$4715 & 179.9 & 189.8 \\
            J0613$-$0200 & 10.8 & 16.4 \\
            J1600$-$3053 & 4.1 & 4.9 \\
            J1614$-$2230 & \textbf{8.9} & 3.3 \\
            J1640+2224 & 2.2 & \textbf{7.9} \\
            J1643$-$1224 & \textbf{7.2} & 1.3 \\
            J1713+0747 & 8.3 & 8.5 \\
            J1843$-$1113 & \textbf{4.2} & 1.7 \\
            J1910+1256 & \textbf{5.8} & 0.2 \\
            J1918$-$0642 & 2.6 & \textbf{4.4} \\
            J2124$-$3358 & 1.1 & \textbf{4.7} \\
            J2317+1439 & 8.7 & 5.3 \\
            \hline \hline
        \end{tabular}
    \end{center}
    \vspace{-\baselineskip}
\end{table}

Finally, we apply an Anderson-Darling test \citep{AndersonDarling1952} on our whitened, noise-subtracted timing residuals to test if the leftover timing residuals show any departures from a unit normal distribution, as is assumed in the likelihood, Eq.~\eqref{eqn:likelihood2}. In the PTA context, \citet{goncharov+2021, Chalumeau+2022, pptadr3:noise, MPTADR2_noise, inpta-dr2-noise-2025} have previously used this test to assess the Gaussianity of whitened timing residuals after single pulsar noise analyses. Following the implementation in \citet{goncharov+2021}, values of the Anderson-Darling statistic (ADS) $> 3.9$ indicate $p < 0.01$ in rejection of Gaussianity (with zero mean, unit variance), while values of the ADS $< 3.9$ are more consistent with Gaussianity. This threshold is only approximately valid, as the exact null distribution of the ADS may vary for individual pulsar datasets. Nonetheless the threshold is suitable for comparisons of each pulsar's dataset under two different models, enabling a test of the fidelity of our noise models to describe all structures within the timing residual data. By comparing the values of the ADS under our CNMs vs the standard DMX model, we can determine if we may have missed any important structure in the DM variations model (via an increase to the ADS) or if our custom model has newly accounted for unmodeled structure (via a decrease to the ADS). Note the ADS is not used as a model selection criterion as it includes no penalty for overfitting.

We compute these first by subtracting realizations of every noise process under the model (including ECORR) using the MAP parameter value from the MCMC chain, and rescaling the TOA errors using the MAP EFAC and EQUAD values. While the MAP estimates obtained from MCMC are not as reliable as those obtained from gradient descent optimization, these estimates are still useful for a broad-level comparison. We use a unit normal with zero mean as our test distribution for the ADS.

Fig.~\ref{fig:ADS} displays the distribution of ADS values from each pulsar's whitened residuals under our CNMs and under the standard DMX models. We find broad consistency of the whitened residuals with a unit normal under both models, and no significant differences in the distributions under either model. We find there are only 10 out of 67 pulsars whose whitened timing residuals show departures from normality under either model. However, the exact set of 10 pulsars which lie above the threshold varies depending on the model, as shown in Table~\ref{tab:ADS}. PSR J0437$-$4715 has an especially large ADS under both models, either 189.8 under DMX or 179.9 under the CNM. Large values of the ADS have also previously been reported for PSR J0437$-$4715 by the PPTA \citep{goncharov+2021}. Overall, these results show the majority of pulsar timing residuals in the NANOGrav 15 year dataset are highly consistent with normality under either choice of chromatic noise model. Alternatively stated, both noise models contains sufficient number of components to fit our timing residuals.

\section{Discussion \& Outlook}
\label{sec:discussion}

We have carried out a per-pulsar customization of chromatic noise models across the full NANOGrav 15 yr dataset. This shifts the analysis away from a single global prescription and toward a model set that can separate propagation-induced structure from genuinely achromatic red processes on a pulsar-by-pulsar basis. The practical consequence is the mitigation of chromatic misspecification that can leak into achromatic red noise and ECORR, and therefore bias the noise budgets used in nanohertz GW searches.

A primary outcome is a re-interpretation of loud achromatic red noise (RN) previously reported under baseline DMX analyses. For many pulsars, signals previously absorbed by achromatic RN are better explained by chromatic variability within the model set explored here. Among the 12 pulsars classified as intrinsic RN-dominated under the standard DMX model, 9 show significant changes in their achromatic RN posteriors after applying our custom noise models, and in 5 cases the evidence for achromatic RN is no longer significant. Where residual achromatic RN persists, it is consistent with plausible physical mechanisms. PSR B1937+21 retains a steep-spectrum achromatic component even when complex chromatic structure is favored, consistent with an intrinsically achromatic origin. Two spider systems, PSRs J0610$-$2100 and J1745+1017, also retain residual RN that may reflect unmodeled orbital or intrabinary contributions.

The pulsars with the most severe scattering environments (e.g., PSRs J1643$-$1224, B1937+21, and J1903+0327) are natural targets for mitigation techniques such as cyclic spectroscopy~\citep{Demorest2011, Dolch2021, turner2024cycspec}, which can address scattering delays at the signal-processing level prior to TOA generation. In the strongest scatterers, the propagation imprint appears difficult to treat purely through parameterized timing-model components, and upstream corrections may be required \citep{Geiger+2024, cordes2026fundamental_noise}. For PSR J1643$-$1224, the presence of a surrounding \textsc{Hii} region provides a straightforward physical explanation for complex and time-variable scattering behavior \citep{Mall2022, Ocker+2024}.

Several pulsars exhibit complex noise budgets where strong chromatic components coexist with significant residual achromatic structure, often indicative of dynamic binary environments. PSR J1705$-$1903, an eclipsing binary, likely exhibits both achromatic orbital perturbations and chromatic delays induced by local material in the system \citep{Morello+2019, ng15data, MPTADR2_noise}. PSR J1903+0327, with a main-sequence companion, may similarly require a tailored treatment of line-of-sight material in the binary. The noise budget of PSR J1012+5307 remains difficult to interpret despite extensive study across multiple PTA datasets \citep{Chalumeau+2022, ng15detchar, ipta3p+2024, Larsen+2024}, underscoring that the standard Fourier-series representation of achromatic RN may not be adequate for every pulsar. 

A robust solar-wind (SW) model is essential for high-precision timing \citep{tiburzi+2021}. We deployed a hybrid framework that combines a binned, deterministic component, capturing the global evolution of the SW electron density over the solar cycle, with per-pulsar SWGP perturbations that account for short-timescale variability along individual lines of sight \citep{hazboun+2022sw, susarla+24sw}. The inferred 180-day binned time series tracks the expected 11-year cycle as traced by F10.7\,cm solar flux measurements (Fig.~\ref{fig:sw_ne_series}), supporting the physical interpretation of the deterministic component. The recovery of DM spectral indices consistent with Kolmogorov-like turbulence ($\gamma_{\rm DM} \approx 8/3$) after including the fine-grained SWGP terms indicates that this model family can better separate interplanetary and interstellar contributions to dispersion variability. This separation matters for GW inference because unmodeled SW variability introduces spatially correlated structure that can project onto common-spectrum signals \citep{tiburzi+2016} and floods our achromatic red noise models with high amplitude and shallow spectral index noise \citep{DiMarco2025}. Future PTA analyses should refine these models by allowing the large-scale solar-wind structure to vary with sky direction \citep{susarla+24sw}, rather than assuming a simple spherical profile, and by continuing to compare the models with direct spacecraft measurements that show significant short-timescale variability in the solar wind, see e.g. \cite{Tiburzi2019}.

We detect significant non-dispersive chromatic noise in 21 pulsars, with 6 representing first detections in PTA timing residuals. The inferred chromatic indices span a broad range, and only a subset are consistent with simple thin-screen expectations. Interpreting these indices requires care: recent theory shows that when scattering delays are comparable to intrinsic pulse widths, the effective chromatic index inferred from TOA residuals can differ from the underlying scattering-delay scaling \citep{Geiger+2024} and that non-Gaussian delay statistics can produce apparent indices larger than naive physical expectations \citep{Kulkarni+2025}. The diversity of inferred indices, and any discrepancies with overlapping analyses in other PTA datasets, continues to motivate coordinated noise-model comparisons in future IPTA releases \citep{ipta3p+2024}.

The refined noise models presented here have direct implications for PTA science goals. By reducing leakage of chromatic power into achromatic RN, the custom models change the inferred RN spectra for a subset of the most influential pulsars in the array. PSR J1713+0747 provides a clear example: incorporating models for its known chromatic/profile events substantially alters the inferred RN amplitude and spectral shape, as well as the distribution of free-spectral power. Because the GWB inference is driven by the lowest frequencies and by a number of high-quality pulsars, chromatic model misspecification in a subset of pulsars can still drive changes in the recovery of common GW signals across the whole PTA, within astrophysically consequential ranges \citep{ng15smbbh, Goncharov2025nature}. Importantly, robust chromatic mitigation is further expected to improve PTA sensitivity to the cross-correlations of the GWB \citep{DiMarco2024}, enhancing the quality of the PTA for tests of general relativity. This motivates continued cross-checks of GWB posteriors under alternate, physically motivated chromatic model families, especially for pulsars that dominate the sensitivity. Precise characterization of RN spectra also supports continuous-wave searches by sharpening the distinction between deterministic low-frequency power and intrinsic stochastic variability. Beyond GWs, the decomposition of noise components provides a framework for studies of the turbulent ISM, the interplanetary medium, and pulsar magnetospheric variability.

Selecting among high-dimensional GP models poses nontrivial computational challenges. Within the model set explored here, importance nested sampling via \texttt{nautilus} performed robustly and provided stable evidence estimates across diverse pulsar noise budgets. Nonetheless, the sheer number of importance nested sampling analyses required for robust exploration of the model space remains a major challenge for evidence-based comparisons. A recurring difficulty is also that chromatic hyperparameter posteriors can depend on basis size, complicating reweighting or product-space sampling strategies. Future analyses could address these issues through transdimensional sampling methods \citep{EllisCornish2016} or model-averaging approaches \citep{vanHaasteren2025ma}, both of which can reduce reliance on any single discrete model choice. The time required to evaluate the likelihood itself is also an important computational limitation, as the matrix inversions required using the Woodbury identity scale unfavorably as the cube of the number of basis elements, which can and do become very large for our chromatic models. Continued improvements in computational efficiency, including accelerated inference frameworks such as \texttt{JaxNS} \citep{jaxns} or \texttt{blackjax} \citep{blackjax}, as well as the extension of next-generation PTA likelihood evaluation methods for chromatic noise \citep{Vallisneri+2025, Laal+2025, Gundersen+2025}, will be important for scaling similar analyses to larger datasets and richer model families.

A central methodological result is that time-domain Gaussian process models introduced in \citet{ng12p5_cnm} offer distinct advantages for modeling chromatic noise in PTA data, and are favored for roughly half of the pulsars. The linear interpolation basis avoids aliasing artifacts that can arise when Fourier representations sample above an effective Nyquist frequency for quasi-periodic signals such as SW variations, and it naturally accommodates sharp transitions without Gibbs artifacts. In addition, diagonal-covariance constructions (such as Ridge-type kernels) can be computationally efficient while remaining flexible. These advantages are particularly pronounced for the SW problem, where the time-domain SWGP models implemented here provide a practical alternative to Fourier-based implementations. The squared-exponential kernel adopted throughout this work is a convenient default and matches choices in recent PTA analyses, but it is not unique. Alternatives such as Mat\'ern kernels may better represent processes with rougher time-domain structure. Rapid kernel comparisons could be performed using spectral-refit methods \citep{Lamb2023}.

Our results also emphasize that fixed-basis models are not always optimal. The choice of basis size can materially affect the inferred chromatic structure, suggesting that adaptive representations may provide better control of model complexity. Candidate directions include Bayesian blocking, or Fourier-basis approaches that explicitly suppress higher-order coefficients when the signal becomes noise-dominated \citep{eptadr2_2:noise}. Alternative approaches may include more physically-motivated selections for basis size or more efficient GP formulations that bypass the need for rank-reduction altogether (e.g., \citealt{Foreman-Mackey+2017}). For the time-domain SW interpolation model used here, future work could cluster interpolation nodes around solar conjunctions, where SW-driven dispersion variations are most pronounced, thereby preserving resolution while reducing model rank. Combining diagonal and non-diagonal chromatic covariance constructions may also allow simultaneous sensitivity to multiple turbulence scales in the ISM.

\section{Conclusions}
\label{sec:conclusions}
After developing and customizing chromatic noise models for all 67 pulsars in the NG15 dataset and comparing the characterization of the pulsar datasets to those used in the fiducial NG15 GW analyses a number of important conclusions can be drawn. These models materially alter the inferred achromatic red and white noise budgets for a substantial fraction of pulsars, demonstrating that chromatic misspecification under baseline prescriptions can bias the apparent strength and spectral shape of achromatic processes. We detect significant non-dispersive chromatic noise in 21 pulsars, 6 of which were not previously identified in PTA residuals, and we find that the inferred chromatic indices span a broad range that is plausibly influenced by TOA-regime effects. We demonstrate that our solar wind framework, combining a global deterministic component with per-pulsar GP perturbations, both improves the separation of interplanetary and interstellar dispersion structure, and reduces a known source of spatially correlated variability relevant to GWB inference.

The analyses herein demonstrate the importance of these customized chromatic noise models to GW analyses across the nanohertz spectrum. The differences in achromatic red noise recovery (see Figs.~\ref{fig:RN_params} and \ref{fig:freespec_BFs}) will have a large effect on the recovery of Hellings-Downs correlations and the spectral characterization of the stochastic GWB. Better characterization of the background, along with the overall improvements in sensitivity across the band (see Figs.~\ref{fig:sensitivity_comp} and \ref{fig:sensitivity_quad_fig}) and the improvements in white noise characterization will make the NANOGrav PTA more sensitive to resolvable single-sources of GWs. All of this is achieved with no changes to the underlying NG15 dataset, only the analysis techniques. Work is currently underway that will corroborate these statements on the NG15 dataset for a selection of GW analyses. 

\section*{Acknowledgments}
We thank Aaron Johnson, Aman Srivastava, and Rutger van Haasteren for useful discussions on Bayesian model selection during this project.
{We thank the anonymous reviewer for providing detailed feedback which improved the quality of the manuscript.}
We also thank Michael Lam, Stella Ocker, Timothy Dolch, and other members of the NANOGrav Noise Budget, Detection, and Timing Working Groups for constructive comments that improved the quality of the manuscript.

The NANOGrav collaboration receives support from National Science Foundation (NSF) Physics Frontiers Center award \#2020265, the Gordon and Betty Moore Foundation, an NSERC Discovery Grant, and CIFAR.
The Arecibo Observatory is a facility of the NSF operated under cooperative agreement (AST-1744119) by the University of Central Florida (UCF) in alliance with Universidad Ana G. M\'endez (UAGM) and Yang Enterprises (YEI), Inc.
The Green Bank Observatory is a facility of the NSF operated under cooperative agreement by Associated Universities, Inc.
The National Radio Astronomy Observatory is a facility of the NSF operated under cooperative agreement by Associated Universities, Inc.
J.B. was supported in part through NASA and Oregon Space Grant Consortium, cooperative agreement 80NSSC20M0035.
J.B., J.S.H., and K.W.\ acknowledge support from NSF CAREER award \#2339728.
Y.C. was supported in part by NASA CT Space Grant PTE Federal Award Number 80NSSC20M0129, as well as the the Yale College Dean’s Office and the STARS Program.
C.M.F.M.\ was supported in part by the National Science Foundation under Grants No.\ NSF PHY-1748958,  AST-2106552, and NASA LPS 80NSSC24K0440. C.M.F.M. also thanks the Center for Computational Astrophysics (CCA) of the Flatiron Institute for support. The Flatiron Institute is supported by the Simons Foundation.
J.S.\ is supported by an NSF Astronomy and Astrophysics Postdoctoral Fellowship under award AST-2202388, and acknowledges previous support by the NSF under award 1847938.
P.R.B.\ is supported by the Science and Technology Facilities Council, grant number ST/W000946/1.
Pulsar research at UBC is supported by an NSERC Discovery Grant and by CIFAR.
K.C.\ is supported by a UBC Four Year Fellowship (6456).
T.D.\ and M.T.L.\ received support by an NSF Astronomy and Astrophysics Grant (AAG) award number 2009468 during this work.
E.cf.\ is supported by NASA under award number 80GSFC24M0006.
D.C.G.\ is supported by NSF Astronomy and Astrophysics Grant (AAG) award \#2406919.
D.R.L.\ and M.A.M.\ are supported by NSF \#1458952.
M.A.M.\ is supported by NSF \#2009425.
The Dunlap Institute is funded by an endowment established by the David Dunlap family and the University of Toronto.
T.T.P.\ acknowledges support from the Extragalactic Astrophysics Research Group at E\"{o}tv\"{o}s Lor\'{a}nd University, funded by the E\"{o}tv\"{o}s Lor\'{a}nd Research Network (ELKH), which was used during the development of this research.
H.A.R.\ is supported by NSF Partnerships for Research and Education in Physics (PREP) award No.\ 2216793.
S.M.R.\ and I.H.S.\ are CIFAR Fellows.
Portions of this work performed at the U.S. Naval Research Laboratory were supported by ONR 6.1 funding.

\section*{Author Contributions}
All authors contributed to the activities of the NANOGrav collaboration leading to the work presented here and reviewed the manuscript, text, and figures prior to the paper’s submission. Additional specific contributions to this paper are as follows. J.G.B. and B.L. co-led the project, carried out the analyses, interpreted the results, and led the writing of the manuscript. J.S.H., C.M.F.M., and J.S. conceived of and advised on the project, and contributed to the interpretation of the analyses. Y.C., K.W., M.T.M., and D.J.O. contributed to carrying out analyses and interpreting results.

The $15$-year data set was produced through a combination of observations, arrival time calculations, data checks and refinements, and timing-model development and analysis which were performed by the following authors: G.A., A.A., A.M.A., Z.A., P.T.B., P.R.B., H.T.C., K.C., M.E.D., P.B.D., T.D., E.cf., W.F., E.F., G.E.F., N.G., P.A.G., J.G., D.C.G., J.S.H., R.J.J., M.L.J., D.L.K., M.K., M.T.L., D.R.L., J.L., R.S.L., A.M., M.A.M., N.M., B.W.M., C.N., D.J.N., T.T.P., B.B.P.P., N.S.P., H.A.R., S.M.R., P.S.R., A.S., C.S., B.J.S., I.H.S., K.S., A.S., J.K.S., H.M.W..

\facilities{Arecibo, GBT, VLA}

\software{
    \texttt{astropy} \citep{astropy},
    \texttt{numpy} \citep{numpy},
    \texttt{scipy} \citep{scipy},
    \texttt{pandas} \citep{pandas},
    \texttt{enterprise} \citep{enterprise},
    \texttt{enterprise\_extensions} \citep{enterprise_ext},
    \texttt{ptmcmcsampler} \citep{ptmcmcsampler},
    \texttt{PINT} \citep{PINT},
    \texttt{la\_forge} \citep{laforge2020},
    \texttt{nautilus} \citep{Lange2023},
    \texttt{matplotlib} \citep{matplotlib},
    \texttt{kalepy} \citep{kalepy},
    \texttt{upsetplot} \citep{upsetplot},
    \texttt{corner} \citep{corner},
    \texttt{mermaid} \citep{mermaid},
    \texttt{hasasia} \citep{hazboun:2019has}
}

\bibliographystyle{aasjournal}
\bibliography{hazgrav}

\appendix

\section{Tables of Detailed Analysis Priors \& Results}
\label{appendix:tables}

\textbf{Priors:} Table~\ref{tab:priors} shows all deterministic parameters and GP hyperparameters we sample during our analyses alongside their corresponding prior distributions. \textbf{Posteriors:} Table~\ref{tab:FD_params} shows posterior red and DM noise parameters for pulsar that favored Fourier GPs, alongside Savage-Dickey Bayes Factors and pulsar timespans. Table~\ref{tab:TD_params} shows the same for pulsars that favored TD or Ridge GPs. For pulsars that favor FC noise, Table~\ref{tab:FC_params} {shows} posterior FC parameters alongside the nominal DM values of the pulsars. Tables~\ref{tab:SW_params1} and~\ref{tab:SW_params2} show posterior SW parameters and Bayes Factors alongside their ecliptic latitudes. Table~\ref{tab:det_params} gives posteriors on pulsar-specific deterministic model parameters.

\begin{table}
    \centering
    \begin{tabular}{ c | c c c c @{}}
        \hline\hline Model Component & Parameter & Prior & Units & Corresponding Signal Type \\
        \hline\hline & EFAC & $\mathcal{U}(0.01, 10.0)$ & & Measurement Noise \\
        WN & EQUAD & $\log_{10}\mathcal{U}(10^{-8.5}, 10^{-5})$ & sec & Uncorrelated Noise \\
        & ECORR & $\log_{10}\mathcal{U}(10^{-8.5}, 10^{-5})$ & sec & Kernel Noise \\
        \hline \multirow{2}{*}{RN, DM, \& FC} & $A$ & $\log_{10}\mathcal{U}(10^{-20}, 10^{-11})$ & & GP (Fourier) \\
        & $\gamma$ & $\mathcal{U}(0, 7)$ & & GP (Fourier) \\
        \hline \multirow{2}{*}{DM \& FC} & $\sigma$ & $\log_{10}\mathcal{U}(10^{-10}, 10^{-4.5})$ & sec & GP (TD, Ridge) \\
        & $\ell$ & $\log_{10}\mathcal{U}(10,T_{\rm psr})$ & day & GP (TD) \\
        \hline & $\Gamma_p$ & $\log_{10}\mathcal{U}(10^{-4},10^4)$ & & GP (TD) \\
        \multirow{2}{*}{DM} & $p$ & $\log_{10}\mathcal{U}(10^{-2},10^3)$ & year & GP (TD) \\
        & $\alpha_{\mathrm{wgt}}$ & $\log_{10}\mathcal{U}(10^{-4},10^3)$ & & GP (TD\_RF) \\
        & $\ell_2$ & $\log_{10}\mathcal{U}(1,10^7)$ & MHz & GP (TD\_RF) \\
        \hline FC & $\chi$ & \textsc{UniformGaussUpperTail}(2.5,7,1) & & GP \\
        \hline \multirow{2}{*}{Annual Variations} & $\rho$ & $\log_{10}\mathcal{U}(10^{-10},10^{-2})$ & sec & GP (Annual Fourier) \\
        & $\chi$ & $\mathcal{U}(0,7)$ & & GP (Annual Fourier) \\
        \hline \multirow{2}{*}{Exp. Dip} & $A$ & $\log_{10}\mathcal{U}(10^{-10},10^{-4})$ & sec & Deterministic \\
        & $\tau$ & $\log_{10}\mathcal{U}(10^{1.2},10^{2.5})$ & day & Deterministic \\
        \hline Exp. Dip & $t_0$ & $\mathcal{U}(57050, 57100)$ & MJD & Deterministic \\
        PSR J1643$-$1224 & $\chi$ & $\mathcal{U}(-3, 0)$ & & Deterministic \\
        \hline Exp. Dip & $t_0$ & $\mathcal{U}(54740, 54780)$ & MJD & Deterministic \\
        PSR J1713+0747 (1st) & $\chi$ & $\mathcal{U}(0, 5)$ & & Deterministic \\
        \hline Exp. Dip & $t_0$ & $\mathcal{U}(57506, 57514)$ & MJD & Deterministic \\
        PSR J1713+0747 (2nd) & $\chi$ & $\mathcal{U}(0.9, 1.7)$ & & Deterministic \\
        \hline Exp. Dip & $t_0$ & $\mathcal{U}(56300, 56350)$ & MJD & Deterministic \\ 
        PSR J2145$-$0750 & $\chi$ & $\mathcal{U}(-1.5, 1.5)$ & & Deterministic \\
        \hline & $A$ & $\log_{10}\mathcal{U}(10^{-10}, 10^{-4})$ & sec & Deterministic \\
        Gaussian bump & $\tau$ & $\log_{10}\mathcal{U}(10, 10^3)$ & day & Deterministic \\
        PSR J1600$-$3053 & $t_0$ & $\mathcal{U}(57500, 58000)$ & MJD & Deterministic \\
        & $\chi$ & $\log_{10}\mathcal{U}(0, 7)$ & & Deterministic \\
        \hline & $n_{E,i}$ & $\mathcal{U}(0,30)$ & cm$^{-3}$ & Deterministic \\
        \multirow{2}{*}{SW} & $A_{n_E}$ & $\log_{10}\mathcal{U}(10^{-12}, 1)$ & cm$^{-3}$ & GP (Fourier) \\
        & $\gamma_{n_E}$ & $\mathcal{U}(-6, 5)$ & & GP (Fourier) \\
        & $\sigma_{n_E}$ & $\log_{10}\mathcal{U}(10^{-4}, 10^{3})$ & cm$^{-3}$ & GP (Ridge) \\
        \hline RN (Free Spectral) & $\rho_i$ & $\log_{10}\mathcal{U}(10^{-10}, 10^{-4})$ & sec & GP (Fourier) \\
        \hline\hline
    \end{tabular}
    \caption{Priors and units on noise model parameters during MCMC analyses. The first column indicates which component of the model the parameter belongs to, and the last column labels the more  detailed implementation used for the model (e.g. a deterministic signal vs a GP). See \S\ref{sec:noise_models} for further details on the descriptions of the different types of signals, as well as the definition of the custom prior on the FC index $\chi$.}
    \label{tab:priors}
\end{table}

\begin{table}
    \centering
    \begin{tabular}{ c | c c c | c c c c | c @{}}
        \hline\hline & \multicolumn{3}{c|}{Red Noise} & \multicolumn{4}{c|}{DM GP} & \\
        \hline Pulsar & $\log_{10}A$ & $\gamma$ & $\mathcal{B}^{\rm{RN}}_\oslash$ & $\log_{10}A$ & $\gamma$ & $\mathcal{B}^{\rm{DM}}_\oslash$ & $N_f$ & $T_{\rm psr}$ (year) \\
        \hline B1855+09 & $-13.8_{-0.3}^{+0.3}$ & $3.5_{-0.7}^{+0.9}$ & $>1000$ & $-13.44_{-0.05}^{+0.05}$ & $2.7_{-0.2}^{+0.3}$ & $>1000$ & $200$ & 15.6 \\
        $^*$B1953+29 & $-12.8^{95\%}$ & $ - $ & $\mathbf{2.5}$ & $-12.9^{95\%}$ & $ - $ & $1.3$ & $100$ & 11.1 \\
        J0023+0923 & $\mathbf{-13.2_{-0.2}^{+0.1}}$ & $1.1_{-0.7}^{+0.9}$ & $\mathbf{138.2}$ & $-13.9_{-0.1}^{+0.1}$ & $2.4_{-0.3}^{+0.4}$ & $>1000$ & $200$ & 9.0 \\
        J0030+0451 & $-14.3_{-0.5}^{+0.4}$ & $4.4_{-0.9}^{+1.1}$ & $>1000$ & $-14.7_{-0.9}^{+0.4}$ & $3.6_{-1.0}^{+1.8}$ & $>1000$ & $200$ & 15.5 \\
        J0340+4130 & $-13.1^{95\%}$ & $ - $ & $0.7$ & $-13.0_{-0.1}^{+0.1}$ & $1.9_{-0.3}^{+0.3}$ & $>1000$ & $100$ & 8.1 \\
        J0645+5158 & $-13.3^{95\%}$ & $ - $ & $0.9$ & $-13.7_{-0.1}^{+0.1}$ & $2.1_{-0.4}^{+0.5}$ & $>1000$ & $50$ & 8.9 \\
        J0740+6620 & $-13.3^{95\%}$ & $ - $ & $0.6$ & $-13.6_{-0.2}^{+0.1}$ & $1.9_{-0.4}^{+0.9}$ & $>1000$ & $100$ & 6.3 \\
        J1024$-$0719 & $-13.3^{95\%}$ & $ - $ & $0.6$ & $-13.6_{-0.2}^{+0.1}$ & $2.7_{-0.5}^{+0.8}$ & $>1000$ & $50$ & 10.5 \\
        J1600$-$3053 & $-13.4^{95\%}$ & $ - $ & $\mathbf{1.7}$ & $-13.13_{-0.05}^{+0.05}$ & $2.3_{-0.2}^{+0.2}$ & $>1000$ & $150$ & 12.5 \\
        J1614$-$2230 & $-15.0_{-0.8}^{+0.9}$ & $5.0_{-1.8}^{+1.4}$ & $492.7$ & $-13.15_{-0.04}^{+0.04}$ & $1.9_{-0.2}^{+0.2}$ & $>1000$ & $100$ & 11.5 \\
        J1630+3734 & $-12.6^{95\%}$ & $ - $ & $0.8$ & $-12.9^{95\%}$ & $ - $ & $1.2$ & $50$ & 3.5 \\
        J1705$-$1903 & $\mathbf{-12.2_{-0.2}^{+0.2}}$ & $0.4_{-0.3}^{+0.7}$ & $184.1$ & $-12.2_{-0.1}^{+0.1}$ & $2.0_{-0.3}^{+0.3}$ & $>1000$ & $200$ & 3.7 \\
        J1713+0747 & $\mathbf{-14.7_{-0.4}^{+0.3}}$ & $\mathbf{3.7_{-0.8}^{+1.2}}$ & $>1000$ & $-13.89_{-0.05}^{+0.05}$ & $1.7_{-0.2}^{+0.2}$ & $>1000$ & $200$ & 15.5 \\
        J1738+0333 & $-13.2^{95\%}$ & $ - $ & $\mathbf{0.7}$ & $-13.0_{-0.1}^{+0.1}$ & $2.8_{-0.4}^{+0.5}$ & $>1000$ & $50$ & 10.7 \\
        J1741+1351 & $-13.4^{95\%}$ & $ - $ & $1.7$ & $-13.8_{-0.1}^{+0.1}$ & $2.6_{-0.4}^{+0.4}$ & $>1000$ & $50$ & 11.0 \\
        J1744$-$1134 & $-13.5^{95\%}$ & $ - $ & $\mathbf{1.2}$ & $-13.7_{-0.7}^{+0.2}$ & $2.2_{-0.6}^{+1.7}$ & $>1000$ & $100$ & 15.7 \\
        J1832$-$0836 & $-13.2^{95\%}$ & $ - $ & $1.8$ & $-12.8_{-0.1}^{+0.1}$ & $2.5_{-0.3}^{+0.3}$ & $>1000$ & $100$ & 7.1 \\
        J1843$-$1113 & $-12.8^{95\%}$ & $ - $ & $0.9$ & $-12.8_{-0.1}^{+0.1}$ & $2.3_{-0.4}^{+0.5}$ & $>1000$ & $100$ & 3.5 \\
        J1853+1303 & $-13.3_{-0.2}^{+0.1}$ & $1.3_{-0.6}^{+0.8}$ & $2512.5$ & $-13.5_{-0.1}^{+0.1}$ & $2.1_{-0.3}^{+0.4}$ & $>1000$ & $150$ & 9.1 \\
        J1903+0327 & $\mathbf{-12.4_{-0.1}^{+0.1}}$ & $1.1_{-0.6}^{+0.6}$ & $>1000$ & $-12.0_{-0.1}^{+0.1}$ & $2.9_{-0.3}^{+0.4}$ & $>1000$ & $50$ & 10.7 \\
        J1910+1256 & $-13.1^{95\%}$ & $ - $ & $0.9$ & $-13.3_{-0.3}^{+0.2}$ & $2.8_{-0.5}^{+0.8}$ & $>1000$ & $50$ & 11.4 \\
        J1911+1347 & $-13.4^{95\%}$ & $ - $ & $0.6$ & $-13.7_{-0.2}^{+0.1}$ & $2.5_{-0.7}^{+0.9}$ & $>1000$ & $50$ & 7.0 \\
        J1918$-$0642 & $-14.1_{-0.7}^{+0.5}$ & $3.3_{-1.3}^{+1.6}$ & $178.3$ & $-13.5_{-0.1}^{+0.1}$ & $2.6_{-0.3}^{+0.4}$ & $>1000$ & $100$ & 15.5 \\
        J1923+2515 & $-13.5^{95\%}$ & $ - $ & $0.7$ & $-13.8_{-0.2}^{+0.1}$ & $2.5_{-0.6}^{+0.8}$ & $>1000$ & $50$ & 9.0 \\
        J1946+3417 & $-12.6^{95\%}$ & $ - $ & $\mathbf{0.9}$ & $-12.3_{-0.1}^{+0.1}$ & $2.0_{-0.3}^{+0.3}$ & $>1000$ & $150$ & 5.7 \\
        J2010$-$1323 & $-13.3^{95\%}$ & $ - $ & $1.0$ & $-13.4_{-0.1}^{+0.1}$ & $2.1_{-0.2}^{+0.5}$ & $>1000$ & $150$ & 10.5 \\
        J2017+0603 & $-13.2^{95\%}$ & $ - $ & $1.0$ & $-13.2_{-0.1}^{+0.1}$ & $2.3_{-0.5}^{+0.5}$ & $>1000$ & $50$ & 8.3 \\
        J2033+1734 & $-13.0^{95\%}$ & $ - $ & $0.7$ & $-13.2_{-0.2}^{+0.1}$ & $1.8_{-0.5}^{+1.0}$ & $>1000$ & $100$ & 7.0 \\
        J2214+3000 & $-13.2^{95\%}$ & $ - $ & $0.7$ & $-13.6_{-0.7}^{+0.3}$ & $3.0_{-1.2}^{+2.0}$ & $188.9$ & $50$ & 8.4 \\
        J2229+2643 & $-13.2^{95\%}$ & $ - $ & $0.8$ & $-14.4_{-0.4}^{+0.4}$ & $5.0_{-1.1}^{+1.1}$ & $>1000$ & $50$ & 7.0 \\
        J2234+0611 & $-14.3_{-0.7}^{+0.7}$ & $4.4_{-2.1}^{+1.7}$ & $837.5$ & $-13.7_{-0.1}^{+0.1}$ & $3.1_{-0.6}^{+0.8}$ & $>1000$ & $50$ & 6.5 \\
        J2234+0944 & $-13.2^{95\%}$ & $ - $ & $0.7$ & $-13.0_{-0.1}^{+0.1}$ & $1.5_{-0.3}^{+0.3}$ & $>1000$ & $150$ & 7.1 \\
        J2302+4442 & $-12.7^{95\%}$ & $ - $ & $0.7$ & $-13.3_{-0.1}^{+0.1}$ & $2.4_{-0.6}^{+0.8}$ & $>1000$ & $50$ & 8.1 \\
        J2322+2057 & $-13.3^{95\%}$ & $ - $ & $0.6$ & $-14.0_{-0.3}^{+0.2}$ & $3.3_{-0.8}^{+1.1}$ & $>1000$ & $50$ & 5.4 \\
        \hline\hline
    \end{tabular}
    \caption{Posterior red and DM noise parameters and Savage-Dickey Bayes Factors $\mathcal{B}^{\rm{RN/DM}}_\oslash$ for 34 pulsars favoring the Fourier GP basis. The number of frequencies in the DMGP basis $N_f$ is selected on a per-pulsar basis, while $N_f = 30$ for RN. Where $\mathcal{B}^{\rm{RN/DM}}_\oslash > 10$, we report medians and 68\% Bayesian credible intervals on all parameters. {Otherwise, we report only the 95\% one-sided Bayesian credible interval on $\log_{10}A$ after reweighting the amplitude prior on $A$ in Table~\ref{tab:priors} from $\log_{10}$-uniform to uniform, assuming uniform priors on $\gamma$.} PSR B1953+29 is noted to show severe parameter covariance between red and DM noise processes, see \S{\ref{sec:noise_covariance}} for details. Bolded RN parameters have changed by $>$$1\sigma$ versus the standard noise model, and bolded Bayes factors have changed by over an order of magnitude versus the standard noise model (cf. \S\ref{sec:achromatic-red-noise-comparison}).}
    \label{tab:FD_params}
\end{table}

\begin{table}
    \centering
    \begin{tabular}{ c | c c c | c c c c c c | c @{}}
        \hline\hline & \multicolumn{3}{c|}{Red Noise} & \multicolumn{6}{c|}{DM GP} & \\
        \hline Pulsar & $\log_{10}A$ & $\gamma$ & $\mathcal{B}^{\rm{RN}}_\oslash$ & $\log_{10}\sigma$ & $\log_{10}\ell$ & $\log_{10}\alpha_{\rm wgt}$ & $\log_{10}\ell_2$ & $\mathcal{B}^{\rm{DM}}_\oslash$ & $dt$ (day) & $T_{\rm psr}$ (year) \\
        \hline\hline $^\dagger$B1937+21 & $-13.5_{-0.1}^{+0.1}$ & $3.8_{-0.3}^{+0.3}$ & $>1000$ & $-5.7_{-0.1}^{+0.1}$ & $2.49_{-0.04}^{+0.04}$ & $ - $ & $ - $ & $>1000$ & $3$ & 15.9 \\
        J0406+3039 & $-12.7^{95\%}$ & $ - $ & $0.7$ & $-6.4^{95\%}$ & $ - $ & $ - $ & $ - $ & $1.0$ & $15$ & 3.6 \\
        J0437$-$4715 & $-12.4^{95\%}$ & $ - $ & $\mathbf{0.9}$ & $-5.7_{-0.1}^{+0.1}$ & $ - $ & $ - $ & $ - $ & $>1000$ & $7$ & 4.8 \\
        J0509+0856 & $-11.6^{95\%}$ & $ - $ & $1.6$ & $-5.6^{95\%}$ & $ - $ & $ - $ & $ - $ & $0.8$ & $3$ & 3.6 \\
        J0557+1551 & $-12.3^{95\%}$ & $ - $ & $3.9$ & $-5.8^{95\%}$ & $ - $ & $ - $ & $ - $ & $0.7$ & $30$ & 4.6 \\
        J0605+3757 & $-12.1^{95\%}$ & $ - $ & $0.8$ & $-5.8^{95\%}$ & $ - $ & $ - $ & $ - $ & $0.7$ & $30$ & 3.4 \\
        J0610$-$2100 & $-12.9_{-0.5}^{+0.3}$ & $3.3_{-1.9}^{+2.1}$ & $64.8$ & $-6.2_{-0.1}^{+0.1}$ & $ - $ & $ - $ & $ - $ & $81.6$ & $30$ & 3.4 \\
        J0613$-$0200 & $-14.1_{-0.4}^{+0.3}$ & $3.9_{-0.8}^{+1.1}$ & $>1000$ & $-6.2_{-0.2}^{+0.2}$ & $2.8_{-0.1}^{+0.1}$ & $ - $ & $ - $ & $>1000$ & $3$ & 15.0 \\
        J0636+5128 & $\mathbf{-14.2_{-0.5}^{+0.6}}$ & $4.5_{-1.9}^{+1.6}$ & $\mathbf{63.9}$ & $-6.8_{-0.1}^{+0.1}$ & $ - $ & $ - $ & $ - $ & $>1000$ & $3$ & 6.3 \\
        J0709+0458 & $-12.0^{95\%}$ & $ - $ & $1.2$ & $-5.4^{95\%}$ & $ - $ & $ - $ & $ - $ & $2.7$ & $15$ & 4.6 \\
        J0931$-$1902 & $-13.3^{95\%}$ & $ - $ & $0.7$ & $-6.6^{95\%}$ & $ - $ & $ - $ & $ - $ & $1.2$ & $30$ & 7.1 \\
        J1012+5307 & $\mathbf{-12.9_{-0.1}^{+0.1}}$ & $0.9_{-0.3}^{+0.3}$ & $>1000$ & $-6.43_{-0.05}^{+0.05}$ & $1.5_{-0.1}^{+0.1}$ & $0.2_{-1.0}^{+1.9}$ & $0.6_{-0.5}^{+0.3}$ & $>1000$ & $20$ & 15.5 \\
        J1012$-$4235 & $-12.0^{95\%}$ & $ - $ & $0.8$ & $-5.8^{95\%}$ & $ - $ & $ - $ & $ - $ & $1.0$ & $7$ & 3.4 \\
        J1022+1001 & $-11.9^{95\%}$ & $ - $ & $1.0$ & $-5.6_{-0.1}^{+0.1}$ & $ - $ & $ - $ & $ - $ & $>1000$ & $3$ & 5.6 \\
        J1125+7819 & $-13.0^{95\%}$ & $ - $ & $0.7$ & $-6.1_{-0.1}^{+0.1}$ & $1.6_{-0.1}^{+0.1}$ & $1.4_{-1.1}^{+1.1}$ & $1.5_{-0.3}^{+0.3}$ & $>1000$ & $3$ & 6.3 \\
        J1312+0051 & $-12.5^{95\%}$ & $ - $ & $0.8$ & $-5.8^{95\%}$ & $ - $ & $ - $ & $ - $ & $0.7$ & $30$ & 4.6 \\
        J1453+1902 & $-12.7^{95\%}$ & $ - $ & $0.7$ & $-6.8^{95\%}$ & $ - $ & $ - $ & $ - $ & $0.5$ & $30$ & 7.0 \\
        J1455$-$3330 & $\mathbf{-13.7_{-0.9}^{+0.5}}$ & $2.8_{-1.4}^{+2.0}$ & $\mathbf{717.9}$ & $-6.6_{-0.1}^{+0.1}$ & $ - $ & $ - $ & $ - $ & $45.8$ & $7$ & 15.7 \\
        J1640+2224 & $-13.4^{95\%}$ & $ - $ & $2.1$ & $-7.2_{-0.2}^{+0.4}$ & $1.9_{-0.3}^{+0.6}$ & $ - $ & $ - $ & $>1000$ & $3$ & 15.5 \\
        J1643$-$1224 & $\mathbf{-12.6_{-0.1}^{+0.1}}$ & $\mathbf{1.5_{-0.4}^{+0.4}}$ & $>1000$ & $-5.67_{-0.04}^{+0.04}$ & $1.91_{-0.03}^{+0.03}$ & $-1.2_{-0.2}^{+0.4}$ & $2.1_{-0.4}^{+0.6}$ & $>1000$ & $3$ & 15.7 \\
        J1719$-$1438 & $-12.5^{95\%}$ & $ - $ & $0.7$ & $-6.2^{95\%}$ & $ - $ & $ - $ & $ - $ & $1.6$ & $30$ & 3.4 \\
        J1730$-$2304 & $-12.7^{95\%}$ & $ - $ & $1.9$ & $-6.6^{95\%}$ & $ - $ & $ - $ & $ - $ & $0.7$ & $7$ & 3.4 \\
        J1745+1017 & $-11.8_{-0.2}^{+0.2}$ & $2.3_{-0.9}^{+0.9}$ & $>1000$ & $-5.6^{95\%}$ & $ - $ & $ - $ & $ - $ & $0.8$ & $30$ & 4.5 \\
        J1747$-$4036 & $-12.5^{95\%}$ & $ - $ & $\mathbf{8.7}$ & $-5.5_{-0.1}^{+0.1}$ & $1.85_{-0.04}^{+0.05}$ & $-1.6_{-0.2}^{+0.2}$ & $1.4_{-0.5}^{+0.3}$ & $>1000$ & $3$ & 8.1 \\
        J1751$-$2857 & $-12.3^{95\%}$ & $ - $ & $0.8$ & $-5.7_{-0.1}^{+0.1}$ & $ - $ & $ - $ & $ - $ & $225.8$ & $30$ & 3.5 \\
        J1802$-$2124 & $-12.1^{95\%}$ & $ - $ & $\mathbf{1.1}$ & $-5.3_{-0.1}^{+0.2}$ & $2.1_{-0.1}^{+0.1}$ & $-1.8_{-0.4}^{+0.6}$ & $1.1_{-0.4}^{+0.8}$ & $>1000$ & $7$ & 3.5 \\
        J1811$-$2405 & $-13.1^{95\%}$ & $ - $ & $0.7$ & $-6.1_{-0.2}^{+0.2}$ & $2.0_{-0.2}^{+0.1}$ & $ - $ & $ - $ & $>1000$ & $3$ & 3.5 \\
        J1909$-$3744 & $-14.5_{-0.4}^{+0.3}$ & $4.0_{-0.9}^{+1.0}$ & $>1000$ & $-6.8_{-0.1}^{+0.1}$ & $2.00_{-0.05}^{+0.05}$ & $ - $ & $ - $ & $>1000$ & $3$ & 15.5 \\
        J1944+0907 & $\mathbf{-13.1_{-0.2}^{+0.1}}$ & $1.7_{-0.7}^{+0.8}$ & $\mathbf{67.4}$ & $-5.0_{-0.2}^{+0.2}$ & $2.9_{-0.1}^{+0.1}$ & $ - $ & $ - $ & $>1000$ & $20$ & 12.5 \\
        J2043+1711 & $-14.5_{-0.7}^{+0.5}$ & $4.0_{-1.6}^{+1.6}$ & $\mathbf{132.9}$ & $-6.8_{-0.1}^{+0.1}$ & $2.00_{-0.05}^{+0.05}$ & $ - $ & $ - $ & $>1000$ & $7$ & 9.1 \\
        J2124$-$3358 & $-12.7^{95\%}$ & $ - $ & $0.7$ & $-6.4_{-0.1}^{+0.1}$ & $ - $ & $ - $ & $ - $ & $74.9$ & $20$ & 3.5 \\
        $^*$J2145$-$0750 & $-13.0^{95\%}$ & $ - $ & $\mathbf{6.0}$ & $-6.4^{95\%}$ & $ - $ & $ - $ & $ - $ & $1.7$ & $30$ & 15.5 \\
        J2317+1439 & $-13.2^{95\%}$ & $ - $ & $0.9$ & $-6.1_{-0.1}^{+0.1}$ & $2.4_{-0.1}^{+0.1}$ & $ - $ & $ - $ & $>1000$ & $7$ & 15.6 \\
        \hline\hline
    \end{tabular}
    \caption{Posterior red and DM noise parameters for 33 pulsars favoring the TD or Ridge GP basis. The interpolation spacing $dt$ for the DM model is selected on a per-pulsar basis. Where $\mathcal{B}^{\rm{RN/DM}}_\oslash > 10$, we report medians and 68\% Bayesian credible intervals on all parameters. {Otherwise, we report only the 95\% one-sided Bayesian credible interval on $\log_{10}A$ after reweighting the amplitude prior on $A$ in Table~\ref{tab:priors} from $\log_{10}$-uniform to uniform, assuming uniform priors on $\gamma$.} The parameters $\log_{10}\alpha_{\rm wgt}$ and $\log_{10}\ell_2$ are only obtained for the 5 pulsars favoring radio-frequency dependent DM variations in the model. PSR B1937+21 is the only pulsar favoring a QP kernel for DM variations; its 2 additional DM parameters are $\log_{10}\Gamma_p = -0.8_{-0.1}^{+0.1}$ and $\log_{10}p = 0.00_{-0.01}^{+0.01}$. PSR J2145$-$0750 is noted to show severe parameter covariance between red and DM noise processes, see \S{\ref{sec:noise_covariance}} for details. Bolded RN parameters have changed by $>$$1\sigma$ versus the standard noise model, and bolded Bayes factors have changed by over an order of magnitude versus the standard noise model (cf. \S\ref{sec:achromatic-red-noise-comparison}).}
    \label{tab:TD_params}
\end{table}

\begin{table}
    \centering
    \begin{tabular}{ c | c c c | c c c | c | c @{}}
        \hline\hline & \multicolumn{3}{c|}{FC Noise (Fourier)} &\multicolumn{3}{c|}{FC Noise (TD/Ridge)} & \\
        \hline Pulsar & $\log_{10}A$ & $\gamma$ & $N_f$ & $\log_{10}\sigma$ & $\log_{10}\ell$ & $dt$ (day) & $\chi$ & \textsc{DM} (pc/cm$^3$) \\
        \hline B1937+21 & $ - $ & $ - $ & $ - $ & $-7.18_{-0.06}^{+0.05}$ & $0.1_{-0.1}^{+0.1}$ & $3$ & $4.8_{-0.2}^{+0.2}$ & 71.1 \\
        B1953+29 & $-13.2_{-0.1}^{+0.1}$ & $2.2_{-0.2}^{+0.3}$ & $100$ & $ - $ & $ - $ & $ - $ & $2.9_{-0.2}^{+0.2}$ & 104.5 \\
        J0437$-$4715 & $ - $ & $ - $ & $ - $ & $-5.9_{-0.1}^{+0.1}$ & $ - $ & $7$ & $4.1_{-0.3}^{+0.4}$ & 2.6 \\
        J0613$-$0200 & $ - $ & $ - $ & $ - $ & $-7.7_{-0.2}^{+0.2}$ & $1.7_{-0.1}^{+0.1}$ & $3$ & $5.1_{-0.5}^{+0.6}$ & 38.8 \\
        J1012+5307 & $ - $ & $ - $ & $ - $ & $-8.3_{-0.2}^{+0.2}$ & $1.6_{-0.3}^{+0.2}$ & $20$ & $8.2_{-0.8}^{+0.8}$ & 8.9 \\
        J1022+1001 & $ - $ & $ - $ & $ - $ & $-8.7_{-0.5}^{+0.6}$ & $ - $ & $3$ & $7.6_{-1.1}^{+0.9}$ & 9.4 \\
        J1125+7819 & $ - $ & $ - $ & $ - $ & $-7.8_{-0.2}^{+0.2}$ & $1.6_{-0.1}^{+0.1}$ & $3$ & $9.1_{-0.7}^{+0.7}$ & 11.2 \\
        J1600$-$3053 & $-14.1_{-0.2}^{+0.2}$ & $1.5_{-0.2}^{+0.2}$ & $150$ & $ - $ & $ - $ & $ - $ & $5.6_{-0.5}^{+0.6}$ & 52.3 \\
        J1640+2224 & $ - $ & $ - $ & $ - $ & $-7.1_{-0.9}^{+0.3}$ & $2.7_{-0.7}^{+0.2}$ & $3$ & $2.8_{-0.2}^{+1.1}$ & 18.5 \\
        J1643$-$1224 & $ - $ & $ - $ & $ - $ & $-7.3_{-0.2}^{+0.2}$ & $1.9_{-0.1}^{+0.1}$ & $3$ & $6.9_{-0.6}^{+0.6}$ & 62.3 \\
        J1705$-$1903 & $-13.0_{-0.1}^{+0.1}$ & $0.3_{-0.2}^{+0.3}$ & $200$ & $ - $ & $ - $ & $ - $ & $4.1_{-0.3}^{+0.3}$ & 57.5 \\
        J1713+0747 & $-15.8_{-0.2}^{+0.2}$ & $0.9_{-0.2}^{+0.2}$ & $200$ & $ - $ & $ - $ & $ - $ & $9.5_{-0.7}^{+0.7}$ & 16.0 \\
        J1738+0333 & $-14.0_{-0.7}^{+0.4}$ & $3.1_{-0.9}^{+1.4}$ & $50$ & $ - $ & $ - $ & $ - $ & $6.5_{-1.6}^{+1.3}$ & 33.8 \\
        J1744$-$1134 & $-14.4_{-0.7}^{+0.4}$ & $1.1_{-0.3}^{+0.3}$ & $100$ & $ - $ & $ - $ & $ - $ & $4.6_{-1.1}^{+2.1}$ & 3.1 \\
        J1747$-$4036 & $ - $ & $ - $ & $ - $ & $-6.5_{-0.2}^{+0.1}$ & $0.3_{-0.2}^{+0.3}$ & $3$ & $5.1_{-0.4}^{+0.5}$ & 153.0 \\
        J1802$-$2124 & $ - $ & $ - $ & $ - $ & $-7.2_{-0.3}^{+0.4}$ & $2.1_{-0.2}^{+0.1}$ & $7$ & $7.6_{-1.0}^{+0.9}$ & 149.6 \\
        J1903+0327 & $-12.4_{-0.1}^{+0.1}$ & $2.6_{-0.3}^{+0.3}$ & $50$ & $ - $ & $ - $ & $ - $ & $7.6_{-0.4}^{+0.4}$ & 297.6 \\
        J1944+0907 & $ - $ & $ - $ & $ - $ & $-7.1_{-0.9}^{+0.4}$ & $1.9_{-0.1}^{+0.1}$ & $20$ & $3.5_{-0.8}^{+1.7}$ & 24.4 \\
        J1946+3417 & $-12.7_{-0.1}^{+0.1}$ & $0.9_{-0.3}^{+0.3}$ & $150$ & $ - $ & $ - $ & $ - $ & $5.9_{-0.7}^{+0.8}$ & 110.2 \\
        J2145$-$0750 & $ - $ & $ - $ & $ - $ & $-8.0_{-0.3}^{+0.3}$ & $ - $ & $30$ & $7.1_{-1.0}^{+0.9}$ & 8.9 \\
        J2317+1439 & $ - $ & $ - $ & $ - $ & $-7.6_{-0.4}^{+0.2}$ & $1.7_{-0.1}^{+0.1}$ & $7$ & $2.9_{-0.3}^{+0.5}$ & 21.9 \\
        \hline\hline
    \end{tabular}
    \caption{Medians and 68\% credible intervals of posterior FC noise parameters for all 21 pulsars with significant FC noise. If the Fourier GP was favored for the pulsar, the parameters and number of frequencies $N_f$ are listed in the 2nd column. If instead the TD or Ridge GP was favored for the pulsar, the parameters and interpolation spacing $dt$ are listed in the 3rd column. The Savage-Dickey Bayes Factor $\mathcal{B}^{\rm{FC}}_\oslash$ is much larger than $10^3$ in all cases. Nominal DM values of each pulsar are also listed for reference. Posteriors on the FC indices $\chi$ can also be found in Fig.~\ref{fig:chrom_idxs} along with the prior.}
    \label{tab:FC_params}
\end{table}

\begin{table}
    \centering
    \begin{tabular}{ c | c c c | c c | c | c @{}}
        \hline\hline & \multicolumn{3}{c|}{SWGP (Fourier)} &\multicolumn{2}{c|}{SWGP (Ridge)} & \\
        \hline Pulsar & $\log_{10}A_{n_E}$ & $\gamma_{n_E}$ & $N_f$ & $\log_{10}\sigma_{n_E}$ & $dt$ & $\mathcal{B}^{\rm{SWGP}}_\oslash$ & \textsc{Elat} ($^\circ$) \\
        \hline B1855+09 & $-9.4_{-0.8}^{+1.1}$ & $-3.9_{-1.5}^{+2.2}$ & $200$ & $ - $ & $ - $ & 365.8 & 32.3 \\
        B1937+21 & $ - $ & $ - $ & $ - $ & $0.89_{-0.04}^{+0.04}$ & $3$ days & $>1000$ & 42.3 \\
        J0023+0923 & $-6.6_{-0.1}^{+0.1}$ & $1.9_{-0.2}^{+0.2}$ & $200$ & $ - $ & $ - $ & $>1000$ & 6.3 \\
        J0030+0451 & $-6.6_{-0.1}^{+0.1}$ & $2.2_{-0.2}^{+0.2}$ & $200$ & $ - $ & $ - $ & $>1000$ & 1.4 \\
        J0340+4130 & $ - $ & $ - $ & $ - $ & $1.1^{95\%}$ & $1$ year & 0.9 & 21.3 \\
        J0406+3039 & $ - $ & $ - $ & $ - $ & $1.3^{95\%}$ & $1$ year & 1.1 & 9.6 \\
        J0509+0856 & $ - $ & $ - $ & $ - $ & $1.7_{-0.1}^{+0.1}$ & $3$ days & 237.3 & -13.9 \\
        J0557+1551 & $ - $ & $ - $ & $ - $ & $1.6^{95\%}$ & $1$ year & 0.9 & -7.6 \\
        J0605+3757 & $ - $ & $ - $ & $ - $ & $1.7^{95\%}$ & $1$ year & 0.9 & 14.5 \\
        J0613$-$0200 & $ - $ & $ - $ & $ - $ & $0.46_{-0.06}^{+0.06}$ & $3$ days & $>1000$ & -25.4 \\
        J0636+5128 & $ - $ & $ - $ & $ - $ & $1.0^{95\%}$ & $1$ year & 0.8 & 28.2 \\
        J0645+5158 & $ - $ & $ - $ & $ - $ & $0.8^{95\%}$ & $1$ year & 0.9 & 28.9 \\
        J0709+0458 & $ - $ & $ - $ & $ - $ & $2.0^{95\%}$ & $1$ year & 1.4 & -17.4 \\
        J0931$-$1902 & $ - $ & $ - $ & $ - $ & $1.1^{95\%}$ & $1$ year & 0.8 & -31.8 \\
        J1022+1001 & $ - $ & $ - $ & $ - $ & $1.2_{-0.2}^{+0.2}$ & $1$ year & $>1000$ & -0.1 \\
        J1024$-$0719 & $ - $ & $ - $ & $ - $ & $0.9^{95\%}$ & $1$ year & 1.5 & -16.0 \\
        J1312+0051 & $ - $ & $ - $ & $ - $ & $1.9^{95\%}$ & $1$ year & 0.9 & 7.9 \\
        J1453+1902 & $ - $ & $ - $ & $ - $ & $1.2^{95\%}$ & $1$ year & 0.8 & 33.9 \\
        J1455$-$3330 & $ - $ & $ - $ & $ - $ & $0.6^{95\%}$ & $1$ year & 0.7 & -16.0 \\
        J1600$-$3053 & $ - $ & $ - $ & $ - $ & $0.6^{95\%}$ & $1$ year & 0.8 & -10.1 \\
        J1614$-$2230 & $ - $ & $ - $ & $ - $ & $0.3^{95\%}$ & $1$ year & 0.7 & -1.3 \\
        J1640+2224 & $ - $ & $ - $ & $ - $ & $0.52_{-0.06}^{+0.05}$ & $3$ days & $>1000$ & 44.1 \\
        J1643$-$1224 & $ - $ & $ - $ & $ - $ & $0.8^{95\%}$ & $1$ year & 0.8 & 9.8 \\
        J1705$-$1903 & $ - $ & $ - $ & $ - $ & $1.3^{95\%}$ & $1$ year & 0.8 & 3.8 \\
        J1713+0747 & $-10.3_{-0.4}^{+0.7}$ & $-4.9_{-0.8}^{+1.3}$ & $200$ & $ - $ & $ - $ & 90.4 & 30.7 \\
        \hline\hline
    \end{tabular}
    \caption{Posterior SWGP parameters for 25 out of 50 pulsars including GP perturbations to $n_E(t)$. Where $\mathcal{B}^{\rm{SWGP}}_\oslash > 10$, we report medians and 68\% Bayesian credible intervals on SWGP parameters, otherwise we report only the 95\% one-sided Bayesian credible interval on the amplitude parameter after reweighting the prior in Table~\ref{tab:priors} from $\log_{10}$-uniform to uniform. Where $dt = 1$ year is indicated for the pulsar, the triangular GP basis modeling annual perturbations to $n_E(t)$ is adopted, which is done by default for all pulsars where $|$\textsc{Elat}$|< 35^\circ$ regardless of the Savage-Dickey Bayes Factor $\mathcal{B}^{\rm{SWGP}}_\oslash$ since the SW signal is always present. The Fourier basis or the fine-grained Ridge GP basis is adopted instead for specific pulsars, where preferred by our model selection procedure. 2 pulsars (B1937+21, J1640+2224) notably prefer a SWGP component outside $|$\textsc{Elat}$|> 35^\circ$. Table is continued in Table~\ref{tab:SW_params2}. Note that priors on SWGP hyperparameters are model-dependent, so Bayes Factors may not be directly comparable across different pulsars using different models for SWGP.}
    \label{tab:SW_params1}
\end{table}

\begin{table}
    \centering
    \begin{tabular}{ c | c c c | c c | c | c @{}}
        \hline\hline & \multicolumn{3}{c|}{SWGP (Fourier)} &\multicolumn{2}{c|}{SWGP (Ridge)} & \\
        \hline Pulsar & $\log_{10}A_{n_E}$ & $\gamma_{n_E}$ & $N_f$ & $\log_{10}\sigma_{n_E}$ & $dt$ & $\mathcal{B}^{\rm{SWGP}}_\oslash$ & \textsc{Elat} ($^\circ$) \\
        \hline J1719$-$1438 & $ - $ & $ - $ & $ - $ & $1.3^{95\%}$ & $1$ year & 0.7 & 8.5 \\
        J1730$-$2304 & $ - $ & $ - $ & $ - $ & $1.06_{-0.08}^{+0.08}$ & $7$ days & $>1000$ & 0.2 \\
        J1738+0333 & $ - $ & $ - $ & $ - $ & $1.1^{95\%}$ & $1$ year & 0.8 & 26.9 \\
        J1744$-$1134 & $ - $ & $ - $ & $ - $ & $0.5^{95\%}$ & $1$ year & 1.3 & 11.8 \\
        J1745+1017 & $ - $ & $ - $ & $ - $ & $2.09_{-0.09}^{+0.09}$ & $30$ days & $>1000$ & 33.7 \\
        J1747$-$4036 & $ - $ & $ - $ & $ - $ & $1.6^{95\%}$ & $1$ year & 1.1 & -17.2 \\
        J1751$-$2857 & $ - $ & $ - $ & $ - $ & $1.6^{95\%}$ & $1$ year & 0.9 & -5.5 \\
        J1802$-$2124 & $ - $ & $ - $ & $ - $ & $1.5^{95\%}$ & $1$ year & 1.3 & 2.0 \\
        J1811$-$2405 & $ - $ & $ - $ & $ - $ & $0.89_{-0.09}^{+0.10}$ & $3$ days & $>1000$ & -0.7 \\
        J1832$-$0836 & $ - $ & $ - $ & $ - $ & $1.0^{95\%}$ & $1$ year & 0.8 & 14.6 \\
        J1843$-$1113 & $ - $ & $ - $ & $ - $ & $1.3^{95\%}$ & $1$ year & 0.8 & 11.8 \\
        J1903+0327 & $ - $ & $ - $ & $ - $ & $1.9^{95\%}$ & $1$ year & 1.0 & 25.9 \\
        J1909$-$3744 & $ - $ & $ - $ & $ - $ & $0.43_{-0.03}^{+0.03}$ & $3$ days & $>1000$ & -15.2 \\
        J1918$-$0642 & $ - $ & $ - $ & $ - $ & $0.6^{95\%}$ & $1$ year & 0.8 & 15.4 \\
        J1944+0907 & $ - $ & $ - $ & $ - $ & $0.98_{-0.06}^{+0.06}$ & $20$ days & $>1000$ & 29.9 \\
        J2010$-$1323 & $ - $ & $ - $ & $ - $ & $0.1^{95\%}$ & $1$ year & 0.7 & 6.5 \\
        J2017+0603 & $ - $ & $ - $ & $ - $ & $1.0^{95\%}$ & $1$ year & 0.8 & 25.0 \\
        J2043+1711 & $ - $ & $ - $ & $ - $ & $0.30_{-0.06}^{+0.06}$ & $7$ days & $>1000$ & 34.0 \\
        J2124$-$3358 & $ - $ & $ - $ & $ - $ & $1.3^{95\%}$ & $1$ year & 0.8 & -17.8 \\
        J2145$-$0750 & $ - $ & $ - $ & $ - $ & $0.7^{95\%}$ & $1$ year & 5.8 & 5.3 \\
        J2229+2643 & $ - $ & $ - $ & $ - $ & $0.9_{-0.2}^{+0.2}$ & $1$ year & 46.0 & 33.3 \\
        J2234+0611 & $ - $ & $ - $ & $ - $ & $0.6^{95\%}$ & $1$ year & 0.7 & 14.1 \\
        J2234+0944 & $ - $ & $ - $ & $ - $ & $1.3^{95\%}$ & $1$ year & 0.8 & 17.3 \\
        J2317+1439 & $ - $ & $ - $ & $ - $ & $0.46_{-0.04}^{+0.04}$ & $7$ days & $>1000$ & 17.7 \\
        J2322+2057 & $ - $ & $ - $ & $ - $ & $1.0^{95\%}$ & $1$ year & 0.9 & 22.9 \\
        \hline\hline
    \end{tabular}
    \caption{Continuation of Table~\ref{tab:SW_params1} for remaining 25 pulsars that include GP perturbations to $n_E(t)$ in their noise models. Blank columns for the Fourier GP basis are to show explicitly that Fourier perturbations were unfavored over TD or annual perturbations to $n_E(t)$ for these pulsars. Note that priors on SWGP hyperparameters are model-dependent, so Bayes Factors may not be directly comparable across different pulsars using different models for SWGP.}
    \label{tab:SW_params2}
\end{table}

\begin{table}
    \centering
    \begin{tabular}{ c | c c c | c @{}}
        \hline\hline Pulsar/Event & $\log_{10}A_{\rm exp}$ & $\log_{10}\tau_{\rm exp}$ & $t_{0,\rm exp}$ & $\chi_{\rm exp}$ \\
        \hline J1643$-$1224 & $ -5.1^{+0.1}_{-0.1} $ & $ 2.0^{+0.1}_{-0.1} $ & $ 57090.3^{+6.6}_{-6.7} $ & $ -1.6^{+0.3}_{-0.3} $ \\
        J1713+0747 (1st) & $ -6.1^{+0.1}_{-0.1} $ & $ 2.5^{+0.1}_{-0.1} $ & $ 54758.5^{+4.8}_{-4.9} $ & $ 3.2^{+0.2}_{-0.2} $ \\
        J1713+0747 (2nd) & $ -5.86^{+0.03}_{-0.03} $ & $ 1.6^{+0.1}_{-0.1} $ & $ 57510.4^{+1.3}_{-1.4} $ & $ 1.3^{+0.1}_{-0.1} $ \\
        J2145$-$0750 & $ -5.6^{+0.1}_{-0.1} $ & $ 2.1^{+0.2}_{-0.2} $ & $ 56322.0^{+7.4}_{-7.4} $ & $ -0.6^{+0.4}_{-0.4} $ \\
        \hline\hline
    \end{tabular}
    \caption{Inferred parameters of the chromatic exponential decay event model, applied for four known events in PSRs J1643$-$1224, J1713+0747, and J2145$-$0750 \citep{Shannon+2016, lam+2018, goncharov+2021}. Reported values are the parameter medians and 68\% credible intervals from the posteriors.}
    \label{tab:det_params}
\end{table}

\section{Rank-Reduced Gaussian Process and Deterministic Signal Definitions}
\label{appendix:GP_kernels}

The GP bases and kernels we use in this work for chromatic signals are comprehensively described by \citet{ng12p5_cnm} and other previous works \citep{vanHaasterenLevin2013, lentati+:2016, hazboun:2019sc, Chalumeau+2022, ng15detchar, Nitu+2024, susarla+24sw}, but we repeat various definitions here for completeness.

The GP design matrices (denoted here as $\mathbf{F}$ for each model) map the TOAs (indexed by $i = [0,\cdots,N_{\rm TOA}-1] \in \mathbb{N}$) to a lower-dimensional latent space with finite basis elements (index by $j = [0,\cdots,N_j-1] \in \mathbb{N}$). The design matrices in general operate using the TOAs as input but are also scaled to account for (free) chromatic processes as
\begin{align}
    \label{eq:chromatic_basis}
    F_{ij} \to F_{ij}\times \left(\frac{\nu_i}{1400\text{ MHz}}\right)^{-\chi},
\end{align}
where $\chi = 2$ for DMGP. SWGP instead uses Eq.~\eqref{eq:sw_dt} such that
\begin{align}
    \label{eq:SW_basis}
    F_{ij} \to F_{ij}\times \frac{e^2}{2\pi m_ec^2\nu_i^2}\frac{\pi - \theta_i}{r_\oplus\sin\theta_i}(1~\rm{AU})^2
\end{align}
so the GP parameters in the latent space are converted into units of cm$^{-3}$ rather than a time delay in seconds (i.e., we place a GP on the $n_E(t)$ variations rather than the time delays). In the remaining definitions we assume the latent space has units of seconds.

The Fourier basis uses the following design matrix,
\begin{align}
    \label{eq:Fourier_basis}
    F^{\rm Fourier}_{ij} = \begin{cases}
        \cos(2\pi f_Jt_i), \quad J=j/2 & \text{for even }j, \\
        \sin(2\pi f_Jt_i), \quad J=(j-1)/2 & \text{for odd }j,
    \end{cases}
\end{align}
where $f_J = (J+1)/T_{\rm NG15}$ and $N_j = 2N_f$. The prior covariance on the latent space parameters (denoted $\bm{\phi}$) 
is physically interpretable under the Fourier basis as a prior power spectrum, for which we most often use the power law model
\begin{align}
    \label{eq:powerlaw}
    \phi^{\rm PL}_{jk} = \frac{A^2}{12\pi^2f^3T_{\rm NG15}}\left(\frac{f_J}{f_{\rm yr}}\right)^{-\gamma}\delta_{jk}\;\rm{s}^2.
\end{align}
By convention this definition follows from standard derivations of the induced TOA delay from a stochastic GWB, with spectral amplitude $A$ and index $\gamma$ hyperparameters. As a diagnostic for achromatic RN, we also use the free spectral prior covariance, in which the power at each frequency is estimated independently
\begin{align}
    \label{eq:free_spectrum}
    \phi^{\rm FS}_{jk} = \rho^2_J\delta_{jk},
\end{align}
where $\rho_J$ is the $J$th spectral amplitude hyperparameter in units of seconds at $f_J$.

In this work we employ a two-column Fourier design matrix to account for annual chromatic signals,
\begin{align}
    \label{eq:annual_basis}
    F^{\rm 1yr}_{ij} = \begin{cases}
        \cos(2\pi f_{\rm yr}t_i) & \text{for }j=0, \\
        \sin(2\pi f_{\rm yr}t_i) & \text{for }j=1,
    \end{cases}
\end{align}
where $N_j = 2$. This model uses a free spectral prior covariance 
\begin{align}
    \label{eq:annual_prior}
    \phi^{\rm 1yr}_{jk} = \rho_{\rm 1yr}^2\delta_{jk},
\end{align}
with the single hyperparameter $\rho_{\rm 1yr}$. The benefit of this formulation is analytic marginalization over the phase of the annual signal.

The TD linear interpolation basis is defined by the following design matrix,
\begin{align}
    \label{eq:TD_Basis}
    F^{\rm TD}_{ij} = \frac{1}{dt}\begin{cases}
        t'_{j+1} - t_i & \text{for }t'_j \le t_i \le t'_{j+1}, \\
        t_i - t'_{j-1} & \text{for }t'_{j-1} \le t_i \le t'_{j}, \\
        0 & \text{elsewhere},
    \end{cases}
\end{align}
where $t'_j = \min(t) + j\cdot dt$ and $N_j = T_{\rm psr}/dt$, with the modification that empty columns of $\mathbf{F}^{\rm{TD}}$, which result when gaps between TOAs are larger than $2dt$, are removed after construction of the basis. The Ridge, SE, and QP kernels now operate on the rank-reduced TOA grid, where the QP kernel yields the prior covariance
\begin{align}
    \label{eq:QP_kernel}
    \phi^{\rm QP}_{jk} = \left(\frac{\sigma}{50000}\right)^2\delta_{jk} + \sigma^2\exp\left(-\frac{|t'_j - t'_k|^2}{2\ell^2}\right)\exp\left(-\Gamma_p\sin^2\left(\frac{\pi|t'_j - t'_k|}{p}\right)\right),
\end{align}
with four hyperparameters $\sigma$, $\ell$, $\Gamma_p$, and $p$. The SE kernel is a special case of the QP kernel when $\Gamma_p = 0$, and the Ridge (diagonal) kernel is a special case of the SE kernel when $\ell = 0$.

The TD\_RF basis (used exclusively for DM variations) uses a ``nearest'' interpolation scheme in time and radio frequency, with the design matrix
\begin{align}
    \label{eq:RF_band_basis}
    F^{\rm TD\_RF}_{ij} = \begin{cases}
        1 & \text{for }t'_{j_t} \le t_i \le t'_{j_t+1} \;\&\; \nu'_{j_\nu} \le \nu_i \le \nu'_{j_\nu+1}, \\
        0 & \text{elsewhere},
    \end{cases}
\end{align}
where $j = j_tj_\nu$, the grid $t'$ (and corresponding time index $j_t$) are defined as previously for the TD basis, and $j_\nu$ iterates through the endpoints of four bands defined between $\vec{\nu}' = [600,1000,1900,3000,5000]$ MHz. The prior covariance uses the average $\bar{t}_j$ and $\bar{\nu}_j$ of the TOAs within the interval given by each $j$ as inputs, rather than the grid points given by $t'_j$ and $\nu'_j$. As previously, empty columns of $\mathbf{F}^{\rm TD\_RF}$ are removed after construction. The prior covariance is then the product of Eq.~\eqref{eq:QP_kernel} (using $\bar{t}$ instead of $t'$) and the RQ kernel,
\begin{align}
    \label{eq:RQ_kernel}
    \phi^{\rm QP\_RF}_{jk} = \phi^{\rm QP}_{jk}\left(1 + \frac{|\bar{\nu}_j - \bar{\nu}_k|^2}{2\ell_2\alpha_{\rm wgt}}\right)^{-\alpha_{\rm wgt}},
\end{align}
where $\alpha_{\rm wgt}$ and $\ell_2$ are the hyperparameters encoding the strength of the decorrelations across radio frequency. Setting $\alpha_{\rm wgt} = 0$ recovers $\bm{\phi}^{\rm QP\_RF} \to \bm{\phi}^{\rm QP}$, i.e., DM variations which are perfectly correlated across the entire band. While we do not use the TD\_RF basis for this purpose, we note that pairing $\bm{\phi}^{\rm{QP}}$ with $\mathbf{F}^{\rm{TD\_RF}}$ basis may still allow constant DM offsets between different bands, which may be used to search for time-independent chromatic offsets beyond what is accounted for by the typical \texttt{FD} timing model parameters \citep{ng15data}.

We use the triangular basis from \citet{Nitu+2024} as one of our options for modeling SW variations and it is defined with the design matrix
\begin{align}
    \label{eq:Nitu_basis}
    F^{\rm{SW}\Lambda}_{ij} = \begin{cases}
        1 - \left|\frac{t_i - T_{c,j}}{1\;\rm{yr}}\right| & \text{for }|t_i - T_{c,j}| < 1\;\rm{yr}, \\
        0 & \rm{elsewhere},
    \end{cases}
\end{align}
where $T_{c,j}$ is the $j$th annual closest approach between the Sun and the pulsar, and $N_j = T_{\rm psr}/\rm{yr}$. In this work, we consider only the diagonal prior covariance (Ridge kernel) to accompany this basis, as is done in \citet{Nitu+2024}.

Finally, the event signals (\S\ref{sec:event_models}) we search for in select pulsars are defined by the deterministic transient decaying exponential ``dip'' function,
\begin{align}
    \label{eq:det_dip}
    s^{\rm exp}(t_i) = -A_{\rm exp}\Theta(t_i - t_{0,\rm{exp}})\exp\left(-\frac{t_i - t_{0,\rm{exp}}}{\tau_{\rm exp}}\right)\left(\frac{\nu_i}{1400\;\rm{MHz}}\right)^{-\chi_{\rm exp}},
\end{align}
where $A_{\rm exp}$, $\tau_{\rm exp}$, $t_{0,\rm{exp}}$, and $\chi_{\rm exp}$ are the free parameters and $\Theta$ is the Heaviside step function \citep{lam+2018, goncharov+2021}. We also search for a Gaussian event signal in PSR J1600$-$3053,
\begin{align}
    \label{eq:gauss_event}
    s^{\rm{G}}(t_i) = A_{\rm G}\exp\left(-\frac{(t_i - t_{0,\rm{G}})^2}{2\tau^2_{\rm G}}\right)\left(\frac{\nu_i}{1400\;\rm{MHz}}\right)^{-\chi_{\rm G}},
\end{align}
where $A_{\rm G}$, $\tau_{\rm G}$, $t_{0,\rm{G}}$, and $\chi_{\rm G}$ are the free parameters \citep{pptadr3:noise, MPTADR2_noise}. Note that we do not keep the Gaussian signal in the final model for PSR J1600$-$3053, nor do we use this signal template to search extensively for noise transients in other pulsars.

\section{Multi-parameter Savage-Dickey Bayes Factor for Nested Kernels}
\label{appendix:TDSDBF}

\begin{table}
    \begin{center}
        \label{tab:TD_significance_testing}
        \caption{Savage-Dickey Bayes Factors computed to assess significance of different \textsc{Model\_TD} kernel components, and the respective parameter ranges where the model is insignificant.}
        \begin{tabular}{c|ccc}
            \hline\hline Component & Parameter Range & $\mathcal{B}$ (DM) & $\mathcal{B}$ (FC) \\
            \hline RF & $\log_{10}\alpha_{\mathrm{wgt}} < -3.5$ & $\mathcal{B}^{\mathrm{QP\_RF}}_{\mathrm{QP}}$ & $-$ \\
            QP & $\log_{10}\Gamma_{p} < -3$ & $\mathcal{B}^{\mathrm{QP\_RF}}_{\mathrm{SE\_RF}}$ & $\mathcal{B}^{\mathrm{QP}}_{\mathrm{SE}}$ \\
            Entire & $\log_{10}\sigma < -9$, $\chi < 4$ & $-$ & $\mathcal{B}^{\mathrm{QP}}_{\oslash}$ \\
            \hline \hline
        \end{tabular}
    \end{center}
    \vspace{-\baselineskip}
\end{table}

Using \textsc{Model\_TD}, evaluating model complexity is tricky because there are four different possible kernels for DM (SE, QP, SE\_RF, and QP\_RF) and three different possible kernels for FC (SE, QP, or None). Thankfully, due to the nested properties of \textsc{Model\_TD}, we can still evaluate the significance of each kernel component using Savage-Dickey Bayes Factors. In each case, we again use a Bayes Factor threshold $\mathcal{B} > 10$ as a criteria to keep the more complicated form of the kernel in the model. The four different Bayes Factors that we can nominally compute for model selection using 1D marginalized parameter posteriors are tabulated in Table~\ref{tab:TD_significance_testing}. 


Computing further Bayes Factors, such as $\mathcal{B}^{\mathrm{QP\_RF}}_{\mathrm{SE}}$ for the DM variations kernel, is possible from a single MCMC analysis using the multiplicative property of Bayes Factors and the condition that one of the component Bayes Factors is computed in the region where the other model component is insignificant, i.e.,
\begin{align}
    \mathcal{B}^{\mathcal{M}_1\mathcal{M}_2}_{\oslash} &= \mathcal{B}^{\mathcal{M}_1\mathcal{M}_2}_{\mathcal{M}_1}\mathcal{B}^{\mathcal{M}_1}_{\oslash} = \frac{\pi(A_2=0)}{\mathcal{P}(A_2=0|d)}\frac{\pi(A_1=0)}{\mathcal{P}(A_1=0|d,A_2=0)}, \label{eq:nested_BF}
\end{align}
where $\mathcal{M}_1$, $\mathcal{M}_2$ represent two independent model components, and when their amplitude parameters $A_1$, $A_2$, equal zero, they turn off the respective model component. In practice these parameters for \textsc{Model\_TD} use log-uniform priors, so we use their posteriors in the regions of parameter space listed in Table~\ref{tab:TD_significance_testing} to understand when the model is insignificant. The second term in Eq.~\ref{eq:nested_BF} is then evaluated by only using posterior samples from the insignificant region, i.e., masking samples where $\log_{10}A_2$ is greater than the threshold.

\begin{figure}
    \centering
    \includegraphics[width=0.35\linewidth]{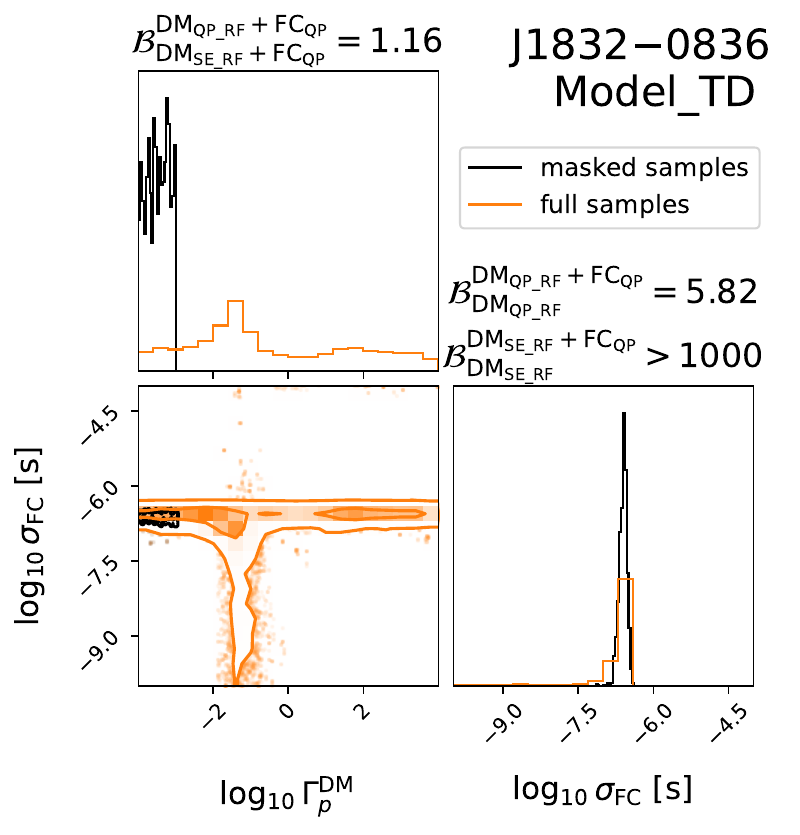}
    \caption{Automated model selection demonstrated in an example case where chromatic model parameters $\log_{10}\Gamma_p$, which amplifies the QP component of the DM kernel, and $\log_{10}(\sigma^{\rm{FC}}/\rm{s})$, which amplifies the FC model component, are covariant with one another. Computing the Savage-Dickey Bayes Factor using the full posterior (orange) indicates neither component is significant. However, computing the Bayes Factor using only samples where the QP kernel is insignificant (black) indicates that the FC component is significant after reducing the QP kernel to the SE kernel}
    \label{fig:param_covariance}
\end{figure}

It is extremely important to perform a masking technique such as this if attempting to evaluate the significance of multiple model components from a single analysis because of \emph{covariances} between noise parameters, which often occur when the data are not informative enough to distinguish between noise processes (cf. Fig.~\ref{fig:param_covariance}; \citealt{eptadr2_2:noise, Ferranti+2025, DR2Lite_2025}). Unfortunately the masking procedure becomes impractical if masking in more than one parameter, as this leaves behind too few independent samples and a Savage-Dickey Bayes Factor cannot be reliably computed. However, since the purpose is to reduce the model down just to the most necessary components whose Bayes Factors all pass the threshold $\mathcal{B} > 10$, a solution is to just assess each combination of insignificant parameters for covariances. In practice we do this by looping through combinations of insignificant model components whose $\mathcal{B} < 10$ and computing the second term of Eq.~\ref{eq:nested_BF}. If two model components are covariant such that $\mathcal{B} > 10$ after one is removed, we decide only the less significant component is removed. Throughout later stages of the model selection, we continuously check if \textsc{Model\_TD} components are still significant in this manner.

An example of this covariance from the first round of our analyses is shown in Fig.~\ref{fig:param_covariance} for PSR J1832$-$0836. In this case, either a QP DM kernel or a FC component is needed to explain some of its timing residual data, but the data cannot distinguish which. Naively using the Savage-Dickey Bayes Factor from Table~\ref{tab:TD_significance_testing} would suggest to (mistakenly) remove both components as neither by itself satisfies $\mathcal{B} > 10$. However, Eq.~\ref{eq:nested_BF} correctly identifies one of the components as significant. In this case, the FC component is more significant than the QP component of DM variations, so the FC component is left in at this stage, while the QP kernel for DM variations is replaced with the simpler SE kernel (although later stages of the model selection process eventually show that FC variations are not truly detected in this pulsar's timing residuals either).

\section{Solar Wind Sensitivity in NANOGrav Pulsars}
\label{appendix:SW_SNR}

Part of what makes the SW so distinct and important to model in PTA datasets is the unique geometric modulation to the base DM it imparts, Eq.~\eqref{eq:sw_dm}, which are strongly dependent on solar impact angle $\theta$. Simultaneously, $\theta$ can never be smaller than the absolute value of the angle \textsc{Elat} between the pulsar and the ecliptic plane, leading to the conventional wisdom that the pulsars which require the most advanced SW modeling are those which are closest to the ecliptic. Nonetheless, other factors such as the observation cadence close near $T_c$ (defined as the closest approach between the Sun and the pulsar), observation frequency $\nu$, and TOA measurement error also impact the sensitivity to the SW. For example, \citet{susarla+24sw} found that variations from the SW can be observed even beyond impact angles $\theta > 45^\circ$ when using ultra-low frequency observations from LOFAR. Given these considerations, in this appendix we quantify and rank each pulsar's theoretical sensitivity to the SW, which we can then relate to our empirical model selection results. As a bonus, we also develop analytic scaling laws for the SW signal to noise ratio (S/N) for a single idealistic pulsar.

One way to define a single pulsar's sensitivity to DM variations from the SW is the SW signal variance divided by the noise variance, summed over all TOAs in the pulsar's dataset,
\begin{align}
    \label{eq:SW_SNR}
    {\rm{S/N}}_{\rm SW}(n_E) &= \sum_{i=0}^{N_{\rm TOA}}\frac{({\rm{DM}}_{\rm SW}(n_E,\theta_i) - \overline{\rm{DM}}_{\rm SW}(n_E))^2}{K_{\rm{DM}}^2\sigma_{{\rm TOA},i}^2\nu_i^4},
\end{align}
where ${\rm{DM}}_{\rm SW}(n_E,\theta_i)$ depends on both $n_E$ and the impact angle $\theta_i$ at the $i$th observation (Eq.~\ref{eq:sw_dm}), $\sigma_{{\rm TOA},i}$ is the $i$th TOA measurement uncertainty, $\nu_i$ is the corresponding radio frequency of the $i$th TOA, and $K_{\rm{DM}} = (2\pi m_ec^2)/e^2$ is the dispersion constant. Eq.~\eqref{eq:SW_SNR} is an optimistic definition as it doesn't include the contribution of other noise terms in the denominator. We subtract the average value $\overline{\rm{DM}}_{\rm SW}$, which will be covariant with the fit for time-independent DM, when defining the DM deviation. The only extrinsic term in Eq.~\eqref{eq:SW_SNR} is the electron density $n_E$, which can be factored out, while all remaining terms can be inferred from each pulsar's observations and are readily computed or given by functions in \texttt{enterprise} and \texttt{enterprise\_extensions}.

\begin{figure}
    \centering
    \includegraphics[width=\linewidth]{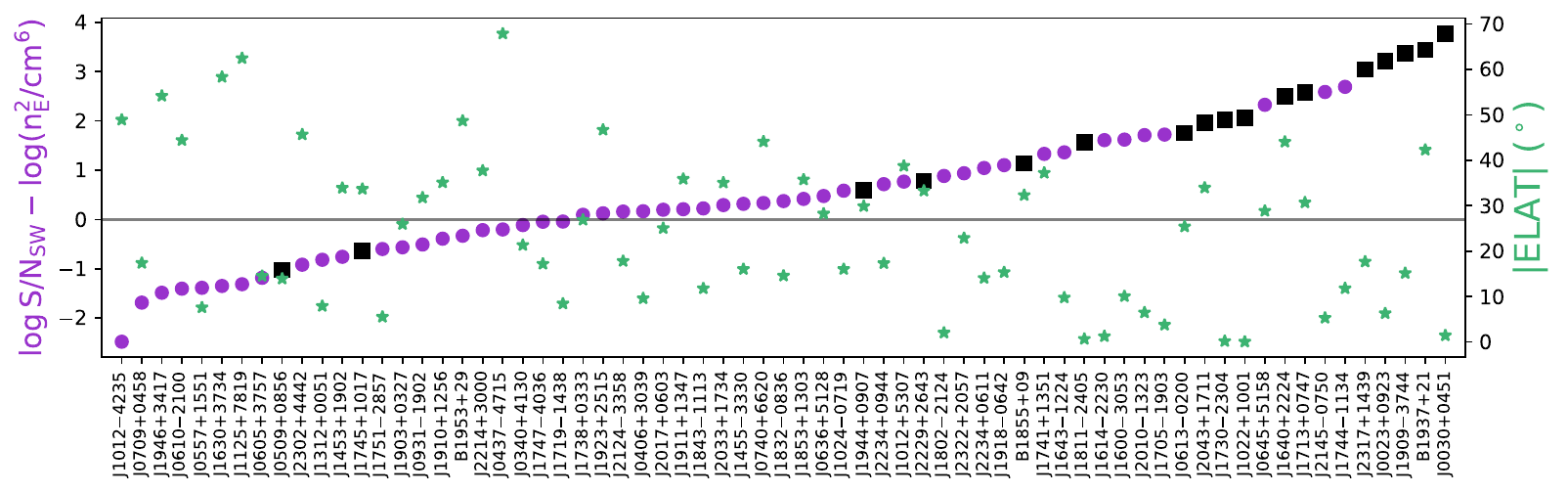}
    \caption{Pulsars are ranked on the $x$-axis according to their theoretical SW sensitivity $\rm{S/N}_{\rm{SW}}$, Eq~\eqref{eq:SW_SNR}, indicated by the purple circles. For pulsars where a significant SWGP is detected, $\rm{S/N}_{\rm{SW}}$ is instead given by the black squares. For comparison, the green stars indicate the absolute value of the pulsar's ecliptic latitude \textsc{Elat}. The horizontal line indicates the nominal value ${\rm{S/N}}_{\rm SW} = 1$ for $n_E = 1$ cm$^{-3}$. Overall, the expected anti-correlation between $\rm{S/N}_{\rm{SW}}$ and \textsc{Elat} is largely washed out by the highly non-uniform properties of each pulsar's dataset in NG15 from pulsar to pulsar, which introduce a large scatter.}
    \label{fig:SW_SNR}
\end{figure}

Fig.~\ref{fig:SW_SNR} shows all 67 pulsars in NG15 ranked from least to most sensitive to the SW based on the value of $\rm{S/N}_{\rm SW}$ for each pulsar. One purpose of this Figure is to cross-check our model selection results, i.e., we expect we are more likely to detect SW $n_E(t)$ variations in pulsars with higher SW sensitivity. The 17 pulsars which showed evidence for a significant SWGP are indicated by the black squares in Fig.~\ref{fig:SW_SNR}. These are mostly clumped towards the high-$\rm{S/N}_{\rm SW}$ end of the spectrum, matching our theoretical expectations overall. There are two pulsars (J0509+0856, J1745+1017) where we found a significant SWGP but their value of ${\rm{S/N}}_{\rm SW}$ lies below the threshold where we expect detections should be possible. This suggests that the SWGP component of these two models could instead be fitting to DM variations sourced by the ISM. As both of these pulsars also have short timespans $T_{\rm psr} \le 4.5$ yr, it may be that these timespans are too short to effectively lift the degeneracy between SW and ISM-induced DM variations.

The five pulsars which are ranked most sensitive to the SW signal are (in order), PSRs J0030+0451, B1937+21, J1909$-$3744, J0023+0923, and J2317+1439. PSR B1937+21 is somewhat of a surprise in the sense that it is not an ecliptic pulsar, with \textsc{Elat}$ = 42.3^\circ$. Nonetheless, Eq.~\eqref{eq:SW_SNR} reveals that it is the 2nd most sensitive pulsar to SW in the dataset, which owes to its disproportionately large number of TOAs, small TOA cadence, and small TOA measurement uncertainties, compared to other pulsars in NG15. This justifies the use of PSR B1937+21's preferred model, which includes a SWGP component. PSR J1640+2224 likewise prefers a SWGP in spite of its large \textsc{Elat} $=44.1^\circ$, but this is explained as it is still ranked the 9th most sensitive pulsar to the SW signal in NG15.

For comparison, Fig.~\ref{fig:SW_SNR} also shows the value of \textsc{Elat} for each pulsar alongside its ${\rm S/N}_{\rm SW}$ value. While there is a slight anti-correlation between \textsc{Elat} and ${\rm S/N}_{\rm SW}$, the scatter is much more pronounced, which is reflective of the highly non-uniform frequency coverage, data volume, and measurement uncertainty distribution across different pulsars in NG15. As such, for future NANOGrav datasets we recommend the use of Eq.~\eqref{eq:SW_SNR} or a similar technique for assessing relative sensitivity to the SW among different pulsars, rather than simply using ecliptic latitude.

While Eq.~\eqref{eq:SW_SNR} is straightforward to compute numerically with real data, we can build intuition for how SW sensitivity scales with different data properties using a few assumptions. First, we assume a continuous TOA distribution, $t_i \to t$. 
Next we compute the impact angle $\theta$ as
\begin{align}
    \theta = \arccos\left(\hat{p}\cdot\hat{r}\right),
\end{align}
where $\hat{p}$ and $\hat{r}$ are the unit vectors from the Earth to the pulsar and the Earth to the Sun respectively. $\hat{p}$ is effectively constant over time, but includes the value \textsc{Elat} that importantly weights the SW's effect on the timing. Without loss of generality in the final expression, we can define $\hat{p}$ so that the ecliptic and the $xy$-plane co-align, and the pulsar lies on the $xz$-plane in the hemisphere above the ecliptic, such that
\begin{align}
    \hat{p} &= [\cos\beta, 0, \sin\beta],
\end{align}
where we define $\beta \equiv \Elat$. Meanwhile, $\hat{r}$ undergoes annual variations,
\begin{align}
    \hat{r} &= [\cos(2\pi(t - T_c)/{\rm{yr}}), \sin(2\pi(t - T_c)/{\rm{yr}}), 0],
\end{align}
where $T_c$ is the time of the closest approach between the pulsar and the Sun during a given year. Using these definitions, $\theta$ reduces to
\begin{align}
    \theta = \arccos(\cos\beta\cos(2\pi(t - T_c)/{\rm yr})).
\end{align}
Using Eq.~\eqref{eq:sw_dm}, the SW-induced DM variation is then
\begin{align}
    {\rm DM}_{\rm SW} = n_E(1\;{\rm{AU}})\frac{\pi - \arccos(\cos\beta\cos(2\pi(t - T_c)/{\rm yr}))}{\sqrt{1 - \cos^2\beta\cos^2(2\pi(t - T_c)/{\rm yr})}}.
\end{align}
Assuming a continuous TOA distribution 
allows us to compute the mean DM as the integral
\begin{align}
    \overline{\rm DM}_{\rm SW} = \frac{n_E(1\;{\rm AU})}{\rm{yr}}\int_{T_c}^{T_c + 1{\rm yr}} dt\frac{\pi - \arccos(\cos\beta\cos(2\pi(t - T_c)/{\rm yr}))}{\sqrt{1 - \cos^2\beta\cos^2(2\pi(t - T_c)/{\rm yr})}}.
\end{align}
Variable substitution yields
\begin{align}
    \overline{\rm DM}_{\rm SW} = \frac{n_E(1\;{\rm AU})}{2\pi}\int_0^{2\pi} dx\frac{\pi - \arccos(k\cos x)}{\sqrt{1 - k^2\cos^2x}},
\end{align}
where $x = 2\pi(t - T_c)/{\rm yr}$, and $k = \cos\beta$. The integral can be simplified by first using the reflective symmetry of the numerator about the angle $\pi$ and to yield an integral over the interval $[0,\pi]$,
\begin{align}
    \overline{\rm DM}_{\rm SW} = \frac{n_E(1\;{\rm AU})}{2\pi}\int_0^{\pi} dx\frac{2\pi - \arccos(k\cos x) - \arccos(-k\cos x)}{\sqrt{1 - k^2\cos^2x}},
\end{align}
and then the identity $\arccos(k\cos x) + \arccos(-k\cos x) = \pi$ yields
\begin{align}
    \overline{\rm DM}_{\rm SW} = \frac{n_E(1\;{\rm AU})}{2}\int_0^{\pi} dx\frac{1}{\sqrt{1 - k^2\cos^2x}},
\end{align}
The remaining integral is related to the complete elliptic integral of the first kind, defined as $K(k) = \int_0^{\pi/2} (1 - k^2\sin^2\theta)^{-1/2} d\theta$ (not to be confused with the dispersion constant $K_{\rm DM}$), which may be expressed as an infinite series of Legendre polynomials $K(k) = (\pi/2)\sum_{n=0}^\infty(P_{2n}(0))^2k^{2n}$. The result is
\begin{align}
    \overline{\rm DM}_{\rm SW} = n_E(1\;{\rm AU})K(\cos\beta).
\end{align}
Eq.~\eqref{eq:SW_SNR} also requires we compute the variance of the SW signal, which we can approximate again assuming the limit of a continuous TOA distribution. The relevant quantity to compute is
\begin{align}
    \overline{{\rm DM}^2}_{\rm SW}(t) &= \int_{T_c}^{T_c + \rm{1yr}}\frac{dt}{\rm{yr}}{\rm DM}^2_{\rm SW}(t) = \frac{n_E^2(1\;{\rm AU})^2}{\rm{yr}}\int_{T_c}^{T_c + 1{\rm yr}} dt\frac{\left(\pi - \arccos(\cos\beta\cos(2\pi(t - T_c)/{\rm yr})\right)^2}{1 - \cos^2\beta\cos^2(2\pi(t - T_c)/{\rm yr})},
\end{align}
which simplifies using the previous substitutions as
\begin{align}
    \overline{{\rm DM}^2}_{\rm SW}(t) &= \frac{n_E^2(1\;{\rm AU})^2}{2\pi}\int_0^{2\pi} dx\frac{(\pi^2 - 2\pi\arccos(k\cos x) + \arccos^2(k\cos x))^2}{1 - k^2\cos^2x}.
\end{align}
Following the previous steps, the first two terms in the numerator cancel after integration, leaving
\begin{align}
    \overline{{\rm DM}^2}_{\rm SW}(t) &= \frac{n_E^2(1\;{\rm AU})^2}{2\pi}\int_0^{2\pi} dx\frac{\arccos^2(k\cos x)}{1 - k^2\cos^2x}.
\end{align}





This second moment does not reduce to standard elliptic integrals, but we can derive the asymptotic behavior analytically in two limits. For pulsars far from the ecliptic ($\beta \to 90^\circ$, $k \to 0$), we have $k = \cos\beta \ll 1$. Expanding the DM expression to leading order in $k$,
\begin{align}
    \arccos(k\cos x) &\approx \frac{\pi}{2} - k\cos x, \\
    \sqrt{1 - k^2\cos^2 x} &\approx 1,
\end{align}
yields
\begin{align}
    {\rm DM}_{\rm SW} \approx n_E\left(\frac{\pi}{2} + k\cos x\right).
\end{align}
The annual mean is $\overline{\rm DM}_{\rm SW} = n_E \pi/2$, since $\overline{\cos x} = 0$ over a full period. The second moment is
\begin{align}
    \overline{{\rm DM}^2}_{\rm SW} = n_E^2 \left(\frac{\pi^2}{4} + k^2\overline{\cos^2 x}\right) = n_E^2 \left(\frac{\pi^2}{4} + \frac{k^2}{2}\right),
\end{align}
giving a variance
\begin{align}
    \sigma_{\rm SW}^2 = \overline{{\rm DM}^2}_{\rm SW} - \overline{{\rm DM}}_{\rm SW}^2 = \frac{n_E^2 k^2}{2} = \frac{n_E^2 \cos^2\beta}{2}.
\end{align}
The standard deviation therefore scales as
\begin{equation}
    \sigma_{\rm SW} = \frac{n_E}{\sqrt{2}}\cos\beta, \qquad \beta \gtrsim 70^\circ.
    \label{eq:sigma_high_lat}
\end{equation}

For pulsars near the ecliptic ($\beta \to 0$, $k \to 1$), the DM variation is dominated by the singularity at solar conjunction ($x = 0$), where the line of sight passes closest to the Sun. For small $\beta$ and small $x$, we have $1 - k^2\cos^2 x \approx \beta^2 + x^2$ and $\arccos(k\cos x) \approx \sqrt{\beta^2 + x^2}$, so the DM reduces to
\begin{align}
    {\rm DM}_{\rm SW} \approx n_E \frac{\pi - \sqrt{\beta^2 + x^2}}{\sqrt{\beta^2 + x^2}}.
\end{align}
The second moment integral is dominated by the region near $x = 0$ where $\sqrt{\beta^2 + x^2} \ll \pi$,
\begin{align}
    \overline{{\rm DM}_{\rm SW}^2} \approx \frac{n_E^2}{2\pi} \int_{-\infty}^{\infty} \frac{\pi^2}{\beta^2 + x^2}\, dx = \frac{n_E^2 \pi^2}{2\pi} \cdot \frac{\pi}{\beta} = \frac{n_E^2 \pi^2}{2\beta}.
\end{align}
Meanwhile, $\overline{\rm DM}_{\rm SW} = n_E K(k) \approx n_E \ln(4/\beta)$ for $k \to 1$, so $\overline{\rm DM}_{\rm SW}^2 \propto \ln^2(1/\beta)$ grows more slowly than $1/\beta$. The standard deviation is therefore dominated by the second moment,
\begin{equation}
    \sigma_{\rm SW} \sim \frac{n_E \pi}{\sqrt{2\beta}}, \qquad \beta \to 0.
    \label{eq:sigma_low_lat}
\end{equation}
These results lead to the final ${\rm S/N}_{\rm SW}$ scalings for small $\beta \to 0$,
\begin{align}
    {\rm S/N}_{\rm SW} \propto \frac{n_E^2}{\sigma_{\rm TOA}^2\nu^4}\beta^{-1}
\end{align}
or for $\beta > 70^\circ$,
\begin{align}
    {\rm S/N}_{\rm SW} \propto \frac{n_E^2}{\sigma_{\rm TOA}^2\nu^4}\cos^2\beta.
\end{align}

\begin{figure*}
    \centering
    \includegraphics[width=\textwidth]{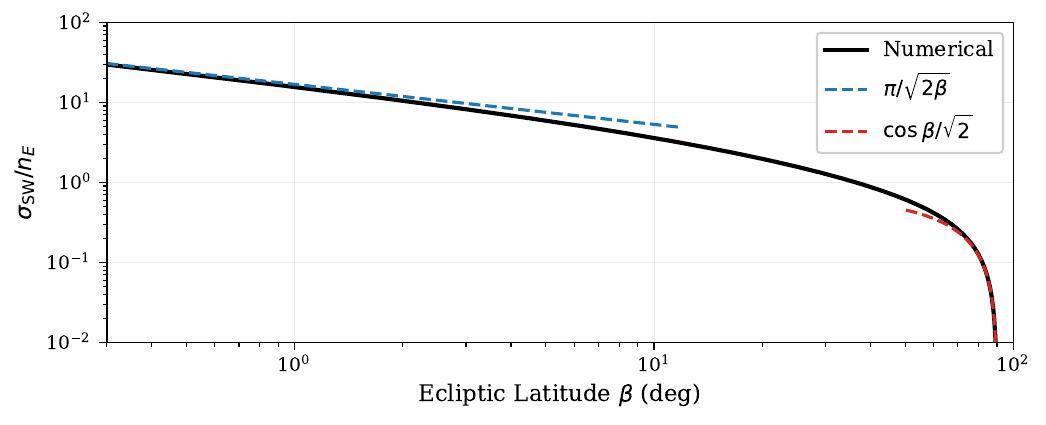}
    \caption{Asymptotic scaling of solar wind sensitivity with ecliptic latitude. The normalized standard deviation $\sigma_{\rm SW}/n_E$ computed numerically (solid) compared to the asymptotic forms $\pi/\sqrt{2\beta}$ (blue dashed; Eq.~\ref{eq:sigma_low_lat}) and $\cos\beta/\sqrt{2}$ (red dashed; Eq.~\ref{eq:sigma_high_lat}). The high-latitude scaling is accurate to $\lesssim 5\%$ for $\beta \gtrsim 70^\circ$.}
    \label{fig:sw_scaling}
\end{figure*}

Figure~\ref{fig:sw_scaling} compares these asymptotic forms to the full numerical integration for the standard deviation $\sigma_{\rm SW}$. The high-latitude approximation (Eq.~\ref{eq:sigma_high_lat}) is accurate to $\lesssim 10\%$ for $\beta \gtrsim 70^\circ$. The low-latitude scaling (Eq.~\ref{eq:sigma_low_lat}) captures the divergence toward the ecliptic, with deviations at moderate $\beta$ arising from the $\ln^2(1/\beta)$ correction to the variance.

\end{document}